\documentclass[%
reprint,
superscriptaddress,
nofootinbib,
amsmath,amssymb,
aps,
prc,
floatfix,
]{revtex4-2}

\usepackage{graphicx}% Include figure files
\usepackage{dcolumn}% Align table columns on decimal point
\usepackage{bm}% bold math
\usepackage{color}
\usepackage{here}
\usepackage{hyperref}% add hypertext capabilities
\usepackage{nicefrac}
\usepackage{longtable}
\usepackage{lipsum}
\usepackage{ulem}
\usepackage{multirow}
\newcommand{\nuc}[2]{\hbox{$^{#1}$#2}}

\newcommand{\reac}{\nuc{94}{Zr}(\nuc{18}{O},\nuc{16}{O})\nuc{96}{Zr} }

\newcommand{\zeroplustwo}{$1581.64(6)$} 	%0+2
\newcommand{\zeroplusthree}{$2695.18(3)$} 	%0+3
\newcommand{\zeroplusfour}{$2925.55(3)$} 	%0+4

\newcommand{\twoplusone}{$1750.497(15)$} 	%2+1
\newcommand{\twoplustwo}{$2225.846(17)$}	%2+2
\newcommand{\twoplusthree}{$2668.82(4)$}	%2+3

\newcommand{\threeminusone}{$1897.158(16)$}	%3-1
\newcommand{\fiveminusone}{$3119.87(3)$}		%5-1
\newcommand{\sevenminusone}{$4234.7(5)$}		%7-1

\newcommand{\fourplusone}{$2857.373(23)$} 	%4+1
\newcommand{\fourplustwo}{$3082.36(3)$} 	%4+2

\newcommand{\sixplusone}{$3309.19(9)$} 	%6+1
\newcommand{\sixplustwo}{$3483.44(9)$} 	%6+2
\newcommand{\sixplusthree}{$3772.2(4)$} 	%6+3

\newcommand{\onesevenfivezero}{$1750.42(2)$}	%2+1 -> 0+1
\newcommand{\onefoursix}{$146.653(10)$}		%3-1 -> 2+1
\newcommand{\threetwoeight}{$328.75(3)$}		%2+2 -> 3-1
\newcommand{\foursevenfive}{$475.33(1)$}		%2+2 -> 2+1
\newcommand{\sixfourfour}{$644.18(6)$}			%2+2 -> 0+2
\newcommand{\oneonezerosix}{$1106.88(2)$}		%4+1 -> 2+1
\newcommand{\ninesixzero}{$960.9(2)$}			%4+1 -> 3-1
\newcommand{\sixthreeone}{$631.45(4)$}			%4+1 -> 2+2
\newcommand{\oneoneeightfive}{$1185.19(3)$}	%4+2 -> 3-1
\newcommand{\sixfourthree}{$643.9(2)$}			%4+2 -> 3+1
\newcommand{\twotwofour}{$224.8$}			%4+2 -> 4+1 no unc given
\newcommand{\onethreethreeone}{$1331.8(2)$}	%4+2 -> 2+1
\newcommand{\eightfivesix}{$856.6(2)$}			%4+2 -> 2+2
\newcommand{\nineonefour}{$914.85(5)$}		%6+3 -> 4+1 from Wis23a
\newcommand{\oneoneonefour}{$1114.6$}		%7-1 -> 5-1 no unc given

\newcommand{\nineoneeightZr}{$918.74(5)$}		%2+1 -> 0+1 94Zr
\newcommand{\oneonethreeeightZr}{$1138.88(10)$}	%3-1 -> 2+1 94Zr
\newcommand{\oneoneonetwoMo}{$1110.77(7)$}	%5-1 -> 4+1 98Mo

\newcommand{\tautwoplusone}{0.63(5)_{\text{stat}}(7)_{\text{syst}}}	%2+1
\newcommand{\tautwoplustwo}{1.53(6)_{\text{stat}}(5)_{\text{syst}}}	%2+2
\newcommand{\tautwoplustwoalt}{1.2(2)_{\text{stat}}(5)_{\text{syst}}}	%2+2 from transition to 3-1
\newcommand{\taufourplusone}{1.22(4)_{\text{stat}}(6)_{\text{syst}}}	%4+1
\newcommand{\taufourplustwo}{1.84(4)_{\text{stat}}(24)_{\text{syst}}}	%4+2

\newcommand{\tauthreeminusoneZr}{114_{-6}^{+5}}	%3-1 94Zr

\newcommand{\Btwoplustwozeroplusone}{0.19(1)} 	%2+2 -> 0+1 in W.u.
\newcommand{\Btwoplustwozeroplustwo}{26(2)}	%2+2 -> 0+2 in W.u.
\newcommand{\Btwoplustwotwoplusone}{0.098(6)}	%2+2 -> 2+1 in muN2
\newcommand{\Bfourplusonetwoplusone}{11.2(8)}	%4+1 -> 2+1 in W.u.
\newcommand{\Bfourplusonetwoplustwo}{39(7)}	%4+1 -> 2+2 in W.u.
\newcommand{\Bfourplustwotwoplusone}{0.30(5)}	%4+2 -> 2+1 in W.u.
\newcommand{\Bfourplustwotwoplustwo}{1.7(4)}	%4+2 -> 2+2 in W.u.
\newcommand{\Bfourplustwofourplusone}{0.20(3)}	%4+2 -> 4+1 in muN2

\newcommand{\Bfourtwo}{1.5(3)}

\newcommand{\BtwoplusonezeroplusoneIBM}{2.57} 		%2+1 -> 0+1 in W.u.
\newcommand{\BtwoplusonezeroplustwoIBM}{1.49} 		%2+1 -> 0+2 in W.u.
\newcommand{\BtwoplustwozeroplusoneIBM}{0.001} 	%2+2 -> 0+1 in W.u.
\newcommand{\BtwoplustwozeroplustwoIBM}{28.8} 		%2+2 -> 0+2 in W.u.
\newcommand{\BfourplusonetwoplusoneIBM}{3.13}		%4+1 -> 2+1 in W.u.
\newcommand{\BfourplusonetwoplustwoIBM}{43.5}		%4+1 -> 2+2 in W.u.
\newcommand{\BtwoplustwotwoplusoneIBM}{0.024}		%2+2 -> 2+1 in muN2

\newcommand{\QtwoplusoneIBM}{5.8}		%Q(2+1) in efm2
\newcommand{\QtwoplustwoIBM}{21.9}		%Q(2+2) in efm2
\newcommand{\QfourplusoneIBM}{37.1}		%Q(4+1) in efm2
\newcommand{\BtwoplusonezeroplusoneIBMmix}{2.90} 		%2+1 -> 0+1 in W.u.
\newcommand{\BtwoplusonezeroplustwoIBMmix}{3.91} 		%2+1 -> 0+2 in W.u.
\newcommand{\BtwoplustwozeroplusoneIBMmix}{0.001} 		%2+2 -> 0+1 in W.u.
\newcommand{\BtwoplustwozeroplustwoIBMmix}{26.0} 		%2+2 -> 0+2 in W.u.
\newcommand{\BfourplusonetwoplusoneIBMmix}{8.57}		%4+1 -> 2+1 in W.u.
\newcommand{\BfourplusonetwoplustwoIBMmix}{37.5}		%4+1 -> 2+2 in W.u.
\newcommand{\BtwoplustwotwoplusoneIBMmix}{0.058}		%2+2 -> 2+1 in muN2

\newcommand{\BtwoplusonezeroplusoneQVSM}{1.57} 		%2+1 -> 0+1 in W.u.
\newcommand{\BtwoplusonezeroplustwoQVSM}{10.1} 		%2+1 -> 0+2 in W.u.
\newcommand{\BtwoplustwozeroplusoneQVSM}{0.08} 		%2+2 -> 0+1 in W.u.
\newcommand{\BtwoplustwozeroplustwoQVSM}{25.7} 		%2+2 -> 0+2 in W.u.
\newcommand{\BfourplusonetwoplusoneQVSM}{15.3}		%4+1 -> 2+1 in W.u.
\newcommand{\BfourplusonetwoplustwoQVSM}{37.7}		%4+1 -> 2+2 in W.u.
\newcommand{\BfourplustwotwoplusoneQVSM}{0.02}		%4+2 -> 2+1 in W.u.
\newcommand{\BfourplustwotwoplustwoQVSM}{0.0}			%4+2 -> 2+2 in W.u.
\newcommand{\BthreeminusonezeroplusoneQVSM}{52.9}	%3-1 -> 0+1 in W.u.

\newcommand{\BtwoplustwotwoplusoneQVSM}{0.097}		%2+2 -> 2+1 in muN2
\newcommand{\BfourplustwofourplusoneQVSM}{0.000018}	%4+2 -> 4+1 in muN2

\newcommand{\QtwoplusoneQVSM}{29.3}		%Q(2+1) in efm2
\newcommand{\QtwoplustwoQVSM}{43.7}		%Q(2+2) in efm2
\newcommand{\QfourplusoneQVSM}{59.3}		%Q(4+1) in efm2
\newcommand{\QfourplustwoQVSM}{-2.71}		%Q(4+2) in efm2

\newcommand{\tautwoplusthree}{0.35_{-0.14}^{+0.46}}		%(2+3)
\newcommand{\tauzeroplusthree}{40(10)}					%0+3
\newcommand{\tauzeroplusfour}{29(20)}					%0+4
\newcommand{\BthreeminusonezeroplusoneNDS}{57(4)}		%3-1 -> 0+1 in W.u.
\newcommand{\BthreeminusonezeroplusoneIskra}{42(3)}		%3-1 -> 0+1 in W.u.

\begin{document}

%\preprint{APS/123-QED}

\title{Sizeable mixing between the coexisting normal and intruder structures of \nuc{96}{Zr} revealed through lifetime measurements}
%\title{First study of the coexisting normal and intruder structures in \nuc{96}{Zr} up to $J=4$}% Force line breaks with \\

%
\author{T.~Beck}
\email[Corresponding author: ]{tobias.beck@kuleuven.be}
%\altaffiliation[Present address: ]{Instituut voor Kern- en Stralingsfysica, KU Leuven, B-3001 Leuven, Belgium}
\affiliation{Technische Universit\"at Darmstadt, Department of Physics, Institute for Nuclear Physics, 64289 Darmstadt, Germany}
\affiliation{Instituut voor Kern- en Stralingsfysica, KU Leuven, B-3001 Leuven, Belgium}
\author{T.~Stetz}
\affiliation{Technische Universit\"at Darmstadt, Department of Physics, Institute for Nuclear Physics, 64289 Darmstadt, Germany}
\author{N. Gavrielov}
\affiliation{Physics Department, Nuclear Research Center-Negev, P.O. Box 9001, Beer-Sheva, Israel}
%\altaffiliation[Present address: ]{Physics Department, Nuclear Research Center-Negev, P.O. Box 9001, Beer-Sheva, Israel}
%\affiliation{Racah Institute of Physics, The Hebrew University, Jerusalem 91904, Israel}
\author{N.~Pietralla}
\affiliation{Technische Universit\"at Darmstadt, Department of Physics, Institute for Nuclear Physics, 64289 Darmstadt, Germany}
\author{V.~Werner}
\affiliation{Technische Universit\"at Darmstadt, Department of Physics, Institute for Nuclear Physics, 64289 Darmstadt, Germany}
\author{K.~Yanase}
\altaffiliation[Present address: ]{Department of Physics, Chiba Institute of Technology, 2-1-1 Shibazono, Narashino, Chiba, 275-0023, Japan}
\affiliation{RIKEN Nishina Center, RIKEN, 2-1 Hirosawa, Wako-shi, 351-0198, Saitama, Japan}
\author{M.~Boromiza}
\affiliation{Horia Hulubei National Institute for R\&D in Physics and Nuclear Engineering, 
Strada Reactorului 30, POB MG6, RO-077125 Bucharest-M\u{a}gurele, Romania}
\author{A.~Coman}
\affiliation{Horia Hulubei National Institute for R\&D in Physics and Nuclear Engineering, 
Strada Reactorului 30, POB MG6, RO-077125 Bucharest-M\u{a}gurele, Romania}
\author{C.~Costache}
\affiliation{Horia Hulubei National Institute for R\&D in Physics and Nuclear Engineering, 
Strada Reactorului 30, POB MG6, RO-077125 Bucharest-M\u{a}gurele, Romania}
\author{N.~M.~Florea}
\affiliation{Horia Hulubei National Institute for R\&D in Physics and Nuclear Engineering, 
Strada Reactorului 30, POB MG6, RO-077125 Bucharest-M\u{a}gurele, Romania}
\author{I.~Gheorghe}
\affiliation{Horia Hulubei National Institute for R\&D in Physics and Nuclear Engineering, 
Strada Reactorului 30, POB MG6, RO-077125 Bucharest-M\u{a}gurele, Romania}
\author{A.~Ionescu}
\affiliation{Horia Hulubei National Institute for R\&D in Physics and Nuclear Engineering, 
Strada Reactorului 30, POB MG6, RO-077125 Bucharest-M\u{a}gurele, Romania}
%\author{\red{R.~Kern}}
%\affiliation{Technische Universit\"at Darmstadt, Department of Physics, Institute for Nuclear Physics, 64289 Darmstadt, Germany}
\author{R.~Lica}
\affiliation{Horia Hulubei National Institute for R\&D in Physics and Nuclear Engineering, 
Strada Reactorului 30, POB MG6, RO-077125 Bucharest-M\u{a}gurele, Romania}
\author{N.~M\u{a}rginean}
\affiliation{Horia Hulubei National Institute for R\&D in Physics and Nuclear Engineering, 
Strada Reactorului 30, POB MG6, RO-077125 Bucharest-M\u{a}gurele, Romania}
\author{R.~M\u{a}rginean}
\affiliation{Horia Hulubei National Institute for R\&D in Physics and Nuclear Engineering, 
Strada Reactorului 30, POB MG6, RO-077125 Bucharest-M\u{a}gurele, Romania}
\author{H.~Mayr}
\affiliation{Technische Universit\"at Darmstadt, Department of Physics, Institute for Nuclear Physics, 64289 Darmstadt, Germany}
\author{C.~Mihai}
\affiliation{Horia Hulubei National Institute for R\&D in Physics and Nuclear Engineering, 
Strada Reactorului 30, POB MG6, RO-077125 Bucharest-M\u{a}gurele, Romania}
\author{R.~E.~Mihai}
\affiliation{Horia Hulubei National Institute for R\&D in Physics and Nuclear Engineering, 
Strada Reactorului 30, POB MG6, RO-077125 Bucharest-M\u{a}gurele, Romania}
\author{A.~Mitu}
\affiliation{Horia Hulubei National Institute for R\&D in Physics and Nuclear Engineering, 
Strada Reactorului 30, POB MG6, RO-077125 Bucharest-M\u{a}gurele, Romania}
\author{C.~R.~Nita}
\affiliation{Horia Hulubei National Institute for R\&D in Physics and Nuclear Engineering, 
Strada Reactorului 30, POB MG6, RO-077125 Bucharest-M\u{a}gurele, Romania}
\author{T.~Otsuka}
\affiliation{Center for Nuclear Study, University of Tokyo, 7-3-1 Hongo, Bunkyo-ku, Tokyo, 113-0033, Japan}
\author{O.~Papst}
\affiliation{Technische Universit\"at Darmstadt, Department of Physics, Institute for Nuclear Physics, 64289 Darmstadt, Germany}
\author{S.~Pascu}
\affiliation{Horia Hulubei National Institute for R\&D in Physics and Nuclear Engineering, 
Strada Reactorului 30, POB MG6, RO-077125 Bucharest-M\u{a}gurele, Romania}
%\author{N.~Shimizu}
%\affiliation{Center for Computational Sciences, University of Tsukuba, 1-1-1, Tenno-dai, Tsukuba, 305-8577, Ibaraki, Japan}
\author{C.~Sotty}
\affiliation{Horia Hulubei National Institute for R\&D in Physics and Nuclear Engineering, 
Strada Reactorului 30, POB MG6, RO-077125 Bucharest-M\u{a}gurele, Romania}
\author{L.~Stan}
\affiliation{Horia Hulubei National Institute for R\&D in Physics and Nuclear Engineering, 
Strada Reactorului 30, POB MG6, RO-077125 Bucharest-M\u{a}gurele, Romania}
\author{S.~Toma}
\affiliation{Horia Hulubei National Institute for R\&D in Physics and Nuclear Engineering, 
Strada Reactorului 30, POB MG6, RO-077125 Bucharest-M\u{a}gurele, Romania}
\author{Y.~Tsunoda}
\affiliation{Center for Computational Sciences, University of Tsukuba, 1-1-1 Tenno-dai, Tsukuba, 305-8577, Ibaraki, Japan}
\affiliation{Center for Nuclear Study, University of Tokyo, 7-3-1 Hongo, Bunkyo-ku, Tokyo, 113-0033, Japan}
\author{A.~Turturic\u{a}}
\affiliation{Horia Hulubei National Institute for R\&D in Physics and Nuclear Engineering, 
Strada Reactorului 30, POB MG6, RO-077125 Bucharest-M\u{a}gurele, Romania}
\author{S.~Ujeniuc}
\affiliation{Horia Hulubei National Institute for R\&D in Physics and Nuclear Engineering, 
Strada Reactorului 30, POB MG6, RO-077125 Bucharest-M\u{a}gurele, Romania}
\author{J.~Wiederhold}
\affiliation{Technische Universit\"at Darmstadt, Department of Physics, Institute for Nuclear Physics, 64289 Darmstadt, Germany}
%\author{\red{W.~Witt}}
%\affiliation{Technische Universit\"at Darmstadt, Department of Physics, Institute for Nuclear Physics, 64289 Darmstadt, Germany}

\date{\today}% It is always \today, today, but any date may be explicitly specified

\begin{abstract}
Lifetimes of excited states of the nucleus \nuc{96}{Zr} were studied with the Doppler-shift attenuation method
following the \reac~two-neutron transfer reaction.
The obtained \mbox{$B(E2;2^+_2\to0^+_2)$} and \mbox{$B(E2;4^+_1\to2^+_2)$} transition strengths,
amounting to $\Btwoplustwozeroplustwo$ and $\Bfourplusonetwoplustwo$\,W.u., respectively,
firmly attribute the $2^+_2$ and $4^+_1$ states to the collective intruder band built on top of the $0^+_2$ state.
The nature of this structure is discussed against predictions of the collective model using the Bohr Hamiltonian, 
the interacting boson model with configuration mixing, and the quasiparticle vacua shell model.
The collective, interband \mbox{$B(E2;4^+_1\to2^+_1)=\Bfourplusonetwoplusone$\,W.u.} transition strength,
refuting the previously assumed purity of both structures, is reproduced in the calculations.
The quasiparticle vacua shell model predicts, in agreement with a simple two-state model based on
experimental $E2$ transition strengths, considerable mixing between the normal and intruder $2^+$ states.
\end{abstract}

\pacs{Valid PACS appear here}% PACS, the Physics and Astronomy Classification Scheme.
%\keywords{Suggested keywords}%Use showkeys class option if keyword display desired
\maketitle

%%%%%%%%%%%%%%%%%%%%%%%%%%%%%%%%%%%%%%%%%%%%%%%%%%%%%%%%%%%%%%%
%	Introduction
%%%%%%%%%%%%%%%%%%%%%%%%%%%%%%%%%%%%%%%%%%%%%%%%%%%%%%%%%%%%%%%

\section{Introduction}

The zirconium isotopic chain has historically played a pivotal role in the understanding of the emergence 
of collective behavior from the single-particle properties of individual nucleons. 
In their seminal work, Federman and Pittel attempted an explanation of nuclear deformation in the $A\approx100$ mass region
from the proton-neutron correlations between spin-orbit partners~\cite{Fed79a}. 
The monopole part of the tensor interaction has subsequently been identified as quintessential for 
these attractive proton-neutron interactions driving the evolution of nuclear structure~\cite{Ots05a,Ots10a,Ots16a}. 
While this behavior has traditionally been associated with the addition of nucleons,
the influence of nucleon configurations within a single nucleus on the nuclear structure 
as a function of excitation energy has been highlighted more recently~\cite{Tog16a}. 
Following the interpretation of the rapid shape change found for zirconium isotopes near neutron number $N=60$ 
in the context of this so-called type-II shell evolution by Togashi~\textit{et al.}, 
the study of zirconium isotopes experienced a renaissance in the past decade; 
e.g. Refs.~\cite{Kre16a,Ans17a,Wit18a,Sin18a,Isk19a,Kar20a,Wis23a,Pfe23a,Wu24a}.\\
The rapid quantum phase transition from spherical ground states for \nuc{90-98}{Zr} to deformed ones for \nuc{100}{Zr}
and heavier isotopes has in the past been described theoretically using, notably, the shell model and 
Hartree-Fock-Bogoliubov methods~\cite{Fed79a}. 
A compilation of further approaches is found in Ref.~\cite{Pet12a}.
More recently, calculations using the Monte-Carlo shell model~\cite{Ots01a,Shi12a} (MCSM) and 
in the interacting boson model with configuration mixing~\cite{Duv81a,Duv82a} (IBM-CM) have shifted
the focus to excited-state (intruder) structures~\cite{Tog16a,Gar19a,Gav19a,Gav20a,Gav22a}. 
While both approaches are qualitatively in agreement for the ground-state structures with respect to excitation energies 
and $E2$ transition strengths, they show deviations for intruder configurations possibly rooted in the inclusion of more than two
configurations in the MCSM calculations. 

\begin{figure*}[t]
\centering
\includegraphics[trim=10 0 0 0,width=1.025\linewidth,clip]{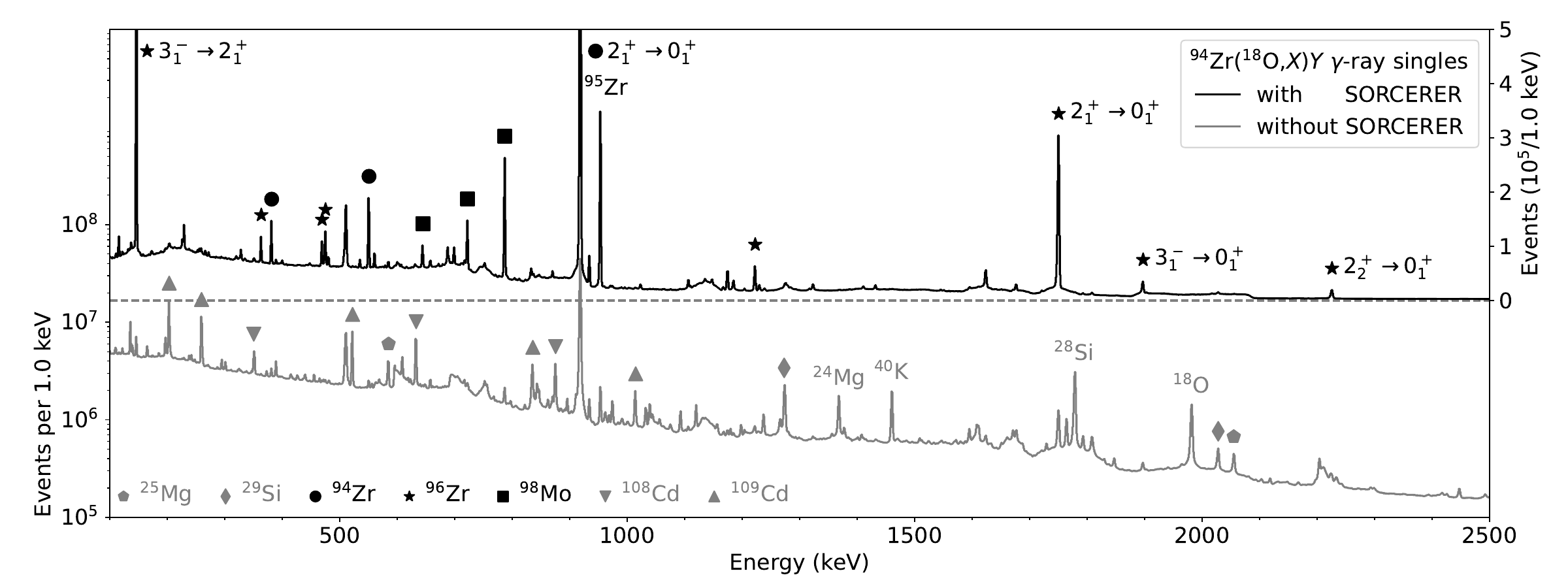}
\caption[]{Compton-suppressed $\gamma$-ray spectra of ROSPHERE with (black, top) and without (grey, bottom)
a condition on the photodiodes of SORCERER. 
The former spectrum, using the linear scale on the right, shows the strong suppression of $\gamma$ rays 
originating from nuclei created in fusion-evaporation reactions.
Known $\gamma$-ray lines of the target \nuc{94}{Zr} as well as \nuc{96}{Zr} and \nuc{98}{Mo}
produced in two-neutron and $\alpha$-transfer reactions are marked by circles,  asterisks, and squares, respectively.
}
\label{fig:spec}
\end{figure*}

For \nuc{96}{Zr}, the influence of the $N=56$ neutron sub-shell closure pushes the $2^+_1$ state to \twoplusone\,keV~\cite{NDS96}, 
leaving the intruder $0^+_2$ state at \zeroplustwo\,keV~\cite{NDS96} as the first excited state. 
In the MCSM it is predicted to be of triaxial nature~\cite{Tog16a} in contrast to the IBM-CM calculations 
where it is characterized by a predominantly spherical, U(5)-like configuration~\cite{Gav19a}. 
Both approaches reproduce the single-particle nature of the transition between the $2^+_1$ and ground states 
characterized by a small transition strength of $B(E2;2^+_1\to0^+_1)=2.3(3)$\,W.u.~\cite{Kre16a}. 
The experimental value for the $E2$ transition strength of the corresponding transition between the intruder $2^+_2$ and 
$0^+_2$ states is about one order of magnitude larger; a value of $39(6)$\,W.u. has recently been proposed~\cite{Zie26a}.
%; though with a relative uncertainty of about $30\%$. 
The minuscule transition strength $B(E2;2^+_2\to0^+_1)=0.26(8)$\,W.u.~\cite{Kre16a} connecting the normal (ground-state) and 
intruder configurations reveals their decoupled nature; a feature reflected by a small mixing between them 
in the calculated IBM-CM wave functions~\cite{Gav22a}.\\\\
The characterization of the intruder configuration requires new precision data, especially for the $E2$ transition
strengths between its members, to facilitate a comparison to theoretical models. 
A new calculation of \nuc{96}{Zr} with the quasiparticle vacua shell model~\cite{Shi21a,Shi25a} (QVSM)
proposes values of $\BtwoplustwozeroplustwoQVSM$ and $\BfourplusonetwoplustwoQVSM$\,W.u.
%for the $B(E2;2^+_2\to0^+_2)$ and $B(E2;4^+_1\to2^+_2)$ transition strengths, respectively, comparable to
for the $2^+\to0^+$ and  $4^+\to2^+$ $E2$ transitions in the intruder band, respectively, comparable to
IBM-CM predictions of $\BtwoplustwozeroplustwoIBM$ and $\BfourplusonetwoplustwoIBM$\,W.u.~\cite{Gav19a}. 
In the literature, the $4^+_1$ state at \fourplusone\,keV~\cite{NDS96} is assigned intruder origin due to its large transition strength 
to the $2^+_2$ state~\cite{Wit19a}. 
The latter has been calculated from the lifetime of $\tau(4^+_1)=860_{-260}^{+670}$\,fs determined with the
Doppler-shift attenuation method (DSAM) after inelastic neutron scattering~\cite{Bel89a}. 
The upper relative uncertainty exceeding $75\%$ leads to large fluctuations in the 
\mbox{$B_{4/2}=B(E2;4^+_1\to2^+_2)/B(E2;2^+_2\to0^+_2)$} fingerprint, 
including the values $B_{4/2}=2.0$ and $1.4$ obtained for a spherical vibrator and a rigid rotor, respectively. 
For the ratio of excitation energies relative to the $0^+_2$ state at \zeroplustwo\,keV \cite{NDS96}, 
a value of \mbox{$R_{4/2}=[E(4^+_1)-E(0^+_2)]/[E(2^+_2)-E(0^+_2)]=1.98$} is found which suggests vibrational character. 
In order to address these open questions on the nature of the intruder structure of \nuc{96}{Zr}, which were put forward in Ref.~\cite{Wit19a}, 
its $4^+$ and, if possible, $6^+$ members have to be identified unambiguously and 
new precision lifetime measurements of the $2^+_2$ and $4^+_1$ states are needed.

%%%%%%%%%%%%%%%%%%%%%%%%%%%%%%%%%%%%%%%%%%%%%%%%%%%%%%%%%%%%%%%
%	Experiment and results
%%%%%%%%%%%%%%%%%%%%%%%%%%%%%%%%%%%%%%%%%%%%%%%%%%%%%%%%%%%%%%%

\section{Experiment and results}

\begin{figure}[tb]
\centering
\includegraphics[trim=0 0 0 0,width=1.025\linewidth,clip]{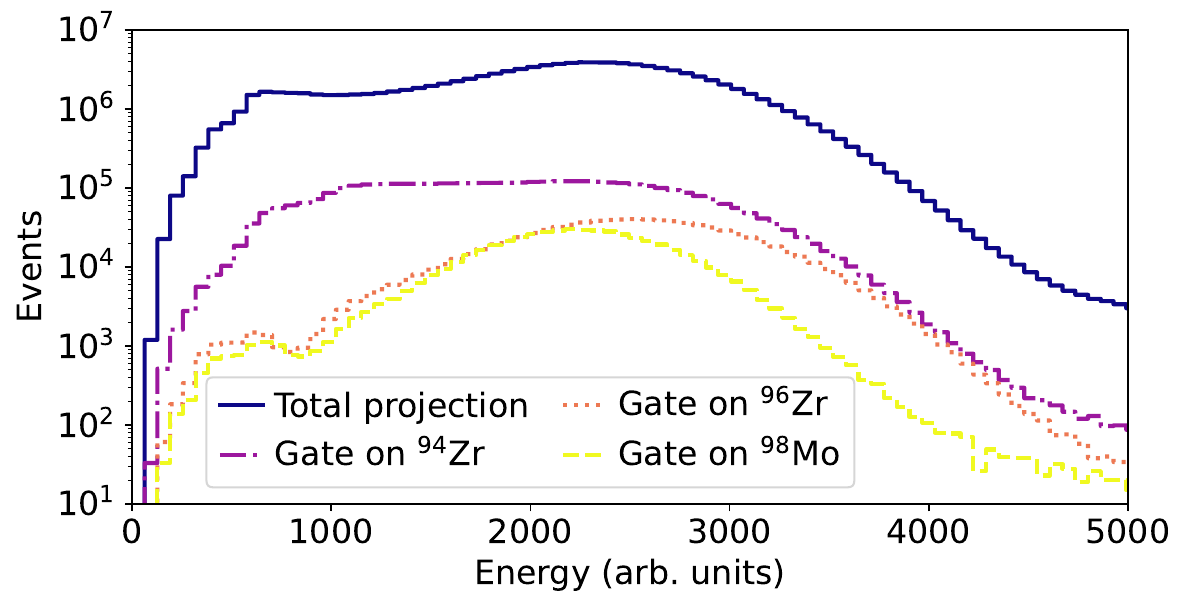}
\caption[]{Energies of reaction products in SORCERER from particle-$\gamma$ correlations.
The spectra for \nuc{94,96}{Zr} and \nuc{98}{Mo} are obtained
from conditions in the $\gamma$-ray spectrum on the $2^+_1\to0^+_1$ transitions 
at $918.74(5)$~\cite{NDS94}, $1750.42(2)$~\cite{NDS96}, 
and $787.372(20)$\,keV~\cite{NDS98}, respectively.
Due to the broad, overlapping distributions, no clean reaction-channel selection
based on the particle energy detected in SORCERER is feasible in the present work.
}
\label{fig:spec_sorcerer}
\end{figure}

\begin{figure*}[t]
\centering
\includegraphics[trim=0 0 0 0,width=1.025\linewidth,clip]{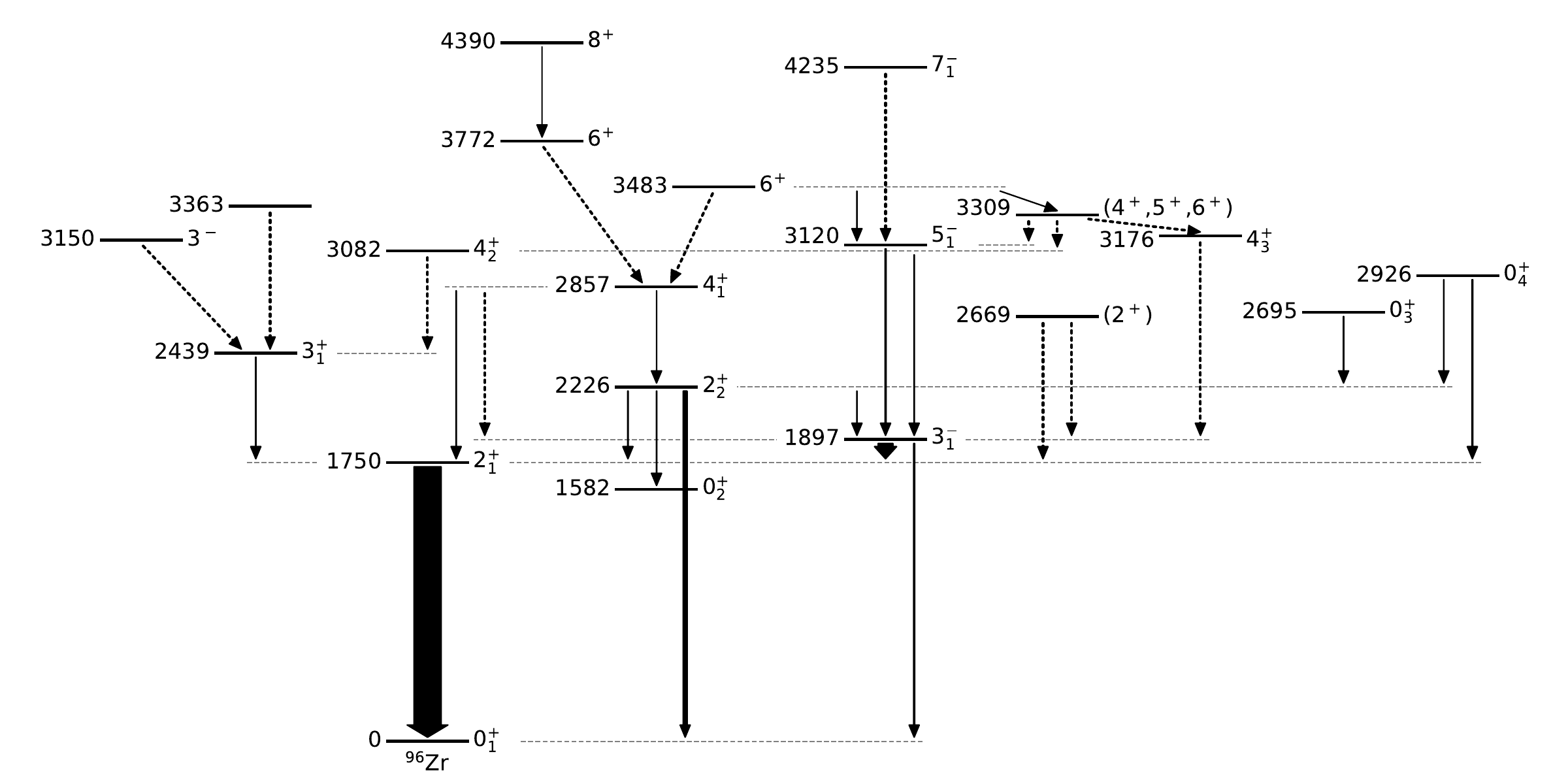}
\caption[]{Level scheme of \nuc{96}{Zr} as populated by the \reac reaction.
Arrow widths reflect peak intensities and are taken relative to the $2^+_1\to0^+_1$ transition
at \onesevenfivezero\,keV~\cite{NDS96}.
Dotted arrows denote transitions only observed in $\gamma\gamma$-coincidence spectra.
All energies, which are rounded to full keV, and quantum numbers are taken from Ref.~\cite{NDS96}.
}
\label{fig:levelscheme}
\end{figure*}

In the present work, the \reac two-neutron transfer reaction was used to populate excited states of \nuc{96}{Zr}. 
The incident beam of \nuc{18}{O} ions was accelerated by the $9$\,MV FN Pelletron tandem accelerator~\cite{Dob09a}
at the Horia Hulubei National Institute of Physics and Nuclear Engineering (IFIN-HH) in Bucharest-M\u{a}gurele, Romania, 
to $49$\,MeV; approximately $2$\,MeV below the Coulomb barrier of the projectile-target system. 
The thickness of the reaction target, enriched to $96.49(1)\%$ in \nuc{94}{Zr}, was $7.22$\,mg/cm$^2$, 
resulting in a stopping of target-like reaction products within the target to allow for the extraction of excited-state lifetimes 
with the Doppler-shift attenuation method~\cite{Ale78a,Nol79a}. 
Projectile-like reaction products, such as \nuc{16}{O} and \nuc{14}{C} after two-neutron and $\alpha$ transfer reactions, respectively, 
were detected with the SORCERER array~\cite{Bec20b} at backward angles with respect to the incident beam direction. 
The prototype of the array used in the present experiment consisted of seven silicon photodiodes in a circular arrangement, 
covering polar angles between approximately $122^{\circ}$ and $164^{\circ}$.
%$121.7^{\circ}$ and $163.5^{\circ}$.
For detection of $\gamma$ rays emitted by the excited reaction products, 
the target was placed in the center of the ROSPHERE array~\cite{Buc16a}, 
consisting of $25$ Compton-suppressed high-purity germanium (HPGe) detectors 
arranged in five rings at polar angles of  $37^{\circ}$, $70^{\circ}$, $90^{\circ}$, $110^{\circ}$, and $143^{\circ}$ 
with respect to the beam axis. 
The summed $\gamma$-ray spectrum of all detectors is shown in Fig.~\ref{fig:spec}. 
Significant contributions to it stem from fusion-evaporation reactions of beam and target nuclei; 
in accordance with calculations using the computer code PACE4~\cite{Gav80a,Tar08a,lise}
$\gamma$ rays of \nuc{108,109}{Cd} are observed most prominently.
Furthermore, a variety of lighter-mass nuclei such as \nuc{24,25}{Mg} or \nuc{28,29}{Si} are produced
from reactions with the mounting structures of the target and SORCERER.
Requiring a coincidence condition with the SORCERER array, these fusion-evaporation events are suppressed~\cite{Bec20b}.
As evident from Fig.~\ref{fig:spec_sorcerer}, the prototype of SORCERER used in the present experiment 
lacked the energy resolution to enable a selection of individual reaction channels.
As a consequence, the spectrum of Fig.~\ref{fig:spec} is dominated by transitions from excited states of \nuc{94}{Zr} 
populated in Coulomb excitation as well as \nuc{96}{Zr} and \nuc{98}{Mo} 
produced by two-neutron and $\alpha$-transfer reactions, respectively. 
For the nucleus of interest, \nuc{96}{Zr}, $31$ transitions between $20$ nuclear levels have been observed and 
identified in the HPGe energy spectra.
Their placement in the known level scheme~\cite{NDS96} is facilitated through energies obtained from HPGe singles spectra 
and the analysis of $\gamma\gamma$ coincidence relationships. 
All observed $\gamma$-ray transitions are placed in the level scheme shown in Fig.~\ref{fig:levelscheme} 
with arrow widths reflecting their relative intensities.

The determination of level lifetimes using DSAM requires precise information 
on the stopping process of ions inside the target material. 
The excited \nuc{96}{Zr} ions produced in the \reac reaction are moving in forward direction 
with respect to the beam axis with an initial velocity of $v/c\approx0.03$ and
are continuously slowed down within the target until at rest, 
resulting in an angular straggling of approximately $7.5^{\circ}$. 
The respective velocity distribution of this process is simulated with the program \texttt{StopSim}~\cite{Sta17a} using
stopping powers calculated with \texttt{SRIM}~\cite{Zie10a}. 
%Following the recipe of Lindhard~\textit{et al.}~\cite{Lin63a}, an empirical downscale factor of $f_{\text{nucl}}=0.7$ 
%is applied to the nuclear stopping power. 
The $\gamma$-ray lineshape analysis is performed individually for the five rings of ROSPHERE,
i.e. the sets of detectors characterized by the same angle $\vartheta$ with respect to the beam axis,
using the program \texttt{APCAD}~\cite{Sta17a}. 
The \texttt{MIGRAD} minimizer was used for all values presented in this work in combination with the \texttt{MINOS} algorithm 
to provide statistical uncertainties~\cite{Jam94a}. 
The calculated stopping powers, which are identified as the main sources of systematic uncertainties, 
were varied in the simulation process by $\pm5\%$ and $\pm10\%$ for the electronic and nuclear components, respectively.
The energy-dependency of the detector resolution was calibrated with a function of the form $f(E)=a+b\sqrt{E}$.
A summary of the results obtained from the experimental data is given in Table~\ref{tab:results}.
\begin{table*}[t]
\caption{Compilation of experimental data for the $2^+_{1,2}$ and $4^+_{1,2}$ states of \nuc{96}{Zr}.
Angular-momentum and parity quantum numbers as well as excitation energies 
%and literature values for branching ratios 
are taken from Ref.~\cite{NDS96}.
Transitions marked with an asterisk are used for the lineshape analysis (cf. Fig.~\ref{fig:lineshapes}).
Reduced transition strengths for radiation character $\lambda L$, which are calculated using branching ratios and, 
if available, multipole mixing ratios from Ref.~\cite{NDS96},
are given in $10^{-5}e^2$fm$^2$ for $E1$, $\mu_N^2$ for $M1$, and W.u. for $E2$
with statistical and systematic uncertainties coming from the lifetime determination added in quadrature.
Predictions of transition strengths obtained from the collective model using the Bohr Hamiltonian~\cite{Mar20b},
the IBM-CM~\cite{Gav19a,Gav20a} with two choices for the mixing parameter $\omega$, and the QVSM are given.
}
\label{tab:results}
\begin{ruledtabular}
\begin{tabular}{ccccccccccc}
$J^{\pi}$	&$E_x$		&$\tau$			&$J^{\pi}_f$	&$E_{\gamma}$				&$\lambda L$		&\multicolumn{5}{c}{$B(\lambda L,J^{\pi}\to J_f^{\pi})$}\\\cline{7-11}
		&(keV)		&(ps)			&			&(keV)						&				&Experiment 				&Coll. model	&\multicolumn{2}{c}{IBM-CM}	&QVSM\\\cline{9-10}
		&			&				&			&							&				&						&			&$\omega=0.02$\,MeV&$\omega=0.04$\,MeV	&\\\colrule
$2^+_1$	&\twoplusone	&$\tautwoplusone$	&$0^+_1$		&\onesevenfivezero$^{\star}$		&$E2$			&$2.9(4)$					&$5.23$		&$\BtwoplusonezeroplusoneIBM$	&$\BtwoplusonezeroplusoneIBMmix$	&$\BtwoplusonezeroplusoneQVSM$\\
%$2^+_1$	&\twoplusone	&$\tautwoplusone$	&$0^+_1$		&\onesevenfivezero$^{\star}$		&$E2$			&$2.9(4)$					&$5.23$		&	&	&\\
&&									&$0^+_2$		&							&$E2$			&$^a$					&			&$\BtwoplusonezeroplustwoIBM$	&$\BtwoplusonezeroplustwoIBMmix$	&$\BtwoplusonezeroplustwoQVSM$\\
$2^+_2$	&\twoplustwo	&$\tautwoplustwo$	&$0^+_1$		&\twoplustwo					&$E2$			&\Btwoplustwozeroplusone	&$0.39$		&$\BtwoplustwozeroplusoneIBM$	&$\BtwoplustwozeroplusoneIBMmix$	&$\BtwoplustwozeroplusoneQVSM$\\
&&									&$2^+_1$		&\foursevenfive$^{\star}$			&$M1$			&$\Btwoplustwotwoplusone$$^b$			&$0.071$		&$\BtwoplustwotwoplusoneIBM	$	&$\BtwoplustwotwoplusoneIBMmix$		&$\BtwoplustwotwoplusoneQVSM$\\
&&									&$0^+_2$		&\sixfourfour					&$E2$			&\Btwoplustwozeroplustwo	&$26.0$		&$\BtwoplustwozeroplustwoIBM$	&$\BtwoplustwozeroplustwoIBMmix$		&$\BtwoplustwozeroplustwoQVSM$\\
&&									&$3^-_1$		&\threetwoeight					&$E1$			&$81(7)$$^c$				&&&&\\
$4^+_1$$^i$	&\fourplusone	&$\taufourplusone$	&$2^+_1$		&\oneonezerosix$^{\star}$			&$E2$			&$11.2(8)$$^e$			&$16.7$		&$\BfourplusonetwoplusoneIBM$	&$\BfourplusonetwoplusoneIBMmix$		&$\BfourplusonetwoplusoneQVSM$\\
&&									&$3^-_1$		&\ninesixzero					&$E1$			&$6(2)$					&&&&\\
&&									&$2^+_2$		&\sixthreeone					&$E2$			&$39(7)$$^f$				&$43.0$		&$\BfourplusonetwoplustwoIBM$	&$\BfourplusonetwoplustwoIBMmix$		&$\BfourplusonetwoplustwoQVSM$\\
$4^+_2$$^i$	&\fourplustwo	&$\taufourplustwo$	&$3^-_1$		&\oneoneeightfive$^{\star}$		&$E1$			&$15(2)$$^g$				&&$^h$		&$^h$	&\\
&&									&$4^+_1$		&\twotwofour					&$M1$			&$0.20(3)$$^k$			&&$^h$		&$^h$	&$\BfourplustwofourplusoneQVSM$\\
&&									&$2^+_1$		&\onethreethreeone				&$E2$			&$0.30(5)$				&&$^h$		&$^h$	&$\BfourplustwotwoplusoneQVSM$\\
&&									&$3^+_1$		&\sixfourthree					&$M1$			&$0.006(1)$$^k$			&&$^h$		&$^h$	&\\
&&									&$2^+_2$		&\eightfivesix					&$E2$			&$1.7(4)$					&&$^h$		&$^h$	&$\BfourplustwotwoplustwoQVSM$\\
\end{tabular}
\end{ruledtabular}
$^a$ Not observed~\cite{NDS96}.\\
$^b$ Calculated using $\delta=-0.09_{-0.02}^{+0.01}$~\cite{NDS96}.\\
$^c$ Calculated using $\delta=-0.02(5)$~\cite{NDS96}.\\
$^d$ Taken from Ref.~\cite{Bel89a}.\\
$^e$ Calculated using $\delta=-0.03(3)$~\cite{NDS96}.\\
$^f$ Calculated using $\delta=-0.02(8)$~\cite{NDS96}.\\
$^g$ Calculated using $\delta=+0.02(3)$~\cite{NDS96}.\\
$^h$ The $4^+_2$ state lies outside the IBM-CM model space.\\
$^i$ The $4^+_1$ state emerges as the $4^+_2$ state in the QVSM calculations and vice versa.\\
$^k$ Assuming pure $M1$.
\end{table*}
In the following, selected states are discussed individually.\\
\begin{figure*}[t]
\centering
\includegraphics[trim=50 0 40 0,width=1.0\linewidth,clip]{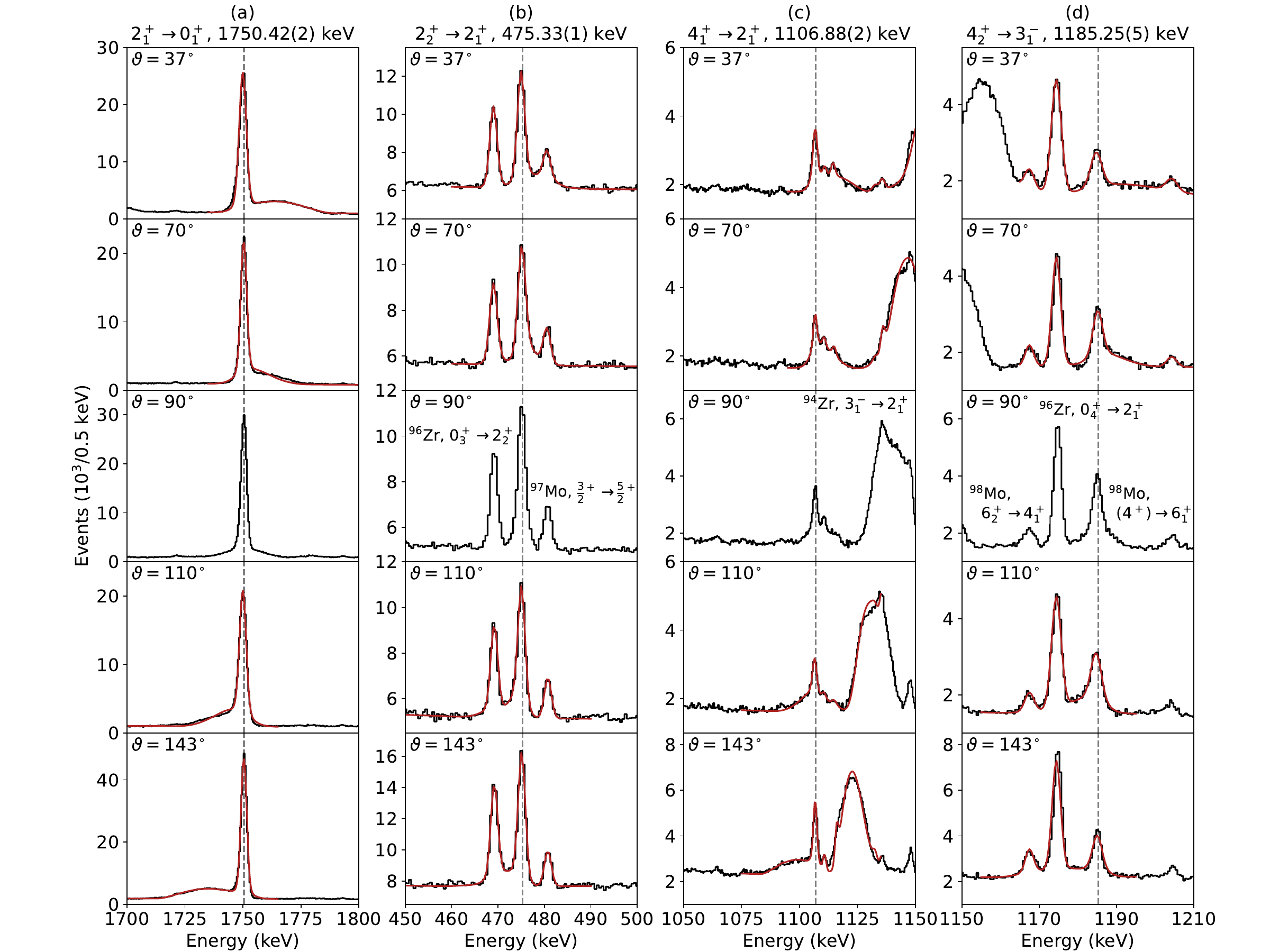}
\caption[]{Overview of particle-gated $\gamma$-ray spectra for transitions depopulating the
$2^+_{1,2}$ and $4^+_{1,2}$ states (columns) at different detection angles of ROSPHERE (lines)
together with adapted lineshapes from \texttt{APCAD}~\cite{Sta17a}.
The \nuc{96}{Zr} transitions of interest are indicated by dashed vertical lines.
For detection angles $\vartheta=90^{\circ}$, where no Doppler shifts are expected,
the spectrum without the simulated lineshape is shown and major contaminant transitions are identified.
The small peaks seen on the high-energy side of the \oneonezerosix-keV transition in column~(c)
are identified as the $7^-_1\to5^-_1$ and $5^-_1\to4^+_1$ transitions of \nuc{96}{Zr} and \nuc{98}{Mo} 
at \oneoneonefour~\cite{NDS96} and \oneoneonetwoMo\,keV~\cite{NDS98}, respectively.
The lifetimes obtained for the individual rings are collected in Fig.~\ref{fig:lifetimes}.
%\red{As soon as I have the individual lifetimes for Fig.~\ref{fig:lifetimes}, I will also add the lineshape fit
%of the \threetwoeight-keV transition for the $2^+_2$ state.}
}
\label{fig:lineshapes}
\end{figure*}
The two most intense $\gamma$ rays of \nuc{96}{Zr} visible in the spectra shown in Fig.~\ref{fig:spec} originate from 
the $3^-_1\to2^+_1$ and $2^+_1\to0^+_1$ transitions at \onefoursix~and \onesevenfivezero\,keV~\cite{NDS96}, respectively.
As evident from the level scheme shown in Fig.~\ref{fig:levelscheme}, the $2^+_1$ state is populated by
a multitude of higher-lying states; some of which without known lifetimes~\cite{NDS96}. 
In the latter case, their lifetime is used as an additional free parameter in the DSAM analysis;
ultimately yielding $\tau(2^+_1)=\tautwoplusone$\,ps [cf. column~(a) of Fig.~\ref{fig:lineshapes}].
This value is in agreement with $0.45(19)$\,ps~\cite{NDS96} obtained in an early Coulomb-excitation experiment
using a \nuc{14}{N} beam
but smaller than the lifetime of $0.82(10)$\,ps found in Ref.~\cite{Kum03a} from DSAM after projectile Coulomb excitation,
for which only feeding from the $3^-_1$ state has been considered.
A more recent result from DSAM after inelastic neutron scattering reported $\tau(2^+_1)=770_{-130}^{+180}$\,fs~\cite{Pet13a},
in agreement with both, the present result and the value of Ref.~\cite{Kum03a}.\\
The dominant feeding contribution to the $2^+_1$ state is attributed to the $3^-_1$ state at \threeminusone\,keV~\cite{NDS96}.
No lineshape is observed, indicating that its depopulating $\gamma$ rays are predominantly emitted at rest, 
consistent with its comparatively long lifetime of $98(6)$\,ps~\cite{NDS96}.\\
For the $2^+_2$ state at \twoplustwo\,keV, the known depopulating transitions to the $0^+_{1,2}$, $2^+_1$, and $3^-_1$ states
have been observed with their determined branching ratios in perfect agreement with literature values~\cite{NDS96}.
An enhancement of the decay path to the $2^+_1$ state, reported as $90(5)\%$ and $70.1(14)\%$ 
in Refs.~\cite{Wis23a} and \cite{Zie26a}, respectively, is not found.
From the known \mbox{$B(E2;2^+_2\to0^+_1)$} transition strength, determined in inelastic electron scattering~\cite{Kre16a},
a lifetime \mbox{$\tau(2^+_2)=1.1_{-0.3}^{+0.5}$\,ps} is deduced~\cite{Wit19a},
significantly improving on the previously known upper limit of $10$\,ps~\cite{Mac90a}.
In the present work, the lifetime of the $2^+_2$ state is determined independently using its depopulating transitions to the 
$3^-_1$ and $2^+_1$ states at \threetwoeight~and \foursevenfive\,keV~\cite{NDS96},
yielding consistent lifetimes of $\tautwoplustwoalt$ and $\tautwoplustwo$\,ps, respectively.
The latter value, which comes with greatly reduced uncertainty when compared to previous results, 
leads to transition strengths \mbox{$B(E2;2^+_2\to0^+_1)=\Btwoplustwozeroplusone$} 
and \mbox{$B(E2;2^+_2\to0^+_2)=\Btwoplustwozeroplustwo$\,W.u.}
with statistical and systematic uncertainties added in quadrature.
These results are, however, smaller than the ones extracted from light ion-induced Coulomb excitation~\cite{Zie26a}.
Reproduction of the latter would require an approximately $35\%$ reduction in the lifetime of the $2^+_2$ state,
corresponding to more than six standard deviations.\\
Sizeable collectivity might be expected for the $(2^+)$ state at \twoplusthree\,keV~\cite{NDS96}
based on its known lifetime of $\tautwoplusthree$\,ps~\cite{Bel89a}.
In the present work, however, a determination of the state's lifetime is not possible
since its dominant decay branch to the $2^+_1$ state is at virtually the same energy as the $2^+_1\to0^+_1$ transition of \nuc{94}{Zr}
and the decay channels to the $2^+_2$ and $3^-_1$ states are too weak for a lineshape analysis.\\
%Based on the known level lifetime $\tautwoplusthree$\,ps~\cite{Bel89a} of the $(2^+)$ state at \twoplusthree\,keV, 
%a potentially collective $(2^+)\to2^+_2$ transition at \fourfourtwo\,keV might place the state as a member
%of the two-phonon multiplet together with the $0^+_3$ and $4^+_1$ states.
%which might show strong collectivity based on the available lifetime ,
%
In the literature, the first $4^+$ state of \nuc{96}{Zr} is postulated at $2750(15)$\,keV~\cite{NDS96}
based on results from particle spectroscopy following the \nuc{100}{Mo}($d$,\nuc{6}{Li})\nuc{96}{Zr} reaction~\cite{vdB84a}.
The authors confirmed a tentative $L=4$ assignment for this state from Flynn \textit{et al.}
who studied inelastic triton scattering off \nuc{92,94,96}{Zr}~\cite{Fly70a}.
In the latter work, which was the first one to report the observation of a level at $2750(15)$\,keV, however, 
its affiliation with the level scheme of \nuc{96}{Zr} is dismissed,
suspecting it to originate from a \nuc{90}{Zr} contaminant found in the scattering target.
Indeed, a $3^-$ state of \nuc{90}{Zr} is known at an excitation energy of $2747.875(16)$\,keV~\cite{NDS90}. 
Neither in the present experiment, nor in any other recent one~\cite{NDS96,Wis23a}, evidence for the existence 
of a potential $4^+$ state of \nuc{96}{Zr} at $2750(15)$\,keV is found.
Following the strategy put forward in Ref.~\cite{Wit19a}, it is omitted in the ensuing discussion.
Hence, the $4^+$ state at \fourplusone\,keV~\cite{NDS96} is, based on the discussion above, 
denoted the first $4^+$ state of \nuc{96}{Zr} throughout this work.
Since it is predominantly populated directly by the two-neutron transfer reaction,
with feeding through the higher-lying $6^+_{2,3}$ states amounting to only about $2\%$,
and a coincidence condition on the flight component of the $2^+_1\to0^+_1$ does not have sufficient statistics,
the DSAM analysis can only be performed on the $4^+_1\to2^+_1$ transition at \oneonezerosix\,keV~\cite{NDS96} 
in the $\gamma$-ray singles spectra.
In these, three contaminant transitions from \nuc{94,96}{Zr} and \nuc{98}{Mo} are observed
in the energy region of the lineshape [cf. column~(c) of Fig.~\ref{fig:lineshapes}].
In forward beam direction, the $7^-_1\to5^-_1$ and $5^-_1\to4^+_1$ transitions
of \nuc{96}{Zr} and \nuc{98}{Mo} at \oneoneonefour~\cite{NDS96} and \oneoneonetwoMo\,keV~\cite{NDS98}, respectively, 
are considered in the lineshape analysis.
For their respective parent states, no lifetime information is available in the literature.
Since a simultaneous description of the $4^+_1\to2^+_1$ transition's lineshape in all detection angles of ROSPHERE 
is only achieved for emission of the contaminant $\gamma$ rays at rest,
this assumption is maintained in the ring-wise analysis.
For detectors placed in backward beam direction, the $3^-_1\to2^+_1$ transition of \nuc{94}{Zr} 
at \oneonethreeeightZr\,keV~\cite{NDS94}, 
which exhibits only a flight component due to its short lifetime of $\tau=127_{-9}^{+10}$\,fs~\cite{Pet13a},
has to be taken into account for a description of the stopped component's high-energy side.
Its lifetime, obtained from DSAM following the $(n,n'\gamma)$ reaction~\cite{Pet13a},
is in good agreement with $\tau=\tauthreeminusoneZr$\,fs determined in the present work.
The combined lineshape analysis of the $4^+_1\to2^+_1$ transition of \nuc{96}{Zr} together with the
above-mentioned contaminants is shown in column~(c) of Fig.~\ref{fig:lineshapes}.
The resulting value $\tau(4^+_1)=\taufourplusone$\,ps [cf. panel~(c) of Fig.~\ref{fig:lifetimes}]
is in agreement with $0.86_{-0.26}^{+0.67}$ reported in Ref.~\cite{Bel89a}
but significantly more precise.
\\
For the $4^+_2$ state at \fourplustwo\,keV~\cite{NDS96} a lower limit on the lifetime of $2$\,ps 
is reported from DSAM measurements after inelastic neutron scattering~\cite{Bel89a}.
%Since the transition to the $3^+_1$ state coincides with the $2^+_2\to0^+_2$ $\gamma$ ray,
%the branching ratio is determined from a coincidence with the $3^+_1\to2^+_1$ transition.
The lineshape analysis based on the dominant decay path to the $3^-_1$ state
yields an adopted lifetime \mbox{$\tau(4^+_2)=\taufourplustwo$\,ps}, in agreement with this limit.
As evident from panel~(d) of Fig.~\ref{fig:lifetimes}, spectra from detectors 
placed at polar angles $\vartheta=70^{\circ}$ and $110^{\circ}$ yielded lifetimes above $2$\,ps 
in contrast to the more extreme forward and backward angles.
An appropriate systematic uncertainty is introduced to account for this discrepancy.\\
Three candidates for $6^+$ states at \sixplusone, \sixplustwo, and \sixplusthree\,keV~\cite{NDS96}
are populated in the present work (cf. Fig.~\ref{fig:levelscheme}).
Based on its decay characteristics, a likely $(5^+)$ assignment has recently been proposed
for the $3309$-keV state~\cite{Wis23a}.
For the state at \sixplustwo\,keV~\cite{NDS96} a lifetime $\tau(6^+)=36(13)$\,ps
is given in the data sheets~\cite{NDS96}.
%\red{While indications of a lineshape might be visible for the $6^+\to5^-_1$ transition at \threesixthree\,keV~\cite{NDS96},
%it is not pronounced enough to perform a reliable DSAM analysis. NO!}
The $6^+_3$ state at \sixplusthree\,keV~\cite{NDS96} is depopulated mostly through a \nineonefour-keV~\cite{Wis23a} transition
to the $4^+_1$ state; it is, however, obfuscated by the dominant $2^+_1\to0^+_1$ transition of \nuc{94}{Zr} at \nineoneeightZr\,keV~\cite{NDS94}
and, thus, not suitable for a lineshape investigation.\\
For $\gamma$-ray transitions of the $0^+_{3,4}$ states at \zeroplusthree~and \zeroplusfour\,keV~\cite{NDS96}, respectively,
no lineshape is observed, suggesting that they predominantly decay at rest. 
Since no feeding transitions are observed, it can be concluded that their lifetimes exceed $10$\,ps
in agreement with literature values of \tauzeroplusthree~and \tauzeroplusfour\,ps~\cite{NDS96}.\\
Of the negative-parity band on top of the $3^-_1$ state at \threeminusone\,keV, also the $5^-_1$ and $7^-_1$ members
at \fiveminusone~and \sevenminusone\,keV~\cite{NDS96}, respectively, are populated. 
While statistics do not suffice for a reliable lifetime analysis of the latter state,
%\red{(there is a value in the MSc thesis, though)},
no lineshape is visible for the $5^-_1$ state despite its reported short lifetime of $\tau(5^-_1)=840_{-300}^{+980}$\,fs~\cite{Bel89a}.
%Hence, the $5^-_1$ state either has a lifetime larger than previously thought
%or is not populated directly by the \reac two-neutron transfer reaction used in the present work.
%\red{Check intensities - this might mean that it is maybe populated through unresolved feeding... not too likely.}\\

%
\begin{figure}[t]
\centering
\includegraphics[trim=0 0 0 0,width=1.025\linewidth,clip]{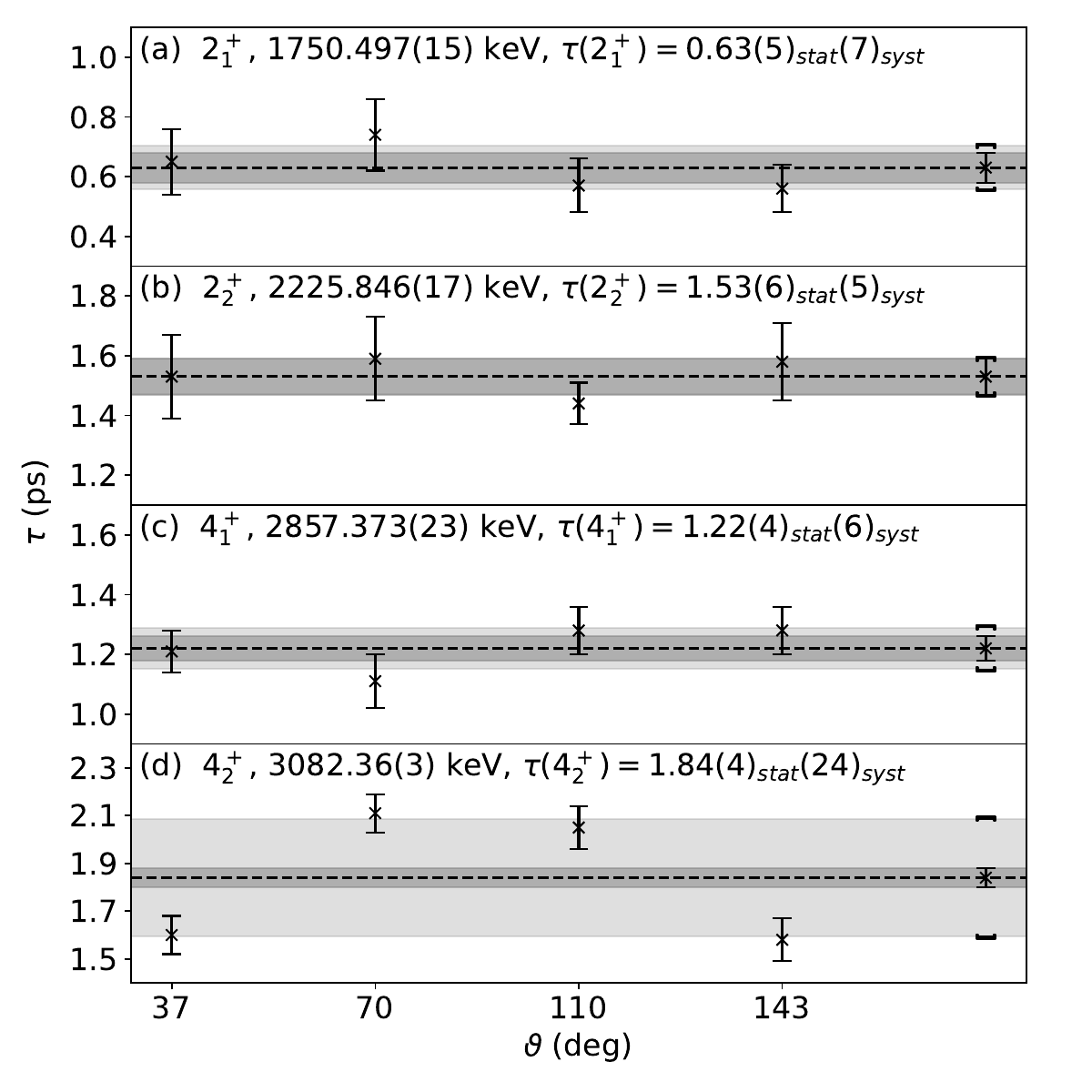}
\caption[]{Lifetimes of the $2^+_{1,2}$ and $4^+_{1,2}$ states determined individually for the ROSPHERE rings 
at polar angles $37^{\circ}$, $70^{\circ}$, $110^{\circ}$, and $143^{\circ}$ (cf. Fig.~\ref{fig:lineshapes}).
The systematic uncertainties of the adopted values, indicated by the data points on the right, 
reflect the spread of results obtained for individual rings.
%\red{Update lifetimes. Maybe also include the values for the $2^+_2$ state from the other transition in panel~(b).}
}
\label{fig:lifetimes}
\end{figure}
%

%%%%%%%%%%%%%%%%%%%%%%%%%%%%%%%%%%%%%%%%%%%%%%%%%%%%%%%%%%%%%%%
%	Discussion
%%%%%%%%%%%%%%%%%%%%%%%%%%%%%%%%%%%%%%%%%%%%%%%%%%%%%%%%%%%%%%%

\section{Discussion}

\begin{figure}[t]
\centering
\includegraphics[trim=0 0 0 7,width=1.025\linewidth,clip]{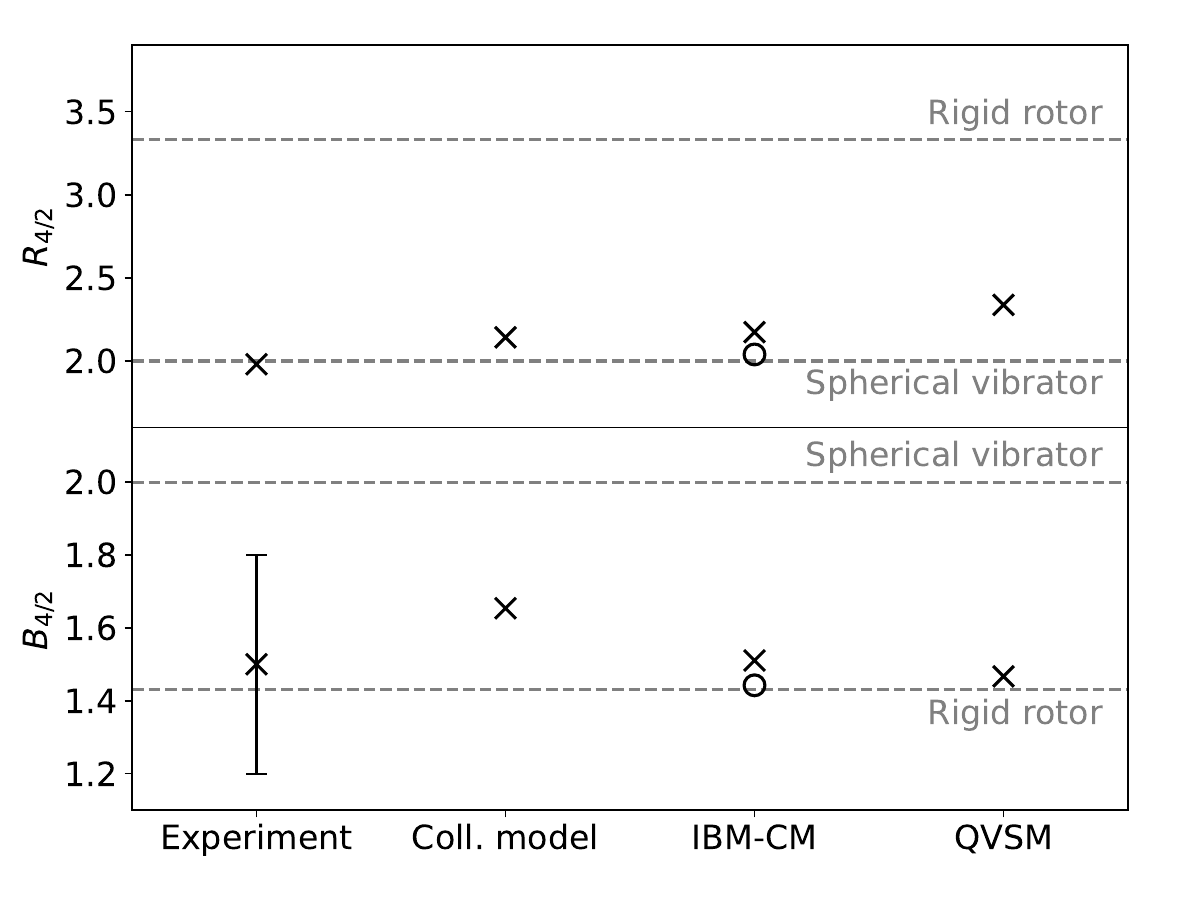}
\caption[]{Comparison of the $R_{4/2}$ and $B_{4/2}$ signatures of the intruder structure of \nuc{96}{Zr}
to predictions from the collective model~\cite{Mar20b}, the interacting boson model with configuration mixing~\cite{Gav19a,Gav20a} (IBM-CM),
and the quasiparticle vacua shell model (QVSM).
The circles represent $R_{4/2}$ and $B_{4/2}$ values obtained from IBM-CM calculations 
with the mixing parameter increased to $\omega=0.04$\,MeV.}
\label{fig:R42}
\end{figure}
\begin{figure*}[t]
\centering
\includegraphics[trim=0 0 0 0,width=1.025\linewidth,clip]{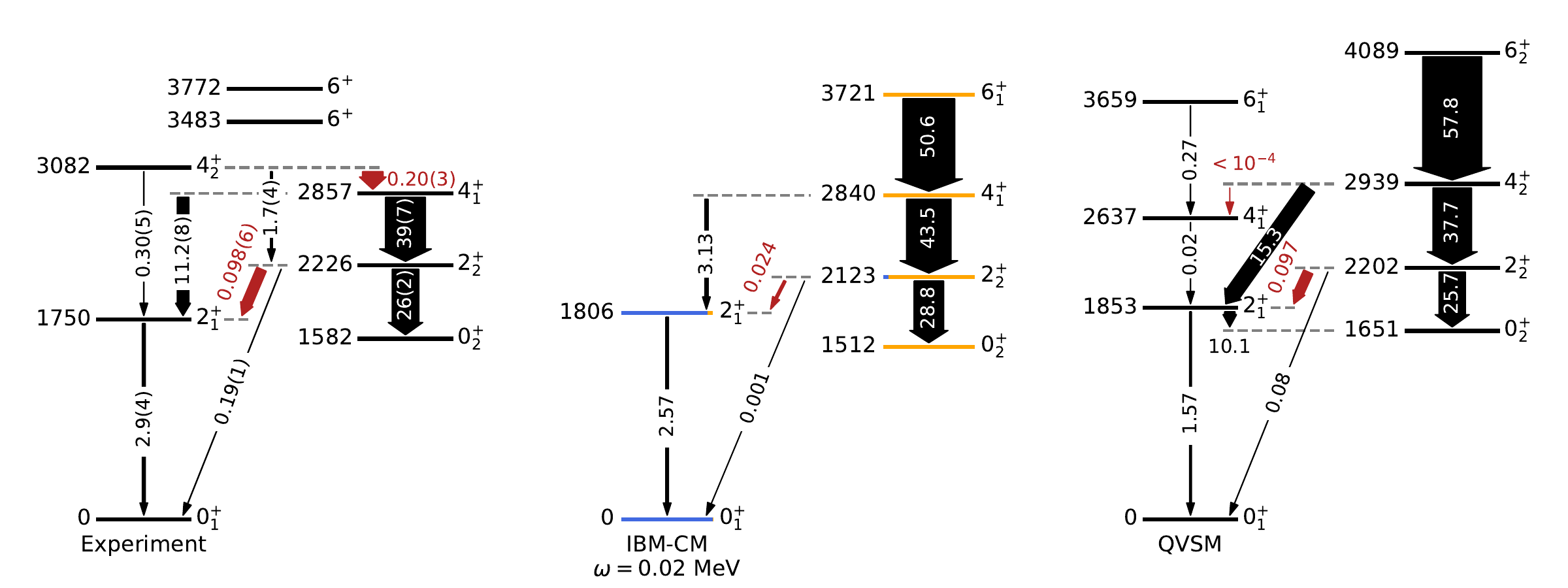}
\caption[]{Comparison of the experimental level scheme with predictions from IBM-CM and QVSM.
Excitation energies are given in keV and the strengths of $M1$ and $E2$ transitions in $\mu_N^2$ and W.u., respectively.
The thickness of the arrows, which are shown in red and black for $M1$ and $E2$ transitions, respectively,
scale with the transition strength. 
The lines marking states obtained from the IBM-CM calculations are color-coded by the fractions of
$[N=3]$ (blue) and $[N=5]$ (orange) configurations in their wave functions.
}
\label{fig:comp_exp_theo}
\end{figure*}
\begin{figure}[tb]
\centering
\includegraphics[trim=0 0 0 0,width=1.025\linewidth,clip]{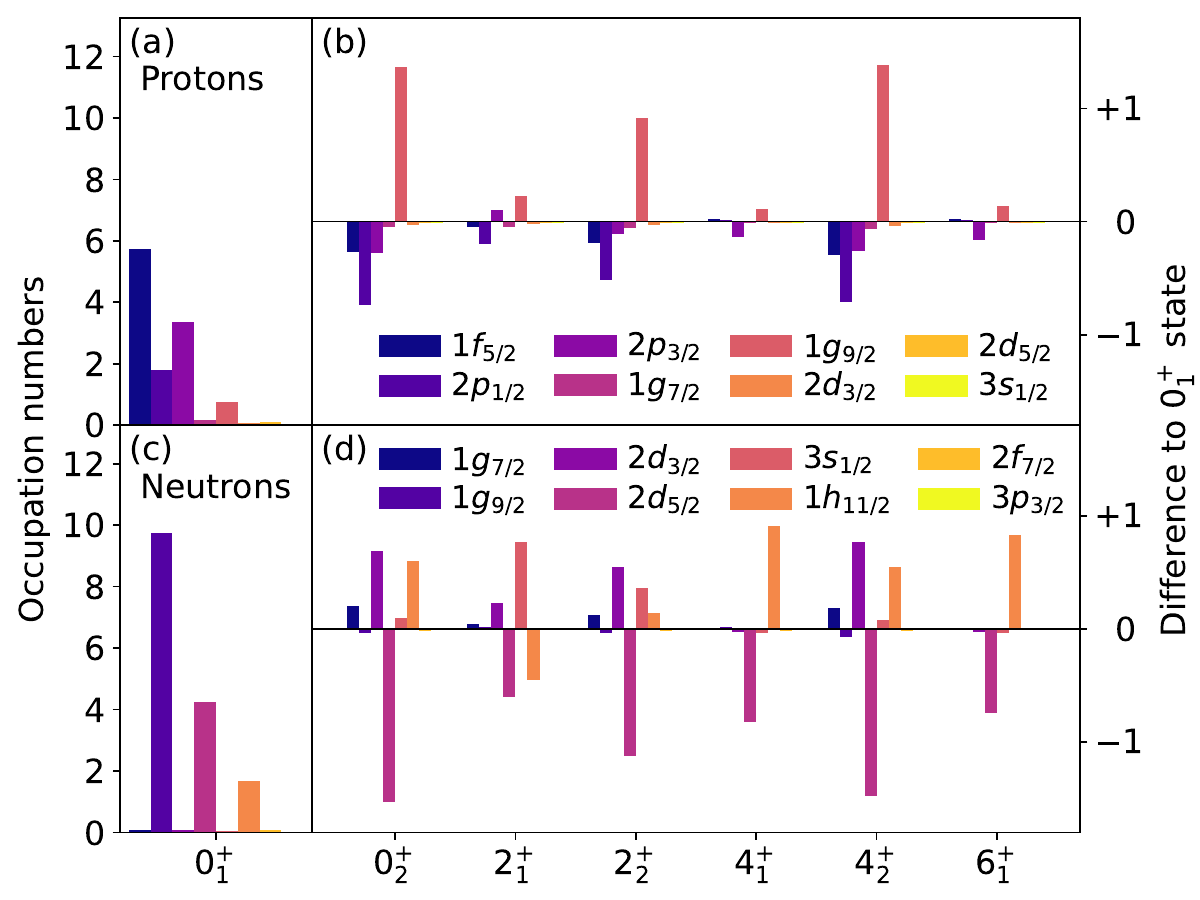}
\caption[]{Occupation numbers of orbitals considered in the QVSM calculations.
Panels~(a) and (c) depict the proton and neutron occupancies found for the ground state, respectively.
The change in occupation numbers for the excited states relative to the $0^+_1$ state is shown in panels (b) and (d).
The $4^+_2$ state in the QVSM calculations corresponds to the experimentally observed $4^+_1$ state and vice versa.}
\label{fig:occupation}
\end{figure}

The new lifetime measurements of the $2^+_2$ and $4^+_{1,2}$ states allow for more detailed insights into the nature
of the intruder structure of \nuc{96}{Zr} built on top of the $0^+_2$ state at \zeroplustwo\,keV~\cite{NDS96}.
%In the literature is has been suggested to be a $4p-4h$ intruder structure, with pairs of protons and neutrons 
%excited across the $Z=40$ proton subshell and $N=56$ neutron subshell closures, respectively. %~\cite{Hey85a}.
The transition strengths collected in Table~\ref{tab:results} confirm its assumed collectivity and 
firmly assign the $2^+_2$ and $4^+_1$ states at \twoplustwo~and \fourplusone\,keV~\cite{NDS96}, respectively, to it.
For the $4^+_1$ state at \fourplusone\,keV, transition strengths \mbox{$B(E2;4^+_1\to2^+_1)=\Bfourplusonetwoplusone$\,W.u.}
and \mbox{$B(E2;4^+_1\to2^+_2)=\Bfourplusonetwoplustwo$\,W.u.} are found experimentally (cf. Table~\ref{tab:results}),
hinting at mixing with the normal structure; likely through the energetically close-lying $4^+_2$ state at \fourplustwo\,keV~\cite{NDS96}
and $B(M1;4^+_2\to4^+_1)=\Bfourplustwofourplusone\,\mu_N^2$.
The experimentally obtained $B(E2;4^+_2\to2^+_1)$ and $B(E2;4^+_2\to2^+_2)$ transition strengths,
amounting to $\Bfourplustwotwoplusone$ and $\Bfourplustwotwoplustwo$\,W.u., respectively, 
support the assignment of the $4^+_2$ state to the normal structure, contrary to the placement in Ref.~\cite{Wis23a}.
The potentially collective nature of the $E2$ transition from the $6^+$ state at \sixplusthree\,keV~\cite{NDS96} 
to the $4^+_1$ state has yet to be proven to affiliate it with the intruder structure.
Notably, in Ref.~\cite{Wis23a} it has, together with the $4^+_1$ state, been placed as part of the ground-state structure.\\
Based on the available data from the $0^+_2$, $2^+_2$, and $4^+_1$ states, the fingerprint quantities
\mbox{$R_{4/2}=[E(4^+_1)-E(0^+_2)]/[E(2^+_2)-E(0^+_2)]=1.98$} and 
\mbox{$B_{4/2}=B(E2;4^+_1\to2^+_2)/B(E2;2^+_2\to0^+_2)=\Bfourtwo$} are obtained; 
the latter with greatly improved uncertainties due to the new lifetimes obtained in the present work.
In the collective picture, the almost equidistant energy steps between the $0^+_2$, $2^+_2$, and $4^+_1$ states 
favor a classification as a spherical vibrator.
The $B_{4/2}$ value, on the other hand, is located close to $10/7\approx1.43$ expected for an idealized rigid rotor~\cite{Boh53a}.
However, also $B_{4/2}=2$ signifying a spherical vibrator~\cite{Sch55a} lies within two standard deviations.\\
In the past, the structure of \nuc{96}{Zr} has been discussed in the context of the collective quadrupole model 
using the Bohr Hamiltonian with~\cite{Mar20b} and without~\cite{Saz19a} inclusion of the $\gamma$ degree of freedom.
%with~\cite{Saz19a} and without~\cite{Mar20b} the assumption of separable $\beta$ and $\gamma$ degrees of freedom.
In both cases, the potential in $\beta$ features spherical and deformed minima separated by a barrier
with the exact characteristics phenomenologically tailored to the known experimental data of low-lying states of \nuc{96}{Zr}.
With the deformed minimum fixed at $\beta=0.24$ through the experimental $B(E2;2^+_2\to0^+_2)$ value,
the collectivity of the intruder structure is reproduced; in Fig.~\ref{fig:R42} the calculated $R_{4/2}$ and $B_{4/2}$ values 
are confronted with their experimental counterparts.
By virtue of the constructed potential barrier, the wave functions of the $2^+_1$ and $4^+_1$ states, 
while mostly confined to the spherical and deformed minima, respectively, 
have enough overlap to allow for a collective transition connecting them.\\\\
While the description in the collective model presented in Refs.~\cite{Saz19a,Mar20b} is limited to \nuc{96}{Zr}, 
theoretical approaches with the IBM-CM and QVSM offer the opportunity to study the evolution of 
nuclear structure in the zirconium isotopic chain with increasing neutron number in a consistent framework.\\
The calculations of the zirconium isotopes 
in the interacting boson model with configuration mixing~\cite{Duv81a,Duv82a} (IBM-CM)
feature valence neutrons above a \nuc{90}{Zr} core and two configurations~\cite{Gav19a,Gav20a,Gav22a}.
The wave function of an eigenstate with angular momentum $L$
is a linear combination of wave functions corresponding to 
two configurations $A$ and $B$ with boson numbers \mbox{$(N,N+2)=(3,5)$} for \nuc{96}{Zr}.
The $[N]$- and $[N+2]$-boson spaces, whose size is calculated with respect to the \nuc{90}{Zr} core,
mimic no active protons above $Z=40$ and proton $2p$-$2h$ states crossing this sub-shell gap, respectively.
%These boson numbers are calculated with respect to the \nuc{90}{Zr} core and
%mimic no active protons above $Z=40$ and proton $2p$-$2h$ states 
%with the $[N]$- and $[N+2]$-boson spaces, respectively.
Consequently, a wave function can be written as
\begin{equation*}
|\Psi;L\rangle = a\,|\psi_A;[N],L\rangle + b\,|\psi_B;[N+2],L\rangle
\end{equation*}
with $a^2+b^2=1$ and $\psi_{A,B}$ summarizing additional quantum numbers for the sake of readability.\\
A comparison of the IBM-CM predictions to the experimental results is depicted in Fig.~\ref{fig:comp_exp_theo},
highlighting the overall good reproduction of the low-energy structure of \nuc{96}{Zr}.
The calculations show a clear separation of the states into two groups, 
corresponding to normal (configuration $A$, $[N]$) and intruder (configuration $B$, $[N+2]$) nature.
In agreement with experimental findings~\cite{Kre16a}, a small mixing parameter $\omega=0.02$\,MeV
is used in the IBM-CM calculation~\cite{Gav19a},
leading to $0^+_{1,2}$ and $2^+_{1,2}$ states with virtually no mixing
as illustrated by wave-function purities of $99.6\%$ and $94.3\%$, respectively. 
Similarly to the $M1$ transition between the lowest $2^+$ states of each configuration,
the underestimated $B(E2;4^+_1\to2^+_1)$ transition strength might, however, hint at the necessity for larger mixing between both structures.
In order to probe this, the IBM-CM calculations have been repeated with an increased mixing parameter $\omega=0.04$\,MeV.
The choice for the value $\omega=0.02$\,MeV was taken along the \nuc{96-110}{Zr} chain for simplicity, 
with emphasis on the \nuc{100}{Zr} isotope near the critical point of the quantum phase transition~\cite{Gav22a}.
However, a larger value was forced for \nuc{92-94}{Zr}, potentially suggesting the same need for \nuc{96}{Zr}, 
where neutrons mainly occupy the $d_{5/2}$ orbit.
As a result, the key $B(M1;2^+_2\to2^+_1)$ and $B(E2;4^+_1\to2^+_1)$ transition strengths are increased 
to $\BtwoplustwotwoplusoneIBMmix\,\mu_N^2$ and $\BfourplusonetwoplusoneIBMmix$\,W.u.,
significantly narrowing the gap to the experimental values of $\Btwoplustwotwoplusone\,\mu_N^2$ and $\Bfourplusonetwoplusone$\,W.u., respectively.
The effect on the remaining $E2$ transition strengths, which are summarized in Table~\ref{tab:results}, is mild or even beneficial.
The wave-function purity remains high for the $0^+$ states but drops to $84.7\%$ for the $2^+_{1,2}$ states.
As for the calculated intruder $4^+_1$ and $6^+_1$ states, the mixing remains weak, contrary to the $2^+_{1,2}$ states.
This mixing analysis is supported by that of the odd-mass Zr isotopes as well~\cite{Gav25a}, and follows the MCSM work~\cite{Kre16a}, 
which suggests neutrons mainly occupy the $d_{5/2}$ orbit and weak mixing with the $g_{9/2}$ proton configuration.\\
Irrespective of the chosen mixing parameter, the IBM-CM predicts a quasi-spherical, U(5)-like intruder band 
built on the $0^+_2$ state and correctly reproduces its collectivity [cf. panel~(b) of Fig.~\ref{fig:R42} and Fig.~\ref{fig:comp_exp_theo}],
whereas it tends to overestimate it in neighboring, heavier \nuc{98}{Zr}~\cite{Kar20a}.\\\\
%\red{What about the energy spacing in Fig.~\ref{fig:R42}?}
%
For a fully microscopic interpretation of the low-energy structure of \nuc{96}{Zr},
large-scale quasiparticle vacua shell-model~\cite{Shi21a,Shi25a} (QVSM) calculations have been performed.
The QVSM is the most advanced version of the MCSM. In particular, it is more effective for heavier mass regions~\cite{Shi21a}.
The calculations employ the same model space as the ones using the MCSM presented in Ref.~\cite{Kre16a};
namely for protons the $2p_{3/2}$, $1f_{5/2}$, and $2p_{1/2}$ orbitals as well as the full $sdg$ shell
and for neutrons the $sdg$ shell furthermore including the $1h_{11/2}$, $2f_{7/2}$, and $3p_{3/2}$ orbitals above.
The employed effective interaction has been modified slightly with respect to the previous MCSM calculations;
the changes are, however, relatively minor and the mechanism of shape transition remains essentially unchanged.
On the other hand, the description of various properties has been improved.
This results in differing occupation numbers, summarized in Fig.~\ref{fig:occupation}, 
for some orbitals when compared to the MCSM calculations~\cite{Kre16a}.
In the latter, the ground state of \nuc{96}{Zr} is characterized by filled proton and neutron orbitals below the $Z=40$ and $N=56$ 
(sub-)shell closures, respectively, whereas those above remain mostly empty.
In the QVSM, on the other hand, especially the proton $1g_{9/2}$ and neutron $1h_{11/2}$ orbitals contain, 
already in the ground state, around one and two nucleons, respectively.
\begin{figure*}[t]
\centering
\includegraphics[trim=150 180 150 250,width=1.025\linewidth,clip]{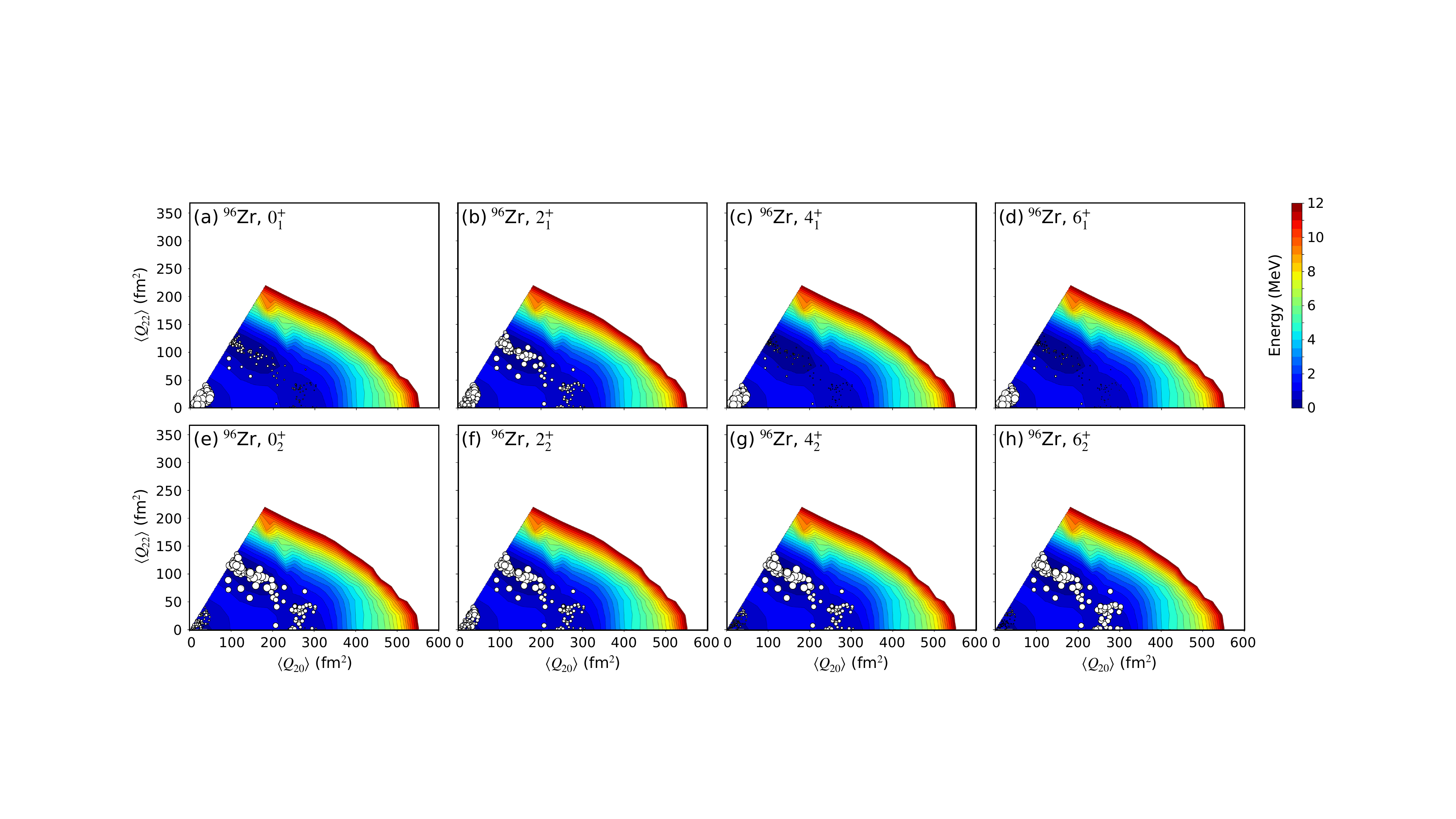}
\caption[]{\textit{T-plots} for the $0^+_{1,2}$, $2^+_{1,2}$, $4^+_{1,2}$, and $6^+_{1,2}$ states of \nuc{96}{Zr}
from the QVSM calculations.
The top and bottom rows contain states associated with the (spherical) normal and (triaxially deformed) intruder structures, respectively.
The circles mark the locations of basis vectors present in the QVSM eigenstate corresponding to the state
in a potential-energy surface spanned by quadrupole moments $Q_{20}$ and $Q_{22}$.
The area of the circle encodes the importance of a basis vector in the eigenstate.
The \textit{T-plots} for the $2^+$ states reveal the existence of sizeable mixing.
The deformed $4^+_2$ state from the QVSM calculations corresponds to the $4^+_1$ state
in the experimental level scheme of \nuc{96}{Zr} and vice versa.
}
\label{fig:tplots}
\end{figure*}
The normal $2^+$ and $4^+$ states are dominated by single-neutron excitations from the $2d_{5/2}$ orbital; 
for the former to the $3s_{1/2}$ and for the latter to the $1h_{11/2}$.
The  deformed $0^+_2$ state has in the past been proposed to be of $4p$-$4h$ nature~\cite{Hey85a,Hey11a} with pairs of protons and neutrons
promoted across the $Z=40$ and $N=56$ (sub-)shell closures, respectively.
In the present QVSM calculations, on average about one proton is promoted from the $2p_{1/2}$ to the $1g_{9/2}$ orbital
and two neutrons are excited from the $2d_{5/2}$ to the $2d_{3/2}$ and $1h_{11/2}$ orbitals.
This is in marked contrast to the MCSM results, which suggested that the $0^+_2$ and $2^+_2$ states contain 
approximately three proton and three neutron particle-hole excitations on average~\cite{Kre16a,Tog16a}.
As a result, their mixing with the spherical $0p$-$0h$ configuration was hindered 
because two-body interactions cannot directly couple these configurations.
The difference in populations implies that near-oblate configurations mix more readily with spherical configurations.
%Indeed, the \textit{T-plots}~\cite{Ots16a} shown in Fig.~\ref{fig:tplots} indicate that the $2^+_2$ state 
%contains not only deformed components but also spherical components.
%This is likely one of the reasons why the present calculation reproduces the interband transition strengths well, 
%whereas they were underestimated in the previous MCSM calculations.
Moreover, the type-II shell evolution still plays an important role.\\
The QVSM calculations agree with the low-lying level scheme of \nuc{96}{Zr} as evident from Fig.~\ref{fig:comp_exp_theo}.
They furthermore reproduce its hallmark octupole collectivity,
predicting the \threeminusone-keV~\cite{NDS96} $3^-_1$ state at $1383$\,keV 
with \mbox{$B(E3;3^-_1\to0^+_1)=\BthreeminusonezeroplusoneQVSM$\,W.u.},
in agreement with the adopted value of $\BthreeminusonezeroplusoneNDS$\,W.u.~\cite{NDS96}
and within $20\%$ of $\BthreeminusonezeroplusoneIskra$\,W.u. reported in Ref.~\cite{Isk19a}.
However, the calculations do not replicate the almost equidistant excitation energies within
the excited-state structure shown in Fig.~\ref{fig:R42}.
While the ordering of the $0^+$ and $2^+$ states is correct, 
%this leads to an inversion of the $4^+$ states in the QVSM;
the calculations do not reproduce the relative ordering of the $4^+_{1,2}$ states
as inferred from their experimentally determined $E2$ transition strengths (cf. Table~\ref{tab:results});
contrasting the evolution of relative excitation energies between collective and non-collective states 
with neutron number toward the shape-phase transition~\cite{Wer19a}.
%\red{Maybe explain with plot.}
In the following, the first (second) $4^+$ state from the QVSM calculations is associated
with the $4^+_2$ ($4^+_1$) state observed at \fourplustwo\,keV [\fourplusone\,keV]~\cite{NDS96}.\\
The nature of the calculated states can be inferred from the \textit{T-plots}~\cite{Ots16a} shown in Fig.~\ref{fig:tplots}
which visualize the importance of QVSM basis states in a potential-energy surface 
spanned by the quadrupole moments $Q_{20}$ and $Q_{22}$.
The $0^+$ states show little mixing, in agreement with the conclusions of Ref.~\cite{Kre16a}.
In contrast, the $2^+$ states, which are predicted at $1853$ and $2202$\,keV, are more strongly mixed
as evident from the sizeable contributions of deformed (spherical) basis states to the wave function 
of the $2^+_1$ ($2^+_2$) state.
For the $4^+$ and $6^+$ states the situation is again similar to the one found for the $0^+$ states, 
showing a clear separation between spherical and triaxially deformed configurations.
Compared to the MCSM calculations, the potential appears somewhat softer in the $\gamma$ degree of freedom
as inferred from the larger spread of the basis-state markers (cf. Fig.~\ref{fig:tplots} and Fig.~3 of Ref.~\cite{Kre16a}).\\
%However, the calculations do not reproduce the relative ordering of the $4^+_{1,2}$ states
%as inferred from their experimentally determined $E2$ transition strengths (cf. Table~\ref{tab:results}).
%In the following, the first (second) $4^+$ state from the QVSM calculations is associated
%with the $4^+_2$ ($4^+_1$) state observed at \fourplustwo\,keV [\fourplusone\,keV]~\cite{NDS96}.\\
%
The predicted $E2$ transition strengths are in close agreement with the experimental results 
(cf. Tab~\ref{tab:results} and Fig.~\ref{fig:comp_exp_theo}).
%By virtue of the more developed mixing found for the $2^+$ states,
%the collective \mbox{$B(M1;2^+_2\to2^+_1)$} and \mbox{$B(E2;4^+_1\to2^+_1)$} interband transitions are reproduced; 
%again assuming an inversion between the $4^+_{1,2}$ states in the QVSM calculations.
The more developed mixing found for the $2^+$ states is likely one of the reasons for the reproduction of
the collective \mbox{$B(M1;2^+_2\to2^+_1)$} and \mbox{$B(E2;4^+_1\to2^+_1)$} interband transitions
which were underestimated in the previous MCSM calculations.
The large \mbox{$B(M1;4^+_2\to4^+_1)=\Bfourplustwofourplusone\,\mu_N^2$} transition strength
is, however, greatly underestimated;
even when correcting for the smaller transition energy in the experimental data (cf. Fig.~\ref{fig:comp_exp_theo}).
In a simplified picture, this might indicate the necessity for more strongly mixed $4^+$ states in the QVSM.\\
In order to probe the wave-function composition of the observed states from the experimental perspective, 
a two-state model (TSM) analysis based on the $E2$ transition rates collected in Table~\ref{tab:results} is appropriate.
The observed states $|J^+_1\rangle$ and $|J^+_2\rangle$ are treated as linear combinations of basis states 
$|J^+_{\mathrm{I}}\rangle$ and $|J^+_{\mathrm{II}}\rangle$, which correspond to 
normal (spherical) and intruder (deformed) configurations $\mathrm{I}$ and $\mathrm{II}$, respectively, yielding
\begin{equation*}
\begin{pmatrix}
|J^+_1\rangle \\ |J^+_2\rangle 
\end{pmatrix}
=
\begin{pmatrix}
\phantom{-}\alpha_J & \beta_J\\ 
-\beta_J & \alpha_J\\
\end{pmatrix}
\begin{pmatrix}
|J^+_{\mathrm{I}}\rangle \\ |J^+_{\mathrm{II}}\rangle
\end{pmatrix}
\end{equation*}
%\begin{equation*}
%\begin{pmatrix}
%|0^+_1\rangle \\ |0^+_2\rangle \\
%|2^+_1\rangle \\ |2^+_2\rangle \\
%|4^+_1\rangle \\ |4^+_2\rangle \\
%\end{pmatrix}
%=
%\begin{pmatrix}
%\alpha_0 & \beta_0 & 0 & 0 & 0 & 0\\ 
%-\beta_0 & \alpha_0 & 0 & 0 & 0 & 0\\
%0 & 0 & \alpha_2 & \beta_2 & 0 & 0\\ 
%0 & 0 & -\beta_2 & \alpha_2 & 0 & 0\\ 
%0 & 0 & 0 & 0 & \alpha_4 & \beta_4 \\
%0 & 0 & 0 & 0 & -\beta_4 & \alpha_4 \\
%\end{pmatrix}
%\begin{pmatrix}
%%|0^+_{\text{sph}}\rangle \\ |0^+_{\text{def}}\rangle \\
%%|2^+_{\text{sph}}\rangle \\ |2^+_{\text{def}}\rangle \\
%%|4^+_{\text{sph}}\rangle \\ |4^+_{\text{def}}\rangle \\
%|0^+_{\mathrm{I}}\rangle \\ |0^+_{\mathrm{II}}\rangle \\
%|2^+_{\mathrm{I}}\rangle \\ |2^+_{\mathrm{II}}\rangle \\
%|4^+_{\mathrm{I}}\rangle \\ |4^+_{\mathrm{II}}\rangle \\
%\end{pmatrix}
%\end{equation*}
%
with the identities $\alpha_J^2+\beta_J^2=1$ and $J\in\lbrace0,2,4\rbrace$.
All interband $E2$ matrix elements between the pure configurations $\mathrm{I}$ and $\mathrm{II}$ are assumed to vanish.
Following the ansatz of Fortune~\cite{For17a}, dimensionless ratios of mixing parameters $\beta_J/\alpha_J$
and $E2$ matrix elements calculated from the transition strengths found in Table~\ref{tab:results} are used.
The phases of the $2^+_1\to0^+_1$ and $2^+_2\to0^+_2$ matrix elements are required to be positive,
whereas all others can also be negative.
If the TSM is limited to the $0^+_{1,2}$ and $2^+_{1,2}$ states, i.e. to the situation discussed in Ref.~\cite{For17a},
wave-function purities of $98.1(3)\%$ and $97.1(4)\%$ are found, respectively. 
If a negative sign is chosen for the small $2^+_2\to0^+_1$ matrix element, 
in contrast to results from a Coulomb-excitation analysis~\cite{Zie26a},
a second solution with purities $99.9(1)\%$ and $89.5(4)\%$ is obtained.
In both cases, the ratio between intrinsic matrix elements in the normal (configuration $\mathrm{I}$) and 
intruder (configuration $\mathrm{II}$) structures, 
\mbox{$r^{20}=\langle2^+_{\mathrm{I}}\|T(E2)\|0^+_{\mathrm{I}}\rangle/\langle2^+_{\mathrm{II}}\|T(E2)\|0^+_{\mathrm{II}}\rangle$}, 
is around $0.3$, implying collectivity in structure~$\mathrm{II}$.
For the hitherto unobserved $2^+_1\to0^+_2$ transition, $E2$ strengths of $0.41(6)$ and $2.9(2)$\,W.u.
are predicted by the TSM, again contingent on the sign of the $2^+_2\to0^+_1$ matrix element.\\
The situation changes considerably when the newly obtained transitions from the $4^+_{1,2}$ states
(cf. Table~\ref{tab:results}) are incorporated.
Notably, the $2^+$ states are now also influenced by the collective $4^+_1\to2^+_1$ transition
with \mbox{$B(E2;4^+_1\to2^+_1)=\Bfourplusonetwoplusone$\,W.u.} (cf. Table~\ref{tab:results}).
The ratio of intrinsic matrix elements between the $4^+$ and $2^+$ states, $r^{42}$, 
is defined analogous to the aforementioned $r^{20}$.
The resulting system comprises five matrix elements with free phases; namely from the $2^+_2\to0^+_1$
and all transitions between the $4^+$ and $2^+$ states.
In the following, only those sign combinations resulting in a close reproduction of the experimental 
matrix elements are considered.
Exemplarily, the results for all positive phases are collected in Table~\ref{tab:tsm}.
\begin{table}[tb]
\caption{Parameters for exemplary TSM analyses, when limited to the $(0^+,2^+)$ states
and for inclusion of the $4^+$ states.
In the latter case, the reproduction of experimental results deteriorates considerably
if equal ratios of intrinsic matrix elements $r^{20}$ and $r^{42}$ are required.
The last two lines state the degree-of-freedom $\nu$ of the considered system
and the obtained goodness-of-fit $\chi^2_{\nu}$	.
The calculations assumed positive phases for all matrix elements.
}
\label{tab:tsm}
\begin{ruledtabular}
\begin{tabular}{clccc}
$J$	&Parameter			&$(0^+,2^+)$		&\multicolumn{2}{c}{$(0^+,2^+,4^+)$}\\
	&					&				&$r^{20}\neq r^{42}$	&$r^{20}=r^{42}$\\\colrule
$0$	&$\alpha_0^2$			&$0.981(3)$		&$0.959(5)$		&$0.981(4)$\\
	&$\beta_0^2$			&$0.019(3)$		&$0.041(5)$		&$0.019(4)$\\
	&$V_{\text{mix}}$ (keV)	&$217(14)$		&$311(20)$		&$217(21)$\\
$2$	&$\alpha_2^2$			&$0.971(4)$		&$0.81(3)$		&$0.67(8)$\\
	&$\beta_2^2$			&$0.029(4)$		&$0.19(3)$		&$0.33(8)$\\
	&$V_{\text{mix}}$ (keV)	&$79(4)$			&$186(11)$		&$220(19)$\\
$4$	&$\alpha_4^2$			&				&$0.969(6)$		&$0.990(6)$\\
	&$\beta_4^2$			&				&$0.031(6)$		&$0.010(6)$\\
	&$V_{\text{mix}}$ (keV)	&				&$38(3)$			&$21(6)$\\\colrule
	&$r^{20}$				&$0.31(3)$		&$0.24(3)$		&$0.10(1)$\\
	&$r^{42}$				&				&$0.001(3)$		&$0.10(1)$\\\colrule
	&$\nu$				&$0$				&$2$				&$3$\\
	&$\chi^2_{\nu}$			&$0.0$			&$2.8$			&$83.7$\\
\end{tabular}
\end{ruledtabular}
\end{table}
Irrespective of the chosen signs, the $0^+$ and $4^+$ states show little mixing,
whereas the wave-function purities of the $2^+$ states drop to around $80\%$,
comparable to the IBM-CM calculations with increased mixing.
The encountered scenario resembles the situation found in the QVSM calculations, 
visualized in the \textit{T-plots} of Fig.~\ref{fig:tplots}.
In these, the $4^+$ (and $6^+$) states appear slightly less mixed than the $0^+$ states.
The fact that the TSM results of Table~\ref{tab:tsm} support this observation might, however, be fortuitous.
The transition strength $B(E2;2^+_1\to0^+_2)=10.1$\,W.u. predicted in the QVSM
is twice as large as $5(1)$\,W.u. obtained from the full TSM based on the experimental transition strengths of Table~\ref{tab:results}.
Notably, the latter value is, in turn, about $70\%$ larger than from the TSM limited to the $(0^+,2^+)$ states
by virtue of the increased mixing in the $2^+$ states.\\
The TSM shows that the necessity for considerably mixed $2^+$ states can be concluded already from the experimental data
under relatively weak assumptions, in close agreement with the predicted structure of the wave functions of both the IBM-CM and QVSM.

%\red{
%Some open questions/to-dos for TB:
%\begin{itemize}
%\item{Quadrupole moments of $2^+_{1,2}$ - different for IBM-CM and similar for QVSM?}
%\item{$2^+_1\to0^+_2$ from IBM-CM?}
%\item{Can we get the mixing in QVSM as "one number" per state for Fig.~\ref{fig:comp_exp_theo}?}
%\item{$0^+_2$ softer in $\gamma$ than from MCSM (compare \textit{T-plots}).}
%\item{In the QVSM the $2^+$ states are closer in energy than the $4^+$ states, in contrast to experiment.}
%\item{Can a comparison of some transfer cross section to the $2^+_1$ and $2^+_2$ states tell us more about their wave-function composition?
%Like in that Fortune paper on it, but with more focus on the $2^+$ states instead of $0^+$.}
%\item{Relate TSM to TSM of \nuc{94}{Zr} from Fortune.}
%\end{itemize}
%}

%%%%%%%%%%%%%%%%%%%%%%%%%%%%%%%%%%%%%%%%%%%%%%%%%%%%%%%%%%%%%%%
%	Summary
%%%%%%%%%%%%%%%%%%%%%%%%%%%%%%%%%%%%%%%%%%%%%%%%%%%%%%%%%%%%%%%

\section{Summary}

In summary, excited states of \nuc{96}{Zr}, populated in the \reac two-neutron transfer reaction, 
have been studied with the Doppler-shift attenuation method.
The measured lifetimes firmly attribute the $2^+_2$ and $4^+_1$ states
at \twoplustwo~and \fourplusone\,keV~\cite{NDS96}, respectively,
to the intruder structure on top of the \zeroplustwo-keV~\cite{NDS96} $0^+_2$ state.
Calculations in the algebraic interacting boson model with configuration mixing and
the quasiparticle vacua shell model are able to successfully reproduce the transition strengths
%but offer different interpretations of the involved structures.
and both oﬀer interpretations of coexisting normal and intruder configurations, 
with some diﬀerence in the shape of the intruder configuration.
Comparison to a simple two-state mixing model highlights the importance of sizeable mixing
in the $2^+$ states for a reproduction of the observed $E2$ transition strengths.
A collective $4^+_2\to4^+_1$ interband $M1$ transition might hint at non-negligible mixing
between the normal and intruder structures also at $J=4$.\\
The present work shows that further data are needed to understand the nature of the excited states of \nuc{96}{Zr}.
The QVSM predicts quadrupole moments of $\QtwoplusoneQVSM$\,$e$fm$^2$ and $\QtwoplustwoQVSM$\,$e$fm$^2$
for the $2^+_1$ and $2^+_2$ states and  $\QfourplusoneQVSM$\,$e$fm$^2$ and $\QfourplustwoQVSM$\,$e$fm$^2$
for the $4^+_1$ and $4^+_2$ states, respectively, reflecting the enhanced mixing found for the $2^+$ states (cf. Fig.~\ref{fig:tplots}).
In contrast, the quadrupole moments put forward by the IBM-CM calculations (for $\omega=0.02$\,MeV) 
are $\QtwoplusoneIBM$\,$e$fm$^2$, $\QtwoplustwoIBM$\,$e$fm$^2$, and $\QfourplusoneIBM$\,$e$fm$^2$
for the $2^+_1$, $2^+_2$, and $4^+_1$ states, respectively, reflecting their more-spherical character.
The small excitation strengths from the ground state to both $2^+$ states might, however, 
make a determination of their quadrupole moments via 'safe' Coulomb excitation~\cite{Cli86a} prohibitively complicated. 
%\red{Magda Zielinska did unsafe Coulex.}
A measurement of the $2^+$ states' (relative) population probability, for instance in the \nuc{94}{Zr}$(t,p)$\nuc{96}{Zr} reaction,
might be used instead to probe the wave-function overlaps between the $2^+_{1,2}$ states of \nuc{96}{Zr}
and the ground state of \nuc{94}{Zr}~\cite{For19a} and, thus, reveal structural differences between the final states of \nuc{96}{Zr}.
In the literature, however, only cross sections for two-neutron transfer to the ground and excited $0^+$ states of \nuc{96}{Zr}
are available to date~\cite{NDS96,Bee69a}.

%%%%%%%%%%%%%%%%%%%%%%%%%%%%%%%%%%%%%%%%%%%%%%%%%%%%%%%%%%%%%%%
%	Acknowledgments
%%%%%%%%%%%%%%%%%%%%%%%%%%%%%%%%%%%%%%%%%%%%%%%%%%%%%%%%%%%%%%%

\begin{acknowledgments}

The authors thank the staff of the IFIN-HH tandem accelerator for providing excellent experimental conditions.
Support by R.~Kern and W.~Witt during data taking is gratefully acknowledged.
T.B. and T.S. are indebted to C.~Stahl for initial help with the \texttt{APCAD} program.
This work was supported by the BMBF under grant Nos. 05P18RDEN9 and 05P18RDFN1/9, 
by the Deutsche Forschungsgemeinschaft (DFG, German Research Foundation) under 
Project-ID 499256822 -- GRK 2891 “Nuclear Photonics”, 
and by the Institute of Atomic Physics, Romania, under Project \mbox{ELI-RO/DFG/2023\_001} \mbox{ARNPhot}.
K.Y. was supported by MEXT as “Program for Promoting Researches on the Supercomputer Fugaku” (JPMXP1020200105 and JPMXP1020230411),
the JSPS KAKENHI under Grant Numbers JP22K14031 and JP25K17404, as well as the general use program in CCS, University of Tsukuba.
K.Y. used computational resources of the supercomputer Fugaku provided by the RIKEN Center for Computational Science 
(Project IDs hp230137 and hp250187) and the supercomputer Pegasus provided by CCS, University of Tsukuba.
This project has received funding from the European Union’s Horizon 2020 
research and innovation programme under Grant Agreement No. 654002 (ENSAR-2)
and the Romanian National Authority for Research through the Nucleu Project No. PN\,23\,21\,01\,02.

\end{acknowledgments}


\begin{thebibliography}{60}%
\makeatletter
\providecommand \@ifxundefined [1]{%
 \@ifx{#1\undefined}
}%
\providecommand \@ifnum [1]{%
 \ifnum #1\expandafter \@firstoftwo
 \else \expandafter \@secondoftwo
 \fi
}%
\providecommand \@ifx [1]{%
 \ifx #1\expandafter \@firstoftwo
 \else \expandafter \@secondoftwo
 \fi
}%
\providecommand \natexlab [1]{#1}%
\providecommand \enquote  [1]{``#1''}%
\providecommand \bibnamefont  [1]{#1}%
\providecommand \bibfnamefont [1]{#1}%
\providecommand \citenamefont [1]{#1}%
\providecommand \href@noop [0]{\@secondoftwo}%
\providecommand \href [0]{\begingroup \@sanitize@url \@href}%
\providecommand \@href[1]{\@@startlink{#1}\@@href}%
\providecommand \@@href[1]{\endgroup#1\@@endlink}%
\providecommand \@sanitize@url [0]{\catcode `\\12\catcode `\$12\catcode
  `\&12\catcode `\#12\catcode `\^12\catcode `\_12\catcode `\%12\relax}%
\providecommand \@@startlink[1]{}%
\providecommand \@@endlink[0]{}%
\providecommand \url  [0]{\begingroup\@sanitize@url \@url }%
\providecommand \@url [1]{\endgroup\@href {#1}{\urlprefix }}%
\providecommand \urlprefix  [0]{URL }%
\providecommand \Eprint [0]{\href }%
\providecommand \doibase [0]{https://doi.org/}%
\providecommand \selectlanguage [0]{\@gobble}%
\providecommand \bibinfo  [0]{\@secondoftwo}%
\providecommand \bibfield  [0]{\@secondoftwo}%
\providecommand \translation [1]{[#1]}%
\providecommand \BibitemOpen [0]{}%
\providecommand \bibitemStop [0]{}%
\providecommand \bibitemNoStop [0]{.\EOS\space}%
\providecommand \EOS [0]{\spacefactor3000\relax}%
\providecommand \BibitemShut  [1]{\csname bibitem#1\endcsname}%
\let\auto@bib@innerbib\@empty
%</preamble>
\bibitem [{\citenamefont {Federman}\ and\ \citenamefont
  {Pittel}(1979)}]{Fed79a}%
  \BibitemOpen
  \bibfield  {author} {\bibinfo {author} {\bibfnamefont {P.}~\bibnamefont
  {Federman}}\ and\ \bibinfo {author} {\bibfnamefont {S.}~\bibnamefont
  {Pittel}},\ }\bibfield  {title} {\bibinfo {title} {Unified shell-model
  description of nuclear deformation},\ }\href
  {https://doi.org/10.1103/PhysRevC.20.820} {\bibfield  {journal} {\bibinfo
  {journal} {Phys. Rev. C}\ }\textbf {\bibinfo {volume} {20}},\ \bibinfo
  {pages} {820} (\bibinfo {year} {1979})}\BibitemShut {NoStop}%
\bibitem [{\citenamefont {Otsuka}\ \emph {et~al.}(2005)\citenamefont {Otsuka},
  \citenamefont {Suzuki}, \citenamefont {Fujimoto}, \citenamefont {Grawe},\
  and\ \citenamefont {Akaishi}}]{Ots05a}%
  \BibitemOpen
  \bibfield  {author} {\bibinfo {author} {\bibfnamefont {T.}~\bibnamefont
  {Otsuka}}, \bibinfo {author} {\bibfnamefont {T.}~\bibnamefont {Suzuki}},
  \bibinfo {author} {\bibfnamefont {R.}~\bibnamefont {Fujimoto}}, \bibinfo
  {author} {\bibfnamefont {H.}~\bibnamefont {Grawe}},\ and\ \bibinfo {author}
  {\bibfnamefont {Y.}~\bibnamefont {Akaishi}},\ }\bibfield  {title} {\bibinfo
  {title} {{E}volution of {N}uclear {S}hells due to the {T}ensor {F}orce},\
  }\href {https://doi.org/10.1103/PhysRevLett.95.232502} {\bibfield  {journal}
  {\bibinfo  {journal} {Phys. Rev. Lett.}\ }\textbf {\bibinfo {volume} {95}},\
  \bibinfo {pages} {232502} (\bibinfo {year} {2005})}\BibitemShut {NoStop}%
\bibitem [{\citenamefont {Otsuka}\ \emph {et~al.}(2010)\citenamefont {Otsuka},
  \citenamefont {Suzuki}, \citenamefont {Honma}, \citenamefont {Utsuno},
  \citenamefont {Tsunoda}, \citenamefont {Tsukiyama},\ and\ \citenamefont
  {Hjorth-Jensen}}]{Ots10a}%
  \BibitemOpen
  \bibfield  {author} {\bibinfo {author} {\bibfnamefont {T.}~\bibnamefont
  {Otsuka}}, \bibinfo {author} {\bibfnamefont {T.}~\bibnamefont {Suzuki}},
  \bibinfo {author} {\bibfnamefont {M.}~\bibnamefont {Honma}}, \bibinfo
  {author} {\bibfnamefont {Y.}~\bibnamefont {Utsuno}}, \bibinfo {author}
  {\bibfnamefont {N.}~\bibnamefont {Tsunoda}}, \bibinfo {author} {\bibfnamefont
  {K.}~\bibnamefont {Tsukiyama}},\ and\ \bibinfo {author} {\bibfnamefont
  {M.}~\bibnamefont {Hjorth-Jensen}},\ }\bibfield  {title} {\bibinfo {title}
  {{N}ovel {F}eatures of {N}uclear {F}orces and {S}hell {E}volution in {E}xotic
  {N}uclei},\ }\href {https://doi.org/10.1103/PhysRevLett.104.012501}
  {\bibfield  {journal} {\bibinfo  {journal} {Phys. Rev. Lett.}\ }\textbf
  {\bibinfo {volume} {104}},\ \bibinfo {pages} {012501} (\bibinfo {year}
  {2010})}\BibitemShut {NoStop}%
\bibitem [{\citenamefont {Otsuka}\ and\ \citenamefont
  {Tsunoda}(2016)}]{Ots16a}%
  \BibitemOpen
  \bibfield  {author} {\bibinfo {author} {\bibfnamefont {T.}~\bibnamefont
  {Otsuka}}\ and\ \bibinfo {author} {\bibfnamefont {Y.}~\bibnamefont
  {Tsunoda}},\ }\bibfield  {title} {\bibinfo {title} {The role of shell
  evolution in shape coexistence},\ }\href
  {https://doi.org/10.1088/0954-3899/43/2/024009} {\bibfield  {journal}
  {\bibinfo  {journal} {J. Phys. G: Nucl. Part. Phys.}\ }\textbf {\bibinfo
  {volume} {43}},\ \bibinfo {pages} {024009} (\bibinfo {year}
  {2016})}\BibitemShut {NoStop}%
\bibitem [{\citenamefont {Togashi}\ \emph {et~al.}(2016)\citenamefont
  {Togashi}, \citenamefont {Tsunoda}, \citenamefont {Otsuka},\ and\
  \citenamefont {Shimizu}}]{Tog16a}%
  \BibitemOpen
  \bibfield  {author} {\bibinfo {author} {\bibfnamefont {T.}~\bibnamefont
  {Togashi}}, \bibinfo {author} {\bibfnamefont {Y.}~\bibnamefont {Tsunoda}},
  \bibinfo {author} {\bibfnamefont {T.}~\bibnamefont {Otsuka}},\ and\ \bibinfo
  {author} {\bibfnamefont {N.}~\bibnamefont {Shimizu}},\ }\bibfield  {title}
  {\bibinfo {title} {{Q}uantum {P}hase {T}ransition in the {S}hape of {Zr}
  isotopes},\ }\href {https://doi.org/10.1103/PhysRevLett.117.172502}
  {\bibfield  {journal} {\bibinfo  {journal} {Phys. Rev. Lett.}\ }\textbf
  {\bibinfo {volume} {117}},\ \bibinfo {pages} {172502} (\bibinfo {year}
  {2016})}\BibitemShut {NoStop}%
\bibitem [{\citenamefont {Kremer}\ \emph {et~al.}(2016)\citenamefont {Kremer},
  \citenamefont {Aslanidou}, \citenamefont {Bassauer}, \citenamefont {Hilcker},
  \citenamefont {Krugmann}, \citenamefont {von Neumann-Cosel}, \citenamefont
  {Otsuka}, \citenamefont {Pietralla}, \citenamefont {Ponomarev}, \citenamefont
  {Shimizu}, \citenamefont {Singer}, \citenamefont {Steinhilber}, \citenamefont
  {Togashi}, \citenamefont {Tsunoda}, \citenamefont {Werner},\ and\
  \citenamefont {Zweidinger}}]{Kre16a}%
  \BibitemOpen
  \bibfield  {author} {\bibinfo {author} {\bibfnamefont {C.}~\bibnamefont
  {Kremer}}, \bibinfo {author} {\bibfnamefont {S.}~\bibnamefont {Aslanidou}},
  \bibinfo {author} {\bibfnamefont {S.}~\bibnamefont {Bassauer}}, \bibinfo
  {author} {\bibfnamefont {M.}~\bibnamefont {Hilcker}}, \bibinfo {author}
  {\bibfnamefont {A.}~\bibnamefont {Krugmann}}, \bibinfo {author}
  {\bibfnamefont {P.}~\bibnamefont {von Neumann-Cosel}}, \bibinfo {author}
  {\bibfnamefont {T.}~\bibnamefont {Otsuka}}, \bibinfo {author} {\bibfnamefont
  {N.}~\bibnamefont {Pietralla}}, \bibinfo {author} {\bibfnamefont {V.~Y.}\
  \bibnamefont {Ponomarev}}, \bibinfo {author} {\bibfnamefont {N.}~\bibnamefont
  {Shimizu}}, \bibinfo {author} {\bibfnamefont {M.}~\bibnamefont {Singer}},
  \bibinfo {author} {\bibfnamefont {G.}~\bibnamefont {Steinhilber}}, \bibinfo
  {author} {\bibfnamefont {T.}~\bibnamefont {Togashi}}, \bibinfo {author}
  {\bibfnamefont {Y.}~\bibnamefont {Tsunoda}}, \bibinfo {author} {\bibfnamefont
  {V.}~\bibnamefont {Werner}},\ and\ \bibinfo {author} {\bibfnamefont
  {M.}~\bibnamefont {Zweidinger}},\ }\bibfield  {title} {\bibinfo {title}
  {{F}irst {M}easurement of {C}ollectivity of {C}oexisting {S}hapes {B}ased on
  {T}ype {II} {S}hell {E}volution: {T}he {C}ase of $^{96}\mathrm{{Z}r}$},\
  }\href {https://doi.org/10.1103/PhysRevLett.117.172503} {\bibfield  {journal}
  {\bibinfo  {journal} {Phys. Rev. Lett.}\ }\textbf {\bibinfo {volume} {117}},\
  \bibinfo {pages} {172503} (\bibinfo {year} {2016})}\BibitemShut {NoStop}%
\bibitem [{\citenamefont {Ansari}\ \emph {et~al.}(2017)\citenamefont {Ansari},
  \citenamefont {R\'egis}, \citenamefont {Jolie}, \citenamefont {Saed-Samii},
  \citenamefont {Warr}, \citenamefont {Korten}, \citenamefont
  {Zieli\ifmmode~\acute{n}\else \'{n}\fi{}ska}, \citenamefont {Salsac},
  \citenamefont {Blanc}, \citenamefont {Jentschel}, \citenamefont {K\"oster},
  \citenamefont {Mutti}, \citenamefont {Soldner}, \citenamefont {Simpson},
  \citenamefont {Drouet}, \citenamefont {Vancraeyenest}, \citenamefont
  {de~France}, \citenamefont {Cl\'ement}, \citenamefont {Stezowski},
  \citenamefont {Ur}, \citenamefont {Urban}, \citenamefont {Regan},
  \citenamefont {Podoly\'ak}, \citenamefont {Larijani}, \citenamefont
  {Townsley}, \citenamefont {Carroll}, \citenamefont {Wilson}, \citenamefont
  {Mach}, \citenamefont {Fraile}, \citenamefont {Paziy}, \citenamefont
  {Olaizola}, \citenamefont {Vedia}, \citenamefont {Bruce}, \citenamefont
  {Roberts}, \citenamefont {Smith}, \citenamefont {Scheck}, \citenamefont
  {Kr\"oll}, \citenamefont {Hartig}, \citenamefont {Ignatov}, \citenamefont
  {Ilieva}, \citenamefont {Lalkovski}, \citenamefont {M\ifmmode~\u{a}\else
  \u{a}\fi{}rginean}, \citenamefont {Otsuka}, \citenamefont {Shimizu},
  \citenamefont {Togashi},\ and\ \citenamefont {Tsunoda}}]{Ans17a}%
  \BibitemOpen
  \bibfield  {author} {\bibinfo {author} {\bibfnamefont {S.}~\bibnamefont
  {Ansari}}, \bibinfo {author} {\bibfnamefont {J.-M.}\ \bibnamefont {R\'egis}},
  \bibinfo {author} {\bibfnamefont {J.}~\bibnamefont {Jolie}}, \bibinfo
  {author} {\bibfnamefont {N.}~\bibnamefont {Saed-Samii}}, \bibinfo {author}
  {\bibfnamefont {N.}~\bibnamefont {Warr}}, \bibinfo {author} {\bibfnamefont
  {W.}~\bibnamefont {Korten}}, \bibinfo {author} {\bibfnamefont
  {M.}~\bibnamefont {Zieli\ifmmode~\acute{n}\else \'{n}\fi{}ska}}, \bibinfo
  {author} {\bibfnamefont {M.-D.}\ \bibnamefont {Salsac}}, \bibinfo {author}
  {\bibfnamefont {A.}~\bibnamefont {Blanc}}, \bibinfo {author} {\bibfnamefont
  {M.}~\bibnamefont {Jentschel}}, \bibinfo {author} {\bibfnamefont
  {U.}~\bibnamefont {K\"oster}}, \bibinfo {author} {\bibfnamefont
  {P.}~\bibnamefont {Mutti}}, \bibinfo {author} {\bibfnamefont
  {T.}~\bibnamefont {Soldner}}, \bibinfo {author} {\bibfnamefont {G.~S.}\
  \bibnamefont {Simpson}}, \bibinfo {author} {\bibfnamefont {F.}~\bibnamefont
  {Drouet}}, \bibinfo {author} {\bibfnamefont {A.}~\bibnamefont
  {Vancraeyenest}}, \bibinfo {author} {\bibfnamefont {G.}~\bibnamefont
  {de~France}}, \bibinfo {author} {\bibfnamefont {E.}~\bibnamefont
  {Cl\'ement}}, \bibinfo {author} {\bibfnamefont {O.}~\bibnamefont
  {Stezowski}}, \bibinfo {author} {\bibfnamefont {C.~A.}\ \bibnamefont {Ur}},
  \bibinfo {author} {\bibfnamefont {W.}~\bibnamefont {Urban}}, \bibinfo
  {author} {\bibfnamefont {P.~H.}\ \bibnamefont {Regan}}, \bibinfo {author}
  {\bibfnamefont {Z.}~\bibnamefont {Podoly\'ak}}, \bibinfo {author}
  {\bibfnamefont {C.}~\bibnamefont {Larijani}}, \bibinfo {author}
  {\bibfnamefont {C.}~\bibnamefont {Townsley}}, \bibinfo {author}
  {\bibfnamefont {R.}~\bibnamefont {Carroll}}, \bibinfo {author} {\bibfnamefont
  {E.}~\bibnamefont {Wilson}}, \bibinfo {author} {\bibfnamefont
  {H.}~\bibnamefont {Mach}}, \bibinfo {author} {\bibfnamefont {L.~M.}\
  \bibnamefont {Fraile}}, \bibinfo {author} {\bibfnamefont {V.}~\bibnamefont
  {Paziy}}, \bibinfo {author} {\bibfnamefont {B.}~\bibnamefont {Olaizola}},
  \bibinfo {author} {\bibfnamefont {V.}~\bibnamefont {Vedia}}, \bibinfo
  {author} {\bibfnamefont {A.~M.}\ \bibnamefont {Bruce}}, \bibinfo {author}
  {\bibfnamefont {O.~J.}\ \bibnamefont {Roberts}}, \bibinfo {author}
  {\bibfnamefont {J.~F.}\ \bibnamefont {Smith}}, \bibinfo {author}
  {\bibfnamefont {M.}~\bibnamefont {Scheck}}, \bibinfo {author} {\bibfnamefont
  {T.}~\bibnamefont {Kr\"oll}}, \bibinfo {author} {\bibfnamefont {A.-L.}\
  \bibnamefont {Hartig}}, \bibinfo {author} {\bibfnamefont {A.}~\bibnamefont
  {Ignatov}}, \bibinfo {author} {\bibfnamefont {S.}~\bibnamefont {Ilieva}},
  \bibinfo {author} {\bibfnamefont {S.}~\bibnamefont {Lalkovski}}, \bibinfo
  {author} {\bibfnamefont {N.}~\bibnamefont {M\ifmmode~\u{a}\else
  \u{a}\fi{}rginean}}, \bibinfo {author} {\bibfnamefont {T.}~\bibnamefont
  {Otsuka}}, \bibinfo {author} {\bibfnamefont {N.}~\bibnamefont {Shimizu}},
  \bibinfo {author} {\bibfnamefont {T.}~\bibnamefont {Togashi}},\ and\ \bibinfo
  {author} {\bibfnamefont {Y.}~\bibnamefont {Tsunoda}},\ }\bibfield  {title}
  {\bibinfo {title} {Experimental study of the lifetime and phase transition in
  neutron-rich $^{98,100,102}\mathrm{Zr}$},\ }\href
  {https://doi.org/10.1103/PhysRevC.96.054323} {\bibfield  {journal} {\bibinfo
  {journal} {Phys. Rev. C}\ }\textbf {\bibinfo {volume} {96}},\ \bibinfo
  {pages} {054323} (\bibinfo {year} {2017})}\BibitemShut {NoStop}%
\bibitem [{\citenamefont {Witt}\ \emph {et~al.}(2018)\citenamefont {Witt},
  \citenamefont {Werner}, \citenamefont {Pietralla}, \citenamefont {Albers},
  \citenamefont {Ayangeakaa}, \citenamefont {Bucher}, \citenamefont
  {Carpenter}, \citenamefont {Cline}, \citenamefont {David}, \citenamefont
  {Hayes}, \citenamefont {Hoffman}, \citenamefont {Janssens}, \citenamefont
  {Kay}, \citenamefont {Kondev}, \citenamefont {Korten}, \citenamefont
  {Lauritsen}, \citenamefont {M\"oller}, \citenamefont {Rainovski},
  \citenamefont {Savard}, \citenamefont {Seweryniak}, \citenamefont {Smith},
  \citenamefont {Stegmann}, \citenamefont {Zhu},\ and\ \citenamefont
  {Wu}}]{Wit18a}%
  \BibitemOpen
  \bibfield  {author} {\bibinfo {author} {\bibfnamefont {W.}~\bibnamefont
  {Witt}}, \bibinfo {author} {\bibfnamefont {V.}~\bibnamefont {Werner}},
  \bibinfo {author} {\bibfnamefont {N.}~\bibnamefont {Pietralla}}, \bibinfo
  {author} {\bibfnamefont {M.}~\bibnamefont {Albers}}, \bibinfo {author}
  {\bibfnamefont {A.~D.}\ \bibnamefont {Ayangeakaa}}, \bibinfo {author}
  {\bibfnamefont {B.}~\bibnamefont {Bucher}}, \bibinfo {author} {\bibfnamefont
  {M.~P.}\ \bibnamefont {Carpenter}}, \bibinfo {author} {\bibfnamefont
  {D.}~\bibnamefont {Cline}}, \bibinfo {author} {\bibfnamefont {H.~M.}\
  \bibnamefont {David}}, \bibinfo {author} {\bibfnamefont {A.}~\bibnamefont
  {Hayes}}, \bibinfo {author} {\bibfnamefont {C.}~\bibnamefont {Hoffman}},
  \bibinfo {author} {\bibfnamefont {R.~V.~F.}\ \bibnamefont {Janssens}},
  \bibinfo {author} {\bibfnamefont {B.~P.}\ \bibnamefont {Kay}}, \bibinfo
  {author} {\bibfnamefont {F.~G.}\ \bibnamefont {Kondev}}, \bibinfo {author}
  {\bibfnamefont {W.}~\bibnamefont {Korten}}, \bibinfo {author} {\bibfnamefont
  {T.}~\bibnamefont {Lauritsen}}, \bibinfo {author} {\bibfnamefont
  {O.}~\bibnamefont {M\"oller}}, \bibinfo {author} {\bibfnamefont
  {G.}~\bibnamefont {Rainovski}}, \bibinfo {author} {\bibfnamefont
  {G.}~\bibnamefont {Savard}}, \bibinfo {author} {\bibfnamefont
  {D.}~\bibnamefont {Seweryniak}}, \bibinfo {author} {\bibfnamefont
  {J.}~\bibnamefont {Smith}}, \bibinfo {author} {\bibfnamefont
  {R.}~\bibnamefont {Stegmann}}, \bibinfo {author} {\bibfnamefont
  {S.}~\bibnamefont {Zhu}},\ and\ \bibinfo {author} {\bibfnamefont {C.~Y.}\
  \bibnamefont {Wu}},\ }\bibfield  {title} {\bibinfo {title} {Sub-shell closure
  and shape coexistence in the transitional nucleus $^{98}\mathrm{Zr}$},\
  }\href {https://doi.org/10.1103/PhysRevC.98.041302} {\bibfield  {journal}
  {\bibinfo  {journal} {Phys. Rev. C}\ }\textbf {\bibinfo {volume} {98}},\
  \bibinfo {pages} {041302} (\bibinfo {year} {2018})}\BibitemShut {NoStop}%
\bibitem [{\citenamefont {Singh}\ \emph {et~al.}(2018)\citenamefont {Singh},
  \citenamefont {Korten}, \citenamefont {Hagen}, \citenamefont {G\"orgen},
  \citenamefont {Grente}, \citenamefont {Salsac}, \citenamefont {Farget},
  \citenamefont {Cl\'ement}, \citenamefont {de~France}, \citenamefont
  {Braunroth}, \citenamefont {Bruyneel}, \citenamefont {Celikovic},
  \citenamefont {Delaune}, \citenamefont {Dewald}, \citenamefont {Dijon},
  \citenamefont {Delaroche}, \citenamefont {Girod}, \citenamefont {Hackstein},
  \citenamefont {Jacquot}, \citenamefont {Libert}, \citenamefont {Litzinger},
  \citenamefont {Ljungvall}, \citenamefont {Louchart}, \citenamefont
  {Gottardo}, \citenamefont {Michelagnoli}, \citenamefont {M\"uller-Gatermann},
  \citenamefont {Napoli}, \citenamefont {Otsuka}, \citenamefont {Pillet},
  \citenamefont {Recchia}, \citenamefont {Rother}, \citenamefont {Sahin},
  \citenamefont {Siem}, \citenamefont {Sulignano}, \citenamefont {Togashi},
  \citenamefont {Tsunoda}, \citenamefont {Theisen},\ and\ \citenamefont
  {Valiente-Dobon}}]{Sin18a}%
  \BibitemOpen
  \bibfield  {author} {\bibinfo {author} {\bibfnamefont {P.}~\bibnamefont
  {Singh}}, \bibinfo {author} {\bibfnamefont {W.}~\bibnamefont {Korten}},
  \bibinfo {author} {\bibfnamefont {T.~W.}\ \bibnamefont {Hagen}}, \bibinfo
  {author} {\bibfnamefont {A.}~\bibnamefont {G\"orgen}}, \bibinfo {author}
  {\bibfnamefont {L.}~\bibnamefont {Grente}}, \bibinfo {author} {\bibfnamefont
  {M.-D.}\ \bibnamefont {Salsac}}, \bibinfo {author} {\bibfnamefont
  {F.}~\bibnamefont {Farget}}, \bibinfo {author} {\bibfnamefont
  {E.}~\bibnamefont {Cl\'ement}}, \bibinfo {author} {\bibfnamefont
  {G.}~\bibnamefont {de~France}}, \bibinfo {author} {\bibfnamefont
  {T.}~\bibnamefont {Braunroth}}, \bibinfo {author} {\bibfnamefont
  {B.}~\bibnamefont {Bruyneel}}, \bibinfo {author} {\bibfnamefont
  {I.}~\bibnamefont {Celikovic}}, \bibinfo {author} {\bibfnamefont
  {O.}~\bibnamefont {Delaune}}, \bibinfo {author} {\bibfnamefont
  {A.}~\bibnamefont {Dewald}}, \bibinfo {author} {\bibfnamefont
  {A.}~\bibnamefont {Dijon}}, \bibinfo {author} {\bibfnamefont {J.-P.}\
  \bibnamefont {Delaroche}}, \bibinfo {author} {\bibfnamefont {M.}~\bibnamefont
  {Girod}}, \bibinfo {author} {\bibfnamefont {M.}~\bibnamefont {Hackstein}},
  \bibinfo {author} {\bibfnamefont {B.}~\bibnamefont {Jacquot}}, \bibinfo
  {author} {\bibfnamefont {J.}~\bibnamefont {Libert}}, \bibinfo {author}
  {\bibfnamefont {J.}~\bibnamefont {Litzinger}}, \bibinfo {author}
  {\bibfnamefont {J.}~\bibnamefont {Ljungvall}}, \bibinfo {author}
  {\bibfnamefont {C.}~\bibnamefont {Louchart}}, \bibinfo {author}
  {\bibfnamefont {A.}~\bibnamefont {Gottardo}}, \bibinfo {author}
  {\bibfnamefont {C.}~\bibnamefont {Michelagnoli}}, \bibinfo {author}
  {\bibfnamefont {C.}~\bibnamefont {M\"uller-Gatermann}}, \bibinfo {author}
  {\bibfnamefont {D.~R.}\ \bibnamefont {Napoli}}, \bibinfo {author}
  {\bibfnamefont {T.}~\bibnamefont {Otsuka}}, \bibinfo {author} {\bibfnamefont
  {N.}~\bibnamefont {Pillet}}, \bibinfo {author} {\bibfnamefont
  {F.}~\bibnamefont {Recchia}}, \bibinfo {author} {\bibfnamefont
  {W.}~\bibnamefont {Rother}}, \bibinfo {author} {\bibfnamefont
  {E.}~\bibnamefont {Sahin}}, \bibinfo {author} {\bibfnamefont
  {S.}~\bibnamefont {Siem}}, \bibinfo {author} {\bibfnamefont {B.}~\bibnamefont
  {Sulignano}}, \bibinfo {author} {\bibfnamefont {T.}~\bibnamefont {Togashi}},
  \bibinfo {author} {\bibfnamefont {Y.}~\bibnamefont {Tsunoda}}, \bibinfo
  {author} {\bibfnamefont {C.}~\bibnamefont {Theisen}},\ and\ \bibinfo {author}
  {\bibfnamefont {J.~J.}\ \bibnamefont {Valiente-Dobon}},\ }\bibfield  {title}
  {\bibinfo {title} {{E}vidence for {C}oexisting {S}hapes through {L}ifetime
  {M}easurements in $^{98}\mathrm{Zr}$},\ }\href
  {https://doi.org/10.1103/PhysRevLett.121.192501} {\bibfield  {journal}
  {\bibinfo  {journal} {Phys. Rev. Lett.}\ }\textbf {\bibinfo {volume} {121}},\
  \bibinfo {pages} {192501} (\bibinfo {year} {2018})}\BibitemShut {NoStop}%
\bibitem [{\citenamefont {Iskra}\ \emph {et~al.}(2019)\citenamefont {Iskra},
  \citenamefont {Broda}, \citenamefont {Janssens}, \citenamefont {Carpenter},
  \citenamefont {Fornal}, \citenamefont {Lauritsen}, \citenamefont {Otsuka},
  \citenamefont {Togashi}, \citenamefont {Tsunoda}, \citenamefont {Walters},\
  and\ \citenamefont {Zhu}}]{Isk19a}%
  \BibitemOpen
  \bibfield  {author} {\bibinfo {author} {\bibfnamefont {{\L}.~W.}\
  \bibnamefont {Iskra}}, \bibinfo {author} {\bibfnamefont {R.}~\bibnamefont
  {Broda}}, \bibinfo {author} {\bibfnamefont {R.~V.~F.}\ \bibnamefont
  {Janssens}}, \bibinfo {author} {\bibfnamefont {M.~P.}\ \bibnamefont
  {Carpenter}}, \bibinfo {author} {\bibfnamefont {B.}~\bibnamefont {Fornal}},
  \bibinfo {author} {\bibfnamefont {T.}~\bibnamefont {Lauritsen}}, \bibinfo
  {author} {\bibfnamefont {T.}~\bibnamefont {Otsuka}}, \bibinfo {author}
  {\bibfnamefont {T.}~\bibnamefont {Togashi}}, \bibinfo {author} {\bibfnamefont
  {Y.}~\bibnamefont {Tsunoda}}, \bibinfo {author} {\bibfnamefont {W.~B.}\
  \bibnamefont {Walters}},\ and\ \bibinfo {author} {\bibfnamefont
  {S.}~\bibnamefont {Zhu}},\ }\bibfield  {title} {\bibinfo {title} {{R}evised
  {B(E3)} transition rate and structure of the $3^-$ level in $^{96}${Zr}},\
  }\href {https://doi.org/https://doi.org/10.1016/j.physletb.2018.10.069}
  {\bibfield  {journal} {\bibinfo  {journal} {Phys. Lett. B}\ }\textbf
  {\bibinfo {volume} {788}},\ \bibinfo {pages} {396} (\bibinfo {year}
  {2019})}\BibitemShut {NoStop}%
\bibitem [{\citenamefont {Karayonchev}\ \emph {et~al.}(2020)\citenamefont
  {Karayonchev}, \citenamefont {Jolie}, \citenamefont {Blazhev}, \citenamefont
  {Dewald}, \citenamefont {Esmaylzadeh}, \citenamefont {Fransen}, \citenamefont
  {H\"afner}, \citenamefont {Knafla}, \citenamefont {Litzinger}, \citenamefont
  {M\"uller-Gatermann}, \citenamefont {R\'egis}, \citenamefont {Schomacker},
  \citenamefont {Vogt}, \citenamefont {Warr}, \citenamefont {Leviatan},\ and\
  \citenamefont {Gavrielov}}]{Kar20a}%
  \BibitemOpen
  \bibfield  {author} {\bibinfo {author} {\bibfnamefont {V.}~\bibnamefont
  {Karayonchev}}, \bibinfo {author} {\bibfnamefont {J.}~\bibnamefont {Jolie}},
  \bibinfo {author} {\bibfnamefont {A.}~\bibnamefont {Blazhev}}, \bibinfo
  {author} {\bibfnamefont {A.}~\bibnamefont {Dewald}}, \bibinfo {author}
  {\bibfnamefont {A.}~\bibnamefont {Esmaylzadeh}}, \bibinfo {author}
  {\bibfnamefont {C.}~\bibnamefont {Fransen}}, \bibinfo {author} {\bibfnamefont
  {G.}~\bibnamefont {H\"afner}}, \bibinfo {author} {\bibfnamefont
  {L.}~\bibnamefont {Knafla}}, \bibinfo {author} {\bibfnamefont
  {J.}~\bibnamefont {Litzinger}}, \bibinfo {author} {\bibfnamefont
  {C.}~\bibnamefont {M\"uller-Gatermann}}, \bibinfo {author} {\bibfnamefont
  {J.-M.}\ \bibnamefont {R\'egis}}, \bibinfo {author} {\bibfnamefont
  {K.}~\bibnamefont {Schomacker}}, \bibinfo {author} {\bibfnamefont
  {A.}~\bibnamefont {Vogt}}, \bibinfo {author} {\bibfnamefont {N.}~\bibnamefont
  {Warr}}, \bibinfo {author} {\bibfnamefont {A.}~\bibnamefont {Leviatan}},\
  and\ \bibinfo {author} {\bibfnamefont {N.}~\bibnamefont {Gavrielov}},\
  }\bibfield  {title} {\bibinfo {title} {Tests of collectivity in
  $^{98}\mathrm{Zr}$ by absolute transition rates},\ }\href
  {https://doi.org/10.1103/PhysRevC.102.064314} {\bibfield  {journal} {\bibinfo
   {journal} {Phys. Rev. C}\ }\textbf {\bibinfo {volume} {102}},\ \bibinfo
  {pages} {064314} (\bibinfo {year} {2020})}\BibitemShut {NoStop}%
\bibitem [{\citenamefont {Wi\ifmmode~\acute{s}\else \'{s}\fi{}niewski}\ \emph
  {et~al.}(2023)\citenamefont {Wi\ifmmode~\acute{s}\else \'{s}\fi{}niewski},
  \citenamefont {Urban}, \citenamefont {Rz{a}ca-Urban}, \citenamefont
  {Sieja}, \citenamefont {Blanc}, \citenamefont {Jentschel}, \citenamefont
  {Micheagnoli}, \citenamefont {Mutti}, \citenamefont {K\"oster}, \citenamefont
  {de~France}, \citenamefont {Simpson},\ and\ \citenamefont {Ur}}]{Wis23a}%
  \BibitemOpen
  \bibfield  {author} {\bibinfo {author} {\bibfnamefont {J.}~\bibnamefont
  {Wi\ifmmode~\acute{s}\else \'{s}\fi{}niewski}}, \bibinfo {author}
  {\bibfnamefont {W.}~\bibnamefont {Urban}}, \bibinfo {author} {\bibfnamefont
  {T.}~\bibnamefont {Rz{a}ca-Urban}}, \bibinfo {author} {\bibfnamefont
  {K.}~\bibnamefont {Sieja}}, \bibinfo {author} {\bibfnamefont
  {A.}~\bibnamefont {Blanc}}, \bibinfo {author} {\bibfnamefont
  {M.}~\bibnamefont {Jentschel}}, \bibinfo {author} {\bibfnamefont
  {C.}~\bibnamefont {Micheagnoli}}, \bibinfo {author} {\bibfnamefont
  {P.}~\bibnamefont {Mutti}}, \bibinfo {author} {\bibfnamefont
  {U.}~\bibnamefont {K\"oster}}, \bibinfo {author} {\bibfnamefont
  {G.}~\bibnamefont {de~France}}, \bibinfo {author} {\bibfnamefont {G.~S.}\
  \bibnamefont {Simpson}},\ and\ \bibinfo {author} {\bibfnamefont {C.~A.}\
  \bibnamefont {Ur}},\ }\bibfield  {title} {\bibinfo {title} {Structure of
  ${N}=56$ isotones with $36\ensuremath{\le}{Z}\ensuremath{\le}42$ protons},\
  }\href {https://doi.org/10.1103/PhysRevC.108.024302} {\bibfield  {journal}
  {\bibinfo  {journal} {Phys. Rev. C}\ }\textbf {\bibinfo {volume} {108}},\
  \bibinfo {pages} {024302} (\bibinfo {year} {2023})}\BibitemShut {NoStop}%
\bibitem [{\citenamefont {Pfeil}\ \emph {et~al.}(2023)\citenamefont {Pfeil},
  \citenamefont {Nomura}, \citenamefont {Gavrielov}, \citenamefont {R\'egis},
  \citenamefont {K\"oster}, \citenamefont {Kim}, \citenamefont {Esmaylzadeh},
  \citenamefont {Harter}, \citenamefont {Jolie}, \citenamefont {Knafla},
  \citenamefont {Ley},\ and\ \citenamefont {Karayonchev}}]{Pfe23a}%
  \BibitemOpen
  \bibfield  {author} {\bibinfo {author} {\bibfnamefont {A.}~\bibnamefont
  {Pfeil}}, \bibinfo {author} {\bibfnamefont {K.}~\bibnamefont {Nomura}},
  \bibinfo {author} {\bibfnamefont {N.}~\bibnamefont {Gavrielov}}, \bibinfo
  {author} {\bibfnamefont {J.-M.}\ \bibnamefont {R\'egis}}, \bibinfo {author}
  {\bibfnamefont {U.}~\bibnamefont {K\"oster}}, \bibinfo {author}
  {\bibfnamefont {Y.~H.}\ \bibnamefont {Kim}}, \bibinfo {author} {\bibfnamefont
  {A.}~\bibnamefont {Esmaylzadeh}}, \bibinfo {author} {\bibfnamefont
  {A.}~\bibnamefont {Harter}}, \bibinfo {author} {\bibfnamefont
  {J.}~\bibnamefont {Jolie}}, \bibinfo {author} {\bibfnamefont
  {L.}~\bibnamefont {Knafla}}, \bibinfo {author} {\bibfnamefont
  {M.}~\bibnamefont {Ley}},\ and\ \bibinfo {author} {\bibfnamefont
  {V.}~\bibnamefont {Karayonchev}},\ }\bibfield  {title} {\bibinfo {title}
  {Lifetime measurements in $^{99}\mathrm{Nb}$ and $^{99}\mathrm{Zr}$:
  Investigation of shape coexistence},\ }\href
  {https://doi.org/10.1103/PhysRevC.108.034310} {\bibfield  {journal} {\bibinfo
   {journal} {Phys. Rev. C}\ }\textbf {\bibinfo {volume} {108}},\ \bibinfo
  {pages} {034310} (\bibinfo {year} {2023})}\BibitemShut {NoStop}%
\bibitem [{\citenamefont {Wu}\ \emph {et~al.}(2024)\citenamefont {Wu},
  \citenamefont {Carpenter}, \citenamefont {Kondev}, \citenamefont {Janssens},
  \citenamefont {Zhu}, \citenamefont {McCutchan}, \citenamefont {Ayangeakaa},
  \citenamefont {Chen}, \citenamefont {Clark}, \citenamefont {Hartley},
  \citenamefont {Lauritsen}, \citenamefont {Pietralla}, \citenamefont {Savard},
  \citenamefont {Seweryniak},\ and\ \citenamefont {Werner}}]{Wu24a}%
  \BibitemOpen
  \bibfield  {author} {\bibinfo {author} {\bibfnamefont {J.}~\bibnamefont
  {Wu}}, \bibinfo {author} {\bibfnamefont {M.~P.}\ \bibnamefont {Carpenter}},
  \bibinfo {author} {\bibfnamefont {F.~G.}\ \bibnamefont {Kondev}}, \bibinfo
  {author} {\bibfnamefont {R.~V.~F.}\ \bibnamefont {Janssens}}, \bibinfo
  {author} {\bibfnamefont {S.}~\bibnamefont {Zhu}}, \bibinfo {author}
  {\bibfnamefont {E.~A.}\ \bibnamefont {McCutchan}}, \bibinfo {author}
  {\bibfnamefont {A.~D.}\ \bibnamefont {Ayangeakaa}}, \bibinfo {author}
  {\bibfnamefont {J.}~\bibnamefont {Chen}}, \bibinfo {author} {\bibfnamefont
  {J.}~\bibnamefont {Clark}}, \bibinfo {author} {\bibfnamefont {D.~J.}\
  \bibnamefont {Hartley}}, \bibinfo {author} {\bibfnamefont {T.}~\bibnamefont
  {Lauritsen}}, \bibinfo {author} {\bibfnamefont {N.}~\bibnamefont
  {Pietralla}}, \bibinfo {author} {\bibfnamefont {G.}~\bibnamefont {Savard}},
  \bibinfo {author} {\bibfnamefont {D.}~\bibnamefont {Seweryniak}},\ and\
  \bibinfo {author} {\bibfnamefont {V.}~\bibnamefont {Werner}},\ }\bibfield
  {title} {\bibinfo {title} {Determination of the spins and parities for the
  ${0}_{4}^{+}$ and ${0}_{5}^{+}$ states in $^{100}\mathrm{Zr}$},\ }\href
  {https://doi.org/10.1103/PhysRevC.109.024314} {\bibfield  {journal} {\bibinfo
   {journal} {Phys. Rev. C}\ }\textbf {\bibinfo {volume} {109}},\ \bibinfo
  {pages} {024314} (\bibinfo {year} {2024})}\BibitemShut {NoStop}%
\bibitem [{\citenamefont {Petrovici}(2012)}]{Pet12a}%
  \BibitemOpen
  \bibfield  {author} {\bibinfo {author} {\bibfnamefont {A.}~\bibnamefont
  {Petrovici}},\ }\bibfield  {title} {\bibinfo {title} {Triple shape
  coexistence and shape evolution in the ${N}=58$ {Sr} and {Zr} isotopes},\
  }\href {https://doi.org/10.1103/PhysRevC.85.034337} {\bibfield  {journal}
  {\bibinfo  {journal} {Phys. Rev. C}\ }\textbf {\bibinfo {volume} {85}},\
  \bibinfo {pages} {034337} (\bibinfo {year} {2012})}\BibitemShut {NoStop}%
\bibitem [{\citenamefont {Otsuka}\ \emph {et~al.}(2001)\citenamefont {Otsuka},
  \citenamefont {Honma}, \citenamefont {Mizusaki}, \citenamefont {Shimizu},\
  and\ \citenamefont {Utsuno}}]{Ots01a}%
  \BibitemOpen
  \bibfield  {author} {\bibinfo {author} {\bibfnamefont {T.}~\bibnamefont
  {Otsuka}}, \bibinfo {author} {\bibfnamefont {M.}~\bibnamefont {Honma}},
  \bibinfo {author} {\bibfnamefont {T.}~\bibnamefont {Mizusaki}}, \bibinfo
  {author} {\bibfnamefont {N.}~\bibnamefont {Shimizu}},\ and\ \bibinfo {author}
  {\bibfnamefont {Y.}~\bibnamefont {Utsuno}},\ }\bibfield  {title} {\bibinfo
  {title} {Monte carlo shell model for atomic nuclei},\ }\href
  {https://doi.org/https://doi.org/10.1016/S0146-6410(01)00157-0} {\bibfield
  {journal} {\bibinfo  {journal} {Prog. Part. Nucl. Phys.}\ }\textbf {\bibinfo
  {volume} {47}},\ \bibinfo {pages} {319 } (\bibinfo {year}
  {2001})}\BibitemShut {NoStop}%
\bibitem [{\citenamefont {Shimizu}\ \emph {et~al.}(2012)\citenamefont
  {Shimizu}, \citenamefont {Abe}, \citenamefont {Tsunoda}, \citenamefont
  {Utsuno}, \citenamefont {Yoshida}, \citenamefont {Mizusaki}, \citenamefont
  {Honma},\ and\ \citenamefont {Otsuka}}]{Shi12a}%
  \BibitemOpen
  \bibfield  {author} {\bibinfo {author} {\bibfnamefont {N.}~\bibnamefont
  {Shimizu}}, \bibinfo {author} {\bibfnamefont {T.}~\bibnamefont {Abe}},
  \bibinfo {author} {\bibfnamefont {Y.}~\bibnamefont {Tsunoda}}, \bibinfo
  {author} {\bibfnamefont {Y.}~\bibnamefont {Utsuno}}, \bibinfo {author}
  {\bibfnamefont {T.}~\bibnamefont {Yoshida}}, \bibinfo {author} {\bibfnamefont
  {T.}~\bibnamefont {Mizusaki}}, \bibinfo {author} {\bibfnamefont
  {M.}~\bibnamefont {Honma}},\ and\ \bibinfo {author} {\bibfnamefont
  {T.}~\bibnamefont {Otsuka}},\ }\bibfield  {title} {\bibinfo {title}
  {{New-generation Monte Carlo shell model for the K computer era}},\ }\href
  {https://doi.org/10.1093/ptep/pts012} {\bibfield  {journal} {\bibinfo
  {journal} {Prog. Theo. Exp. Phys.}\ }\textbf {\bibinfo {volume} {2012}}
  (\bibinfo {year} {2012})},\ \bibinfo {note} {01A205}\BibitemShut {NoStop}%
\bibitem [{\citenamefont {Duval}\ and\ \citenamefont {Barrett}(1981)}]{Duv81a}%
  \BibitemOpen
  \bibfield  {author} {\bibinfo {author} {\bibfnamefont {P.~D.}\ \bibnamefont
  {Duval}}\ and\ \bibinfo {author} {\bibfnamefont {B.~R.}\ \bibnamefont
  {Barrett}},\ }\bibfield  {title} {\bibinfo {title} {Configuration mixing in
  the interacting boson model},\ }\href
  {https://doi.org/https://doi.org/10.1016/0370-2693(81)90321-X} {\bibfield
  {journal} {\bibinfo  {journal} {Phys. Lett. B}\ }\textbf {\bibinfo {volume}
  {100}},\ \bibinfo {pages} {223} (\bibinfo {year} {1981})}\BibitemShut
  {NoStop}%
\bibitem [{\citenamefont {Duval}\ and\ \citenamefont {Barrett}(1982)}]{Duv82a}%
  \BibitemOpen
  \bibfield  {author} {\bibinfo {author} {\bibfnamefont {P.~D.}\ \bibnamefont
  {Duval}}\ and\ \bibinfo {author} {\bibfnamefont {B.~R.}\ \bibnamefont
  {Barrett}},\ }\bibfield  {title} {\bibinfo {title} {Quantitative description
  of configuration mixing in the interacting boson model},\ }\href
  {https://doi.org/https://doi.org/10.1016/0375-9474(82)90061-6} {\bibfield
  {journal} {\bibinfo  {journal} {Nucl. Phys. A}\ }\textbf {\bibinfo {volume}
  {376}},\ \bibinfo {pages} {213} (\bibinfo {year} {1982})}\BibitemShut
  {NoStop}%
\bibitem [{\citenamefont {Garc\'{\i}a-Ramos}\ and\ \citenamefont
  {Heyde}(2019)}]{Gar19a}%
  \BibitemOpen
  \bibfield  {author} {\bibinfo {author} {\bibfnamefont {J.~E.}\ \bibnamefont
  {Garc\'{\i}a-Ramos}}\ and\ \bibinfo {author} {\bibfnamefont {K.}~\bibnamefont
  {Heyde}},\ }\bibfield  {title} {\bibinfo {title} {Quest of shape coexistence
  in {Zr} isotopes},\ }\href {https://doi.org/10.1103/PhysRevC.100.044315}
  {\bibfield  {journal} {\bibinfo  {journal} {Phys. Rev. C}\ }\textbf {\bibinfo
  {volume} {100}},\ \bibinfo {pages} {044315} (\bibinfo {year}
  {2019})}\BibitemShut {NoStop}%
\bibitem [{\citenamefont {Gavrielov}\ \emph
  {et~al.}(2019{\natexlab{a}})\citenamefont {Gavrielov}, \citenamefont
  {Leviatan},\ and\ \citenamefont {Iachello}}]{Gav19a}%
  \BibitemOpen
  \bibfield  {author} {\bibinfo {author} {\bibfnamefont {N.}~\bibnamefont
  {Gavrielov}}, \bibinfo {author} {\bibfnamefont {A.}~\bibnamefont
  {Leviatan}},\ and\ \bibinfo {author} {\bibfnamefont {F.}~\bibnamefont
  {Iachello}},\ }\bibfield  {title} {\bibinfo {title} {Intertwined quantum
  phase transitions in the {Zr} isotopes},\ }\href
  {https://doi.org/10.1103/PhysRevC.99.064324} {\bibfield  {journal} {\bibinfo
  {journal} {Phys. Rev. C}\ }\textbf {\bibinfo {volume} {99}},\ \bibinfo
  {pages} {064324} (\bibinfo {year} {2019}{\natexlab{a}})}\BibitemShut
  {NoStop}%
\bibitem [{\citenamefont {Gavrielov}\ \emph
  {et~al.}(2019{\natexlab{b}})\citenamefont {Gavrielov}, \citenamefont
  {Leviatan},\ and\ \citenamefont {Iachello}}]{Gav20a}%
  \BibitemOpen
  \bibfield  {author} {\bibinfo {author} {\bibfnamefont {N.}~\bibnamefont
  {Gavrielov}}, \bibinfo {author} {\bibfnamefont {A.}~\bibnamefont
  {Leviatan}},\ and\ \bibinfo {author} {\bibfnamefont {F.}~\bibnamefont
  {Iachello}},\ }\bibfield  {title} {\bibinfo {title} {Interplay between
  shape-phase transitions and shape coexistence in the {Zr} isotopes},\ }\href
  {https://doi.org/10.1088/1402-4896/ab456b} {\bibfield  {journal} {\bibinfo
  {journal} {Phys. Scr.}\ }\textbf {\bibinfo {volume} {95}},\ \bibinfo {pages}
  {024001} (\bibinfo {year} {2019}{\natexlab{b}})}\BibitemShut {NoStop}%
\bibitem [{\citenamefont {Gavrielov}\ \emph {et~al.}(2022)\citenamefont
  {Gavrielov}, \citenamefont {Leviatan},\ and\ \citenamefont
  {Iachello}}]{Gav22a}%
  \BibitemOpen
  \bibfield  {author} {\bibinfo {author} {\bibfnamefont {N.}~\bibnamefont
  {Gavrielov}}, \bibinfo {author} {\bibfnamefont {A.}~\bibnamefont
  {Leviatan}},\ and\ \bibinfo {author} {\bibfnamefont {F.}~\bibnamefont
  {Iachello}},\ }\bibfield  {title} {\bibinfo {title} {{Zr} isotopes as a
  region of intertwined quantum phase transitions},\ }\href
  {https://doi.org/10.1103/PhysRevC.105.014305} {\bibfield  {journal} {\bibinfo
   {journal} {Phys. Rev. C}\ }\textbf {\bibinfo {volume} {105}},\ \bibinfo
  {pages} {014305} (\bibinfo {year} {2022})}\BibitemShut {NoStop}%
\bibitem [{\citenamefont {Abriola}\ and\ \citenamefont
  {Sonzogni}(2008)}]{NDS96}%
  \BibitemOpen
  \bibfield  {author} {\bibinfo {author} {\bibfnamefont {D.}~\bibnamefont
  {Abriola}}\ and\ \bibinfo {author} {\bibfnamefont {A.}~\bibnamefont
  {Sonzogni}},\ }\bibfield  {title} {\bibinfo {title} {Nuclear data sheets for
  {A = 96}},\ }\href
  {https://doi.org/https://doi.org/10.1016/j.nds.2008.10.002} {\bibfield
  {journal} {\bibinfo  {journal} {Nucl. Data Sheets}\ }\textbf {\bibinfo
  {volume} {109}},\ \bibinfo {pages} {2501} (\bibinfo {year}
  {2008})}\BibitemShut {NoStop}%
\bibitem [{\citenamefont {Zieli\'nska}\ \emph {et~al.}(2026)\citenamefont
  {Zieli\'nska}, \citenamefont {Garrett}, \citenamefont {Rocchini},
  \citenamefont {Pannu}, \citenamefont {Bergmaier}, \citenamefont {Bildstein},
  \citenamefont {Calder}, \citenamefont {Dworatzek}, \citenamefont {Vedia},
  \citenamefont {Ahmed}, \citenamefont {Andreoiu}, \citenamefont {Annen},
  \citenamefont {Asch}, \citenamefont {Avaa}, \citenamefont {Ball},
  \citenamefont {Buck}, \citenamefont {Coleman}, \citenamefont {{Diaz Varela}},
  \citenamefont {Doherty}, \citenamefont {Dowie}, \citenamefont {Faestermann},
  \citenamefont {Garnsworthy}, \citenamefont {Geerlof}, \citenamefont
  {Greaves}, \citenamefont {Grimes}, \citenamefont {Hackman}, \citenamefont
  {Hertenberger}, \citenamefont {Hicks}, \citenamefont {Kalaydjieva},
  \citenamefont {Karayonchev}, \citenamefont {{La Marca}}, \citenamefont
  {Lange}, \citenamefont {Lenardo}, \citenamefont {Maqungo}, \citenamefont
  {Marchini}, \citenamefont {Marlow}, \citenamefont {Martin}, \citenamefont
  {Mashtakov}, \citenamefont {McCourt}, \citenamefont {Murias}, \citenamefont
  {Murillo-Morales}, \citenamefont {Nannini}, \citenamefont {Peters},
  \citenamefont {Petrache}, \citenamefont {Porzio}, \citenamefont {Richardson},
  \citenamefont {Scheck}, \citenamefont {Spagnoletti}, \citenamefont
  {Stoychev}, \citenamefont {Svensson}, \citenamefont {Taddei}, \citenamefont
  {Tocabens}, \citenamefont {Triambak}, \citenamefont {Vandebrouck},
  \citenamefont {Umashankar}, \citenamefont {Valbuena}, \citenamefont
  {Wrzosek-Lipska}, \citenamefont {Wu},\ and\ \citenamefont {Zidar}}]{Zie26a}%
  \BibitemOpen
  \bibfield  {author} {\bibinfo {author} {\bibfnamefont {M.}~\bibnamefont
  {Zieli\'nska}}, \bibinfo {author} {\bibfnamefont {P.~E.}\ \bibnamefont
  {Garrett}}, \bibinfo {author} {\bibfnamefont {M.}~\bibnamefont {Rocchini}},
  \bibinfo {author} {\bibfnamefont {S.}~\bibnamefont {Pannu}}, \bibinfo
  {author} {\bibfnamefont {A.}~\bibnamefont {Bergmaier}}, \bibinfo {author}
  {\bibfnamefont {V.}~\bibnamefont {Bildstein}}, \bibinfo {author}
  {\bibfnamefont {A.~B.}\ \bibnamefont {Calder}}, \bibinfo {author}
  {\bibfnamefont {S.~R.}\ \bibnamefont {Dworatzek}}, \bibinfo {author}
  {\bibfnamefont {V.}~\bibnamefont {Vedia}}, \bibinfo {author} {\bibfnamefont
  {Z.~T.}\ \bibnamefont {Ahmed}}, \bibinfo {author} {\bibfnamefont
  {C.}~\bibnamefont {Andreoiu}}, \bibinfo {author} {\bibfnamefont {D.~W.}\
  \bibnamefont {Annen}}, \bibinfo {author} {\bibfnamefont {H.}~\bibnamefont
  {Asch}}, \bibinfo {author} {\bibfnamefont {A.~A.}\ \bibnamefont {Avaa}},
  \bibinfo {author} {\bibfnamefont {G.~C.}\ \bibnamefont {Ball}}, \bibinfo
  {author} {\bibfnamefont {S.}~\bibnamefont {Buck}}, \bibinfo {author}
  {\bibfnamefont {R.~J.}\ \bibnamefont {Coleman}}, \bibinfo {author}
  {\bibfnamefont {A.}~\bibnamefont {{Diaz Varela}}}, \bibinfo {author}
  {\bibfnamefont {D.~T.}\ \bibnamefont {Doherty}}, \bibinfo {author}
  {\bibfnamefont {J.~T.~H.}\ \bibnamefont {Dowie}}, \bibinfo {author}
  {\bibfnamefont {T.}~\bibnamefont {Faestermann}}, \bibinfo {author}
  {\bibfnamefont {A.~B.}\ \bibnamefont {Garnsworthy}}, \bibinfo {author}
  {\bibfnamefont {E.~D.}\ \bibnamefont {Geerlof}}, \bibinfo {author}
  {\bibfnamefont {B.}~\bibnamefont {Greaves}}, \bibinfo {author} {\bibfnamefont
  {A.~L.}\ \bibnamefont {Grimes}}, \bibinfo {author} {\bibfnamefont
  {G.}~\bibnamefont {Hackman}}, \bibinfo {author} {\bibfnamefont
  {R.}~\bibnamefont {Hertenberger}}, \bibinfo {author} {\bibfnamefont {S.~F.}\
  \bibnamefont {Hicks}}, \bibinfo {author} {\bibfnamefont {D.}~\bibnamefont
  {Kalaydjieva}}, \bibinfo {author} {\bibfnamefont {V.}~\bibnamefont
  {Karayonchev}}, \bibinfo {author} {\bibfnamefont {T.}~\bibnamefont {{La
  Marca}}}, \bibinfo {author} {\bibfnamefont {S.}~\bibnamefont {Lange}},
  \bibinfo {author} {\bibfnamefont {B.}~\bibnamefont {Lenardo}}, \bibinfo
  {author} {\bibfnamefont {L.}~\bibnamefont {Maqungo}}, \bibinfo {author}
  {\bibfnamefont {N.}~\bibnamefont {Marchini}}, \bibinfo {author}
  {\bibfnamefont {B.}~\bibnamefont {Marlow}}, \bibinfo {author} {\bibfnamefont
  {M.~S.}\ \bibnamefont {Martin}}, \bibinfo {author} {\bibfnamefont {K.~R.}\
  \bibnamefont {Mashtakov}}, \bibinfo {author} {\bibfnamefont {E.}~\bibnamefont
  {McCourt}}, \bibinfo {author} {\bibfnamefont {J.~R.}\ \bibnamefont {Murias}},
  \bibinfo {author} {\bibfnamefont {S.}~\bibnamefont {Murillo-Morales}},
  \bibinfo {author} {\bibfnamefont {A.}~\bibnamefont {Nannini}}, \bibinfo
  {author} {\bibfnamefont {E.~E.}\ \bibnamefont {Peters}}, \bibinfo {author}
  {\bibfnamefont {C.~M.}\ \bibnamefont {Petrache}}, \bibinfo {author}
  {\bibfnamefont {C.}~\bibnamefont {Porzio}}, \bibinfo {author} {\bibfnamefont
  {G.}~\bibnamefont {Richardson}}, \bibinfo {author} {\bibfnamefont
  {M.}~\bibnamefont {Scheck}}, \bibinfo {author} {\bibfnamefont
  {P.}~\bibnamefont {Spagnoletti}}, \bibinfo {author} {\bibfnamefont
  {K.}~\bibnamefont {Stoychev}}, \bibinfo {author} {\bibfnamefont {C.~E.}\
  \bibnamefont {Svensson}}, \bibinfo {author} {\bibfnamefont {E.}~\bibnamefont
  {Taddei}}, \bibinfo {author} {\bibfnamefont {G.}~\bibnamefont {Tocabens}},
  \bibinfo {author} {\bibfnamefont {S.}~\bibnamefont {Triambak}}, \bibinfo
  {author} {\bibfnamefont {M.}~\bibnamefont {Vandebrouck}}, \bibinfo {author}
  {\bibfnamefont {R.}~\bibnamefont {Umashankar}}, \bibinfo {author}
  {\bibfnamefont {S.}~\bibnamefont {Valbuena}}, \bibinfo {author}
  {\bibfnamefont {K.}~\bibnamefont {Wrzosek-Lipska}}, \bibinfo {author}
  {\bibfnamefont {F.}~\bibnamefont {Wu}},\ and\ \bibinfo {author}
  {\bibfnamefont {T.}~\bibnamefont {Zidar}},\ }\bibfield  {title} {\bibinfo
  {title} {Collectivity of the deformed structure in $^{96}${Zr} from
  complementary {C}oulomb-excitation and $\beta$-decay measurements},\ }\href
  {https://doi.org/https://doi.org/10.1016/j.physletb.2026.140481} {\bibfield
  {journal} {\bibinfo  {journal} {Phys. Lett. B}\ }\textbf {\bibinfo {volume}
  {877}},\ \bibinfo {pages} {140481} (\bibinfo {year} {2026})}\BibitemShut
  {NoStop}%
\bibitem [{\citenamefont {Shimizu}\ \emph {et~al.}(2021)\citenamefont
  {Shimizu}, \citenamefont {Tsunoda}, \citenamefont {Utsuno},\ and\
  \citenamefont {Otsuka}}]{Shi21a}%
  \BibitemOpen
  \bibfield  {author} {\bibinfo {author} {\bibfnamefont {N.}~\bibnamefont
  {Shimizu}}, \bibinfo {author} {\bibfnamefont {Y.}~\bibnamefont {Tsunoda}},
  \bibinfo {author} {\bibfnamefont {Y.}~\bibnamefont {Utsuno}},\ and\ \bibinfo
  {author} {\bibfnamefont {T.}~\bibnamefont {Otsuka}},\ }\bibfield  {title}
  {\bibinfo {title} {Variational approach with the superposition of the
  symmetry-restored quasiparticle vacua for nuclear shell-model calculations},\
  }\href {https://doi.org/10.1103/PhysRevC.103.014312} {\bibfield  {journal}
  {\bibinfo  {journal} {Phys. Rev. C}\ }\textbf {\bibinfo {volume} {103}},\
  \bibinfo {pages} {014312} (\bibinfo {year} {2021})}\BibitemShut {NoStop}%
\bibitem [{\citenamefont {Shimizu}\ \emph {et~al.}(2025)\citenamefont
  {Shimizu}, \citenamefont {Yanase}, \citenamefont {Tsunoda},\ and\
  \citenamefont {Otsuka}}]{Shi25a}%
  \BibitemOpen
  \bibfield  {author} {\bibinfo {author} {\bibfnamefont {N.}~\bibnamefont
  {Shimizu}}, \bibinfo {author} {\bibfnamefont {K.}~\bibnamefont {Yanase}},
  \bibinfo {author} {\bibfnamefont {Y.}~\bibnamefont {Tsunoda}},\ and\ \bibinfo
  {author} {\bibfnamefont {T.}~\bibnamefont {Otsuka}},\ }\bibfield  {title}
  {\bibinfo {title} {{M}onte {C}arlo shell model and quasi-particle vacua shell
  model for massively parallel computation},\ }\href
  {https://doi.org/https://doi.org/10.1016/j.jspc.2025.100188} {\bibfield
  {journal} {\bibinfo  {journal} {J. Subat. Part. Cosmol.}\ }\textbf {\bibinfo
  {volume} {4}},\ \bibinfo {pages} {100188} (\bibinfo {year}
  {2025})}\BibitemShut {NoStop}%
\bibitem [{\citenamefont {Witt}\ \emph {et~al.}(2019)\citenamefont {Witt},
  \citenamefont {Pietralla}, \citenamefont {Werner},\ and\ \citenamefont
  {Beck}}]{Wit19a}%
  \BibitemOpen
  \bibfield  {author} {\bibinfo {author} {\bibfnamefont {W.}~\bibnamefont
  {Witt}}, \bibinfo {author} {\bibfnamefont {N.}~\bibnamefont {Pietralla}},
  \bibinfo {author} {\bibfnamefont {V.}~\bibnamefont {Werner}},\ and\ \bibinfo
  {author} {\bibfnamefont {T.}~\bibnamefont {Beck}},\ }\bibfield  {title}
  {\bibinfo {title} {{D}ata on the structural coexistence in the $^{96}${Zr}
  nucleus},\ }\href {https://doi.org/10.1140/epja/i2019-12754-x} {\bibfield
  {journal} {\bibinfo  {journal} {Eur. Phys. J. A}\ }\textbf {\bibinfo {volume}
  {55}},\ \bibinfo {pages} {79} (\bibinfo {year} {2019})}\BibitemShut {NoStop}%
\bibitem [{\citenamefont {Belgya}\ \emph {et~al.}(1989)\citenamefont {Belgya},
  \citenamefont {Moln\'{a}r}, \citenamefont {Fazekas}, \citenamefont {Verbs},
  \citenamefont {Gatenby},\ and\ \citenamefont {Yates}}]{Bel89a}%
  \BibitemOpen
  \bibfield  {author} {\bibinfo {author} {\bibfnamefont {T.}~\bibnamefont
  {Belgya}}, \bibinfo {author} {\bibfnamefont {G.}~\bibnamefont {Moln\'{a}r}},
  \bibinfo {author} {\bibfnamefont {B.}~\bibnamefont {Fazekas}}, \bibinfo
  {author} {\bibfnamefont {{\'A}.}~\bibnamefont {Verbs}}, \bibinfo {author}
  {\bibfnamefont {R.~A.}\ \bibnamefont {Gatenby}},\ and\ \bibinfo {author}
  {\bibfnamefont {S.~W.}\ \bibnamefont {Yates}},\ }\bibfield  {title} {\bibinfo
  {title} {Doppler-shift lifetime measurements in $^{96}${Zr} with the
  inelastic neutron scattering reaction},\ }\href
  {https://doi.org/https://doi.org/10.1016/0375-9474(89)90130-9} {\bibfield
  {journal} {\bibinfo  {journal} {Nucl. Phys. A}\ }\textbf {\bibinfo {volume}
  {500}},\ \bibinfo {pages} {77} (\bibinfo {year} {1989})}\BibitemShut
  {NoStop}%
\bibitem [{\citenamefont {Dobrescu}\ \emph {et~al.}(2009)\citenamefont
  {Dobrescu}, \citenamefont {Mosu}, \citenamefont {Moisa},\ and\ \citenamefont
  {Papureanu}}]{Dob09a}%
  \BibitemOpen
  \bibfield  {author} {\bibinfo {author} {\bibfnamefont {S.}~\bibnamefont
  {Dobrescu}}, \bibinfo {author} {\bibfnamefont {D.~V.}\ \bibnamefont {Mosu}},
  \bibinfo {author} {\bibfnamefont {D.}~\bibnamefont {Moisa}},\ and\ \bibinfo
  {author} {\bibfnamefont {S.}~\bibnamefont {Papureanu}},\ }\bibfield  {title}
  {\bibinfo {title} {The {B}ucharest {FN} {T}andem {A}ccelerator:
  {M}odernization and {D}evelopment},\ }\href
  {https://doi.org/10.1063/1.3120087} {\bibfield  {journal} {\bibinfo
  {journal} {AIP Conf. Proc.}\ }\textbf {\bibinfo {volume} {1099}},\ \bibinfo
  {pages} {51} (\bibinfo {year} {2009})},\ \Eprint
  {https://arxiv.org/abs/https://aip.scitation.org/doi/pdf/10.1063/1.3120087}
  {https://aip.scitation.org/doi/pdf/10.1063/1.3120087} \BibitemShut {NoStop}%
\bibitem [{\citenamefont {Alexander}\ and\ \citenamefont
  {Forster}(1978)}]{Ale78a}%
  \BibitemOpen
  \bibfield  {author} {\bibinfo {author} {\bibfnamefont {T.~K.}\ \bibnamefont
  {Alexander}}\ and\ \bibinfo {author} {\bibfnamefont {J.~S.}\ \bibnamefont
  {Forster}},\ }\bibinfo {title} {Lifetime measurements of excited nuclear
  levels by doppler-shift methods},\ in\ \href@noop {} {\emph {\bibinfo
  {booktitle} {Advances in Nuclear Physics: Volume 10}}},\ \bibinfo {editor}
  {edited by\ \bibinfo {editor} {\bibfnamefont {M.}~\bibnamefont {Baranger}}\
  and\ \bibinfo {editor} {\bibfnamefont {E.}~\bibnamefont {Vogt}}}\ (\bibinfo
  {publisher} {Springer US},\ \bibinfo {address} {Boston, MA},\ \bibinfo {year}
  {1978})\ p.\ \bibinfo {pages} {197}\BibitemShut {NoStop}%
\bibitem [{\citenamefont {Nolan}\ and\ \citenamefont
  {Sharpey-Schafer}(1979)}]{Nol79a}%
  \BibitemOpen
  \bibfield  {author} {\bibinfo {author} {\bibfnamefont {P.~J.}\ \bibnamefont
  {Nolan}}\ and\ \bibinfo {author} {\bibfnamefont {J.~F.}\ \bibnamefont
  {Sharpey-Schafer}},\ }\bibfield  {title} {\bibinfo {title} {The measurement
  of the lifetimes of excited nuclear states},\ }\href
  {https://doi.org/10.1088/0034-4885/42/1/001} {\bibfield  {journal} {\bibinfo
  {journal} {Reports on Progress in Physics}\ }\textbf {\bibinfo {volume}
  {42}},\ \bibinfo {pages} {1 } (\bibinfo {year} {1979})}\BibitemShut {NoStop}%
\bibitem [{\citenamefont {Beck}\ \emph {et~al.}(2020)\citenamefont {Beck},
  \citenamefont {Costache}, \citenamefont {Lic\u{a}}, \citenamefont
  {M\u{a}rginean}, \citenamefont {Mihai}, \citenamefont {Mihai}, \citenamefont
  {Papst}, \citenamefont {Pascu}, \citenamefont {Pietralla}, \citenamefont
  {Sotty}, \citenamefont {Stan}, \citenamefont {Turturic\u{a}}, \citenamefont
  {Werner}, \citenamefont {Wiederhold},\ and\ \citenamefont {Witt}}]{Bec20b}%
  \BibitemOpen
  \bibfield  {author} {\bibinfo {author} {\bibfnamefont {T.}~\bibnamefont
  {Beck}}, \bibinfo {author} {\bibfnamefont {C.}~\bibnamefont {Costache}},
  \bibinfo {author} {\bibfnamefont {R.}~\bibnamefont {Lic\u{a}}}, \bibinfo
  {author} {\bibfnamefont {N.}~\bibnamefont {M\u{a}rginean}}, \bibinfo {author}
  {\bibfnamefont {C.}~\bibnamefont {Mihai}}, \bibinfo {author} {\bibfnamefont
  {R.~E.}\ \bibnamefont {Mihai}}, \bibinfo {author} {\bibfnamefont
  {O.}~\bibnamefont {Papst}}, \bibinfo {author} {\bibfnamefont
  {S.}~\bibnamefont {Pascu}}, \bibinfo {author} {\bibfnamefont
  {N.}~\bibnamefont {Pietralla}}, \bibinfo {author} {\bibfnamefont
  {C.}~\bibnamefont {Sotty}}, \bibinfo {author} {\bibfnamefont
  {L.}~\bibnamefont {Stan}}, \bibinfo {author} {\bibfnamefont {A.~E.}\
  \bibnamefont {Turturic\u{a}}}, \bibinfo {author} {\bibfnamefont
  {V.}~\bibnamefont {Werner}}, \bibinfo {author} {\bibfnamefont
  {J.}~\bibnamefont {Wiederhold}},\ and\ \bibinfo {author} {\bibfnamefont
  {W.}~\bibnamefont {Witt}},\ }\bibfield  {title} {\bibinfo {title}
  {{SORCERER}: {A} novel particle-detection system for transfer-reaction
  experiments at {ROSPHERE}},\ }\href
  {https://doi.org/https://doi.org/10.1016/j.nima.2019.163090} {\bibfield
  {journal} {\bibinfo  {journal} {Nucl. Instrum. Methods Phys. Res. A}\
  }\textbf {\bibinfo {volume} {951}},\ \bibinfo {pages} {163090} (\bibinfo
  {year} {2020})}\BibitemShut {NoStop}%
\bibitem [{\citenamefont {Bucurescu}\ \emph {et~al.}(2016)\citenamefont
  {Bucurescu}, \citenamefont {C\u{a}ta-Danil}, \citenamefont {Ciocan},
  \citenamefont {Costache}, \citenamefont {Deleanu}, \citenamefont {Dima},
  \citenamefont {Filipescu}, \citenamefont {Florea}, \citenamefont {Ghi{\c
  t}\u{a}}, \citenamefont {Glodariu}, \citenamefont {Iva{\c s}cu},
  \citenamefont {Lic{\u a}}, \citenamefont {M{\u a}rginean}, \citenamefont
  {M{\u a}rginean}, \citenamefont {Mihai}, \citenamefont {Negret},
  \citenamefont {Ni{\c t}{\u a}}, \citenamefont {Ol\u{a}cel}, \citenamefont
  {Pascu}, \citenamefont {Sava}, \citenamefont {Stroe}, \citenamefont {{\c
  S}erban}, \citenamefont {{\c S}uv\u{a}il\u{a}}, \citenamefont {Toma},
  \citenamefont {Zamfir}, \citenamefont {C\u{a}ta-Danil}, \citenamefont
  {Gheorghe}, \citenamefont {Mitu}, \citenamefont {Suliman}, \citenamefont
  {Ur}, \citenamefont {Braunroth}, \citenamefont {Dewald}, \citenamefont
  {Fransen}, \citenamefont {Bruce}, \citenamefont {Podoly\'ak}, \citenamefont
  {Regan},\ and\ \citenamefont {Roberts}}]{Buc16a}%
  \BibitemOpen
  \bibfield  {author} {\bibinfo {author} {\bibfnamefont {D.}~\bibnamefont
  {Bucurescu}}, \bibinfo {author} {\bibfnamefont {I.}~\bibnamefont
  {C\u{a}ta-Danil}}, \bibinfo {author} {\bibfnamefont {G.}~\bibnamefont
  {Ciocan}}, \bibinfo {author} {\bibfnamefont {C.}~\bibnamefont {Costache}},
  \bibinfo {author} {\bibfnamefont {D.}~\bibnamefont {Deleanu}}, \bibinfo
  {author} {\bibfnamefont {R.}~\bibnamefont {Dima}}, \bibinfo {author}
  {\bibfnamefont {D.}~\bibnamefont {Filipescu}}, \bibinfo {author}
  {\bibfnamefont {N.}~\bibnamefont {Florea}}, \bibinfo {author} {\bibfnamefont
  {D.}~\bibnamefont {Ghi{\c t}\u{a}}}, \bibinfo {author} {\bibfnamefont
  {T.}~\bibnamefont {Glodariu}}, \bibinfo {author} {\bibfnamefont
  {M.}~\bibnamefont {Iva{\c s}cu}}, \bibinfo {author} {\bibfnamefont
  {R.}~\bibnamefont {Lic{\u a}}}, \bibinfo {author} {\bibfnamefont
  {N.}~\bibnamefont {M{\u a}rginean}}, \bibinfo {author} {\bibfnamefont
  {R.}~\bibnamefont {M{\u a}rginean}}, \bibinfo {author} {\bibfnamefont
  {C.}~\bibnamefont {Mihai}}, \bibinfo {author} {\bibfnamefont
  {A.}~\bibnamefont {Negret}}, \bibinfo {author} {\bibfnamefont
  {C.}~\bibnamefont {Ni{\c t}{\u a}}}, \bibinfo {author} {\bibfnamefont
  {A.}~\bibnamefont {Ol\u{a}cel}}, \bibinfo {author} {\bibfnamefont
  {S.}~\bibnamefont {Pascu}}, \bibinfo {author} {\bibfnamefont
  {T.}~\bibnamefont {Sava}}, \bibinfo {author} {\bibfnamefont {L.}~\bibnamefont
  {Stroe}}, \bibinfo {author} {\bibfnamefont {A.}~\bibnamefont {{\c S}erban}},
  \bibinfo {author} {\bibfnamefont {R.}~\bibnamefont {{\c S}uv\u{a}il\u{a}}},
  \bibinfo {author} {\bibfnamefont {S.}~\bibnamefont {Toma}}, \bibinfo {author}
  {\bibfnamefont {N.~V.}\ \bibnamefont {Zamfir}}, \bibinfo {author}
  {\bibfnamefont {G.}~\bibnamefont {C\u{a}ta-Danil}}, \bibinfo {author}
  {\bibfnamefont {I.}~\bibnamefont {Gheorghe}}, \bibinfo {author}
  {\bibfnamefont {I.~O.}\ \bibnamefont {Mitu}}, \bibinfo {author}
  {\bibfnamefont {G.}~\bibnamefont {Suliman}}, \bibinfo {author} {\bibfnamefont
  {C.~A.}\ \bibnamefont {Ur}}, \bibinfo {author} {\bibfnamefont
  {T.}~\bibnamefont {Braunroth}}, \bibinfo {author} {\bibfnamefont
  {A.}~\bibnamefont {Dewald}}, \bibinfo {author} {\bibfnamefont
  {C.}~\bibnamefont {Fransen}}, \bibinfo {author} {\bibfnamefont {A.~M.}\
  \bibnamefont {Bruce}}, \bibinfo {author} {\bibfnamefont {Z.}~\bibnamefont
  {Podoly\'ak}}, \bibinfo {author} {\bibfnamefont {P.~H.}\ \bibnamefont
  {Regan}},\ and\ \bibinfo {author} {\bibfnamefont {O.~J.}\ \bibnamefont
  {Roberts}},\ }\bibfield  {title} {\bibinfo {title} {The {ROSPHERE}
  $\gamma$-ray spectroscopy array},\ }\href
  {https://doi.org/https://doi.org/10.1016/j.nima.2016.08.052} {\bibfield
  {journal} {\bibinfo  {journal} {Nucl. Instrum. Methods Phys. Res. A}\
  }\textbf {\bibinfo {volume} {837}},\ \bibinfo {pages} {1} (\bibinfo {year}
  {2016})}\BibitemShut {NoStop}%
\bibitem [{\citenamefont {Gavron}(1980)}]{Gav80a}%
  \BibitemOpen
  \bibfield  {author} {\bibinfo {author} {\bibfnamefont {A.}~\bibnamefont
  {Gavron}},\ }\bibfield  {title} {\bibinfo {title} {Statistical model
  calculations in heavy ion reactions},\ }\href
  {https://doi.org/10.1103/PhysRevC.21.230} {\bibfield  {journal} {\bibinfo
  {journal} {Phys. Rev. C}\ }\textbf {\bibinfo {volume} {21}},\ \bibinfo
  {pages} {230} (\bibinfo {year} {1980})}\BibitemShut {NoStop}%
\bibitem [{\citenamefont {Tarasov}\ and\ \citenamefont {Bazin}(2008)}]{Tar08a}%
  \BibitemOpen
  \bibfield  {author} {\bibinfo {author} {\bibfnamefont {O.~B.}\ \bibnamefont
  {Tarasov}}\ and\ \bibinfo {author} {\bibfnamefont {D.}~\bibnamefont
  {Bazin}},\ }\bibfield  {title} {\bibinfo {title} {{LISE++}: {R}adioactive
  beam production with in-flight separators},\ }\href
  {https://doi.org/https://doi.org/10.1016/j.nimb.2008.05.110} {\bibfield
  {journal} {\bibinfo  {journal} {Nucl. Instrum. Methods Phys. Res. B}\
  }\textbf {\bibinfo {volume} {266}},\ \bibinfo {pages} {4657} (\bibinfo {year}
  {2008})},\ \bibinfo {note} {{P}roceedings of the {XVth} {I}nternational
  {C}onference on {E}lectromagnetic {I}sotope {S}eparators and {T}echniques
  {R}elated to their {A}pplications}\BibitemShut {NoStop}%
\bibitem [{\citenamefont {{PACE4 with LISE++}}()}]{lise}%
  \BibitemOpen
  \bibfield  {author} {\bibinfo {author} {\bibnamefont {{PACE4 with LISE++}}},\
  }\href@noop {} {\bibinfo {title}
  {http://lise.nscl.msu.edu/pace4}}\BibitemShut {NoStop}%
\bibitem [{\citenamefont {Abriola}\ and\ \citenamefont
  {Sonzogni}(2006)}]{NDS94}%
  \BibitemOpen
  \bibfield  {author} {\bibinfo {author} {\bibfnamefont {D.}~\bibnamefont
  {Abriola}}\ and\ \bibinfo {author} {\bibfnamefont {A.~A.}\ \bibnamefont
  {Sonzogni}},\ }\bibfield  {title} {\bibinfo {title} {Nuclear data sheets for
  {A = 94}},\ }\href
  {https://doi.org/https://doi.org/10.1016/j.nds.2006.08.001} {\bibfield
  {journal} {\bibinfo  {journal} {Nucl. Data Sheets}\ }\textbf {\bibinfo
  {volume} {107}},\ \bibinfo {pages} {2423} (\bibinfo {year}
  {2006})}\BibitemShut {NoStop}%
\bibitem [{\citenamefont {Chen}\ and\ \citenamefont {Singh}(2020)}]{NDS98}%
  \BibitemOpen
  \bibfield  {author} {\bibinfo {author} {\bibfnamefont {J.}~\bibnamefont
  {Chen}}\ and\ \bibinfo {author} {\bibfnamefont {B.}~\bibnamefont {Singh}},\
  }\bibfield  {title} {\bibinfo {title} {Nuclear data sheets for {A = 98}},\
  }\href {https://doi.org/https://doi.org/10.1016/j.nds.2020.01.001} {\bibfield
   {journal} {\bibinfo  {journal} {Nucl. Data Sheets}\ }\textbf {\bibinfo
  {volume} {164}},\ \bibinfo {pages} {1} (\bibinfo {year} {2020})}\BibitemShut
  {NoStop}%
\bibitem [{\citenamefont {Stahl}\ \emph {et~al.}(2017)\citenamefont {Stahl},
  \citenamefont {Leske}, \citenamefont {Lettmann},\ and\ \citenamefont
  {Pietralla}}]{Sta17a}%
  \BibitemOpen
  \bibfield  {author} {\bibinfo {author} {\bibfnamefont {C.}~\bibnamefont
  {Stahl}}, \bibinfo {author} {\bibfnamefont {J.}~\bibnamefont {Leske}},
  \bibinfo {author} {\bibfnamefont {M.}~\bibnamefont {Lettmann}},\ and\
  \bibinfo {author} {\bibfnamefont {N.}~\bibnamefont {Pietralla}},\ }\bibfield
  {title} {\bibinfo {title} {{APCAD}---{A}nalysis program for the
  continuous-angle {DSAM}},\ }\href
  {https://doi.org/https://doi.org/10.1016/j.cpc.2017.01.009} {\bibfield
  {journal} {\bibinfo  {journal} {Comput. Phys. Commun.}\ }\textbf {\bibinfo
  {volume} {214}},\ \bibinfo {pages} {174} (\bibinfo {year}
  {2017})}\BibitemShut {NoStop}%
\bibitem [{\citenamefont {Ziegler}\ \emph {et~al.}(2010)\citenamefont
  {Ziegler}, \citenamefont {Ziegler},\ and\ \citenamefont {Biersack}}]{Zie10a}%
  \BibitemOpen
  \bibfield  {author} {\bibinfo {author} {\bibfnamefont {J.~F.}\ \bibnamefont
  {Ziegler}}, \bibinfo {author} {\bibfnamefont {M.~D.}\ \bibnamefont
  {Ziegler}},\ and\ \bibinfo {author} {\bibfnamefont {J.~P.}\ \bibnamefont
  {Biersack}},\ }\bibfield  {title} {\bibinfo {title} {{SRIM} -- {T}he stopping
  and range of ions in matter (2010)},\ }\href
  {https://doi.org/https://doi.org/10.1016/j.nimb.2010.02.091} {\bibfield
  {journal} {\bibinfo  {journal} {Nucl. Instrum. Methods Phys. Res. B}\
  }\textbf {\bibinfo {volume} {268}},\ \bibinfo {pages} {1818} (\bibinfo {year}
  {2010})},\ \bibinfo {note} {19th International Conference on Ion Beam
  Analysis}\BibitemShut {NoStop}%
\bibitem [{\citenamefont {James}(1994)}]{Jam94a}%
  \BibitemOpen
  \bibfield  {author} {\bibinfo {author} {\bibfnamefont {F.}~\bibnamefont
  {James}},\ }\bibfield  {title} {\bibinfo {title} {Cern program library long
  writeup d506}} (\bibinfo {year} {1994}),\ \bibinfo {note}
  {https://cds.cern.ch/record/2296388/files/minuit.pdf}\BibitemShut {NoStop}%
\bibitem [{\citenamefont {Mardyban}\ \emph {et~al.}(2020)\citenamefont
  {Mardyban}, \citenamefont {Kolganova}, \citenamefont {Shneidman},
  \citenamefont {Jolos},\ and\ \citenamefont {Pietralla}}]{Mar20b}%
  \BibitemOpen
  \bibfield  {author} {\bibinfo {author} {\bibfnamefont {E.~V.}\ \bibnamefont
  {Mardyban}}, \bibinfo {author} {\bibfnamefont {E.~A.}\ \bibnamefont
  {Kolganova}}, \bibinfo {author} {\bibfnamefont {T.~M.}\ \bibnamefont
  {Shneidman}}, \bibinfo {author} {\bibfnamefont {R.~V.}\ \bibnamefont
  {Jolos}},\ and\ \bibinfo {author} {\bibfnamefont {N.}~\bibnamefont
  {Pietralla}},\ }\bibfield  {title} {\bibinfo {title} {Description of the
  low-lying collective states of $^{96}\mathrm{Zr}$ based on the collective
  {B}ohr {H}amiltonian including the triaxiality degree of freedom},\ }\href
  {https://doi.org/10.1103/PhysRevC.102.034308} {\bibfield  {journal} {\bibinfo
   {journal} {Phys. Rev. C}\ }\textbf {\bibinfo {volume} {102}},\ \bibinfo
  {pages} {034308} (\bibinfo {year} {2020})}\BibitemShut {NoStop}%
\bibitem [{\citenamefont {Kumbartzki}\ \emph {et~al.}(2003)\citenamefont
  {Kumbartzki}, \citenamefont {Benczer-Koller}, \citenamefont {Holden},
  \citenamefont {Jakob}, \citenamefont {Mertzimekis}, \citenamefont {Taylor},
  \citenamefont {Speidel}, \citenamefont {Ernst}, \citenamefont {Stuchbery},
  \citenamefont {Beausang},\ and\ \citenamefont {Kr{\"u}cken}}]{Kum03a}%
  \BibitemOpen
  \bibfield  {author} {\bibinfo {author} {\bibfnamefont {G.}~\bibnamefont
  {Kumbartzki}}, \bibinfo {author} {\bibfnamefont {N.}~\bibnamefont
  {Benczer-Koller}}, \bibinfo {author} {\bibfnamefont {J.}~\bibnamefont
  {Holden}}, \bibinfo {author} {\bibfnamefont {G.}~\bibnamefont {Jakob}},
  \bibinfo {author} {\bibfnamefont {T.}~\bibnamefont {Mertzimekis}}, \bibinfo
  {author} {\bibfnamefont {M.}~\bibnamefont {Taylor}}, \bibinfo {author}
  {\bibfnamefont {K.-H.}\ \bibnamefont {Speidel}}, \bibinfo {author}
  {\bibfnamefont {R.}~\bibnamefont {Ernst}}, \bibinfo {author} {\bibfnamefont
  {A.}~\bibnamefont {Stuchbery}}, \bibinfo {author} {\bibfnamefont
  {C.}~\bibnamefont {Beausang}},\ and\ \bibinfo {author} {\bibfnamefont
  {R.}~\bibnamefont {Kr{\"u}cken}},\ }\bibfield  {title} {\bibinfo {title}
  {Competition between proton and neutron excitations in $^{96}${Zr}},\ }\href
  {https://doi.org/https://doi.org/10.1016/S0370-2693(03)00608-7} {\bibfield
  {journal} {\bibinfo  {journal} {Phys. Lett. B}\ }\textbf {\bibinfo {volume}
  {562}},\ \bibinfo {pages} {193} (\bibinfo {year} {2003})}\BibitemShut
  {NoStop}%
\bibitem [{\citenamefont {Peters}\ \emph {et~al.}(2013)\citenamefont {Peters},
  \citenamefont {Chakraborty}, \citenamefont {Crider}, \citenamefont {Davis},
  \citenamefont {Gnanamani}, \citenamefont {McEllistrem}, \citenamefont
  {Prados-Est\'evez}, \citenamefont {Vanhoy},\ and\ \citenamefont
  {Yates}}]{Pet13a}%
  \BibitemOpen
  \bibfield  {author} {\bibinfo {author} {\bibfnamefont {E.~E.}\ \bibnamefont
  {Peters}}, \bibinfo {author} {\bibfnamefont {A.}~\bibnamefont {Chakraborty}},
  \bibinfo {author} {\bibfnamefont {B.~P.}\ \bibnamefont {Crider}}, \bibinfo
  {author} {\bibfnamefont {B.~H.}\ \bibnamefont {Davis}}, \bibinfo {author}
  {\bibfnamefont {M.~K.}\ \bibnamefont {Gnanamani}}, \bibinfo {author}
  {\bibfnamefont {M.~T.}\ \bibnamefont {McEllistrem}}, \bibinfo {author}
  {\bibfnamefont {F.~M.}\ \bibnamefont {Prados-Est\'evez}}, \bibinfo {author}
  {\bibfnamefont {J.~R.}\ \bibnamefont {Vanhoy}},\ and\ \bibinfo {author}
  {\bibfnamefont {S.~W.}\ \bibnamefont {Yates}},\ }\bibfield  {title} {\bibinfo
  {title} {Level lifetimes in the stable {Zr} nuclei: {E}ffects of chemical
  properties in {D}oppler-shift measurements},\ }\href
  {https://doi.org/10.1103/PhysRevC.88.024317} {\bibfield  {journal} {\bibinfo
  {journal} {Phys. Rev. C}\ }\textbf {\bibinfo {volume} {88}},\ \bibinfo
  {pages} {024317} (\bibinfo {year} {2013})}\BibitemShut {NoStop}%
\bibitem [{\citenamefont {Mach}\ \emph {et~al.}(1990)\citenamefont {Mach},
  \citenamefont {\ifmmode~\acute{C}\else \'{C}\fi{}wiok}, \citenamefont
  {Nazarewicz}, \citenamefont {Fogelberg}, \citenamefont
  {Moszy\ifmmode~\acute{n}\else \'{n}\fi{}ski}, \citenamefont {Winger},\ and\
  \citenamefont {Gill}}]{Mac90a}%
  \BibitemOpen
  \bibfield  {author} {\bibinfo {author} {\bibfnamefont {H.}~\bibnamefont
  {Mach}}, \bibinfo {author} {\bibfnamefont {S.}~\bibnamefont
  {\ifmmode~\acute{C}\else \'{C}\fi{}wiok}}, \bibinfo {author} {\bibfnamefont
  {W.}~\bibnamefont {Nazarewicz}}, \bibinfo {author} {\bibfnamefont
  {B.}~\bibnamefont {Fogelberg}}, \bibinfo {author} {\bibfnamefont
  {M.}~\bibnamefont {Moszy\ifmmode~\acute{n}\else \'{n}\fi{}ski}}, \bibinfo
  {author} {\bibfnamefont {J.}~\bibnamefont {Winger}},\ and\ \bibinfo {author}
  {\bibfnamefont {R.~L.}\ \bibnamefont {Gill}},\ }\bibfield  {title} {\bibinfo
  {title} {Strong octupole and dipole collectivity in $^{96}\mathrm{Zr}$:
  {I}ndication for octupole instability in the ${A}=100$ mass region},\ }\href
  {https://doi.org/10.1103/PhysRevC.42.R811} {\bibfield  {journal} {\bibinfo
  {journal} {Phys. Rev. C}\ }\textbf {\bibinfo {volume} {42}},\ \bibinfo
  {pages} {R811} (\bibinfo {year} {1990})}\BibitemShut {NoStop}%
\bibitem [{\citenamefont {{Van Den Berg}}\ \emph {et~al.}(1984)\citenamefont
  {{Van Den Berg}}, \citenamefont {Saha}, \citenamefont {Jones}, \citenamefont
  {Put},\ and\ \citenamefont {Siemssen}}]{vdB84a}%
  \BibitemOpen
  \bibfield  {author} {\bibinfo {author} {\bibfnamefont {A.~M.}\ \bibnamefont
  {{Van Den Berg}}}, \bibinfo {author} {\bibfnamefont {A.}~\bibnamefont
  {Saha}}, \bibinfo {author} {\bibfnamefont {G.~D.}\ \bibnamefont {Jones}},
  \bibinfo {author} {\bibfnamefont {L.~W.}\ \bibnamefont {Put}},\ and\ \bibinfo
  {author} {\bibfnamefont {R.~H.}\ \bibnamefont {Siemssen}},\ }\bibfield
  {title} {\bibinfo {title} {Study of even-{A} zirconium and strontium isotopes
  with the (d, $^{6}${Li}) reaction},\ }\href
  {https://doi.org/https://doi.org/10.1016/0375-9474(84)90147-7} {\bibfield
  {journal} {\bibinfo  {journal} {Nucl. Phys. A}\ }\textbf {\bibinfo {volume}
  {429}},\ \bibinfo {pages} {1} (\bibinfo {year} {1984})}\BibitemShut {NoStop}%
\bibitem [{\citenamefont {Flynn}\ \emph {et~al.}(1970)\citenamefont {Flynn},
  \citenamefont {Armstrong},\ and\ \citenamefont {Beery}}]{Fly70a}%
  \BibitemOpen
  \bibfield  {author} {\bibinfo {author} {\bibfnamefont {E.~R.}\ \bibnamefont
  {Flynn}}, \bibinfo {author} {\bibfnamefont {D.~D.}\ \bibnamefont
  {Armstrong}},\ and\ \bibinfo {author} {\bibfnamefont {J.~G.}\ \bibnamefont
  {Beery}},\ }\bibfield  {title} {\bibinfo {title} {{I}nelastic {T}riton
  {S}cattering from $^{92,94,96}\mathrm{Zr}$},\ }\href
  {https://doi.org/10.1103/PhysRevC.1.703} {\bibfield  {journal} {\bibinfo
  {journal} {Phys. Rev. C}\ }\textbf {\bibinfo {volume} {1}},\ \bibinfo {pages}
  {703} (\bibinfo {year} {1970})}\BibitemShut {NoStop}%
\bibitem [{\citenamefont {Basu}\ and\ \citenamefont {Mccutchan}(2020)}]{NDS90}%
  \BibitemOpen
  \bibfield  {author} {\bibinfo {author} {\bibfnamefont {S.~K.}\ \bibnamefont
  {Basu}}\ and\ \bibinfo {author} {\bibfnamefont {E.~A.}\ \bibnamefont
  {Mccutchan}},\ }\bibfield  {title} {\bibinfo {title} {Nuclear data sheets for
  {A = 90}},\ }\href
  {https://doi.org/https://doi.org/10.1016/j.nds.2020.04.001} {\bibfield
  {journal} {\bibinfo  {journal} {Nucl. Data Sheets}\ }\textbf {\bibinfo
  {volume} {165}},\ \bibinfo {pages} {1} (\bibinfo {year} {2020})}\BibitemShut
  {NoStop}%
\bibitem [{\citenamefont {Bohr}\ and\ \citenamefont
  {Mottelson}(1953)}]{Boh53a}%
  \BibitemOpen
  \bibfield  {author} {\bibinfo {author} {\bibfnamefont {A.}~\bibnamefont
  {Bohr}}\ and\ \bibinfo {author} {\bibfnamefont {B.~R.}\ \bibnamefont
  {Mottelson}},\ }\bibfield  {title} {\bibinfo {title} {Collective and
  individual-particle aspects of nuclear structure},\ }\href@noop {} {\bibfield
   {journal} {\bibinfo  {journal} {Dan. Mat. Fys. Medd.}\ }\textbf {\bibinfo
  {volume} {27}} (\bibinfo {year} {1953})}\BibitemShut {NoStop}%
\bibitem [{\citenamefont {Scharff-Goldhaber}\ and\ \citenamefont
  {Weneser}(1955)}]{Sch55a}%
  \BibitemOpen
  \bibfield  {author} {\bibinfo {author} {\bibfnamefont {G.}~\bibnamefont
  {Scharff-Goldhaber}}\ and\ \bibinfo {author} {\bibfnamefont {J.}~\bibnamefont
  {Weneser}},\ }\bibfield  {title} {\bibinfo {title} {System of even-even
  nuclei},\ }\href {https://doi.org/10.1103/PhysRev.98.212} {\bibfield
  {journal} {\bibinfo  {journal} {Phys. Rev.}\ }\textbf {\bibinfo {volume}
  {98}},\ \bibinfo {pages} {212 } (\bibinfo {year} {1955})}\BibitemShut
  {NoStop}%
\bibitem [{\citenamefont {Sazonov}\ \emph {et~al.}(2019)\citenamefont
  {Sazonov}, \citenamefont {Kolganova}, \citenamefont {Shneidman},
  \citenamefont {Jolos}, \citenamefont {Pietralla},\ and\ \citenamefont
  {Witt}}]{Saz19a}%
  \BibitemOpen
  \bibfield  {author} {\bibinfo {author} {\bibfnamefont {D.~A.}\ \bibnamefont
  {Sazonov}}, \bibinfo {author} {\bibfnamefont {E.~A.}\ \bibnamefont
  {Kolganova}}, \bibinfo {author} {\bibfnamefont {T.~M.}\ \bibnamefont
  {Shneidman}}, \bibinfo {author} {\bibfnamefont {R.~V.}\ \bibnamefont
  {Jolos}}, \bibinfo {author} {\bibfnamefont {N.}~\bibnamefont {Pietralla}},\
  and\ \bibinfo {author} {\bibfnamefont {W.}~\bibnamefont {Witt}},\ }\bibfield
  {title} {\bibinfo {title} {Description of shape coexistence in
  $^{96}\mathrm{Zr}$ based on the quadrupole-collective {B}ohr {H}amiltonian},\
  }\href {https://doi.org/10.1103/PhysRevC.99.031304} {\bibfield  {journal}
  {\bibinfo  {journal} {Phys. Rev. C}\ }\textbf {\bibinfo {volume} {99}},\
  \bibinfo {pages} {031304} (\bibinfo {year} {2019})}\BibitemShut {NoStop}%
\bibitem [{\citenamefont {Gavrielov}(2025)}]{Gav25a}%
  \BibitemOpen
  \bibfield  {author} {\bibinfo {author} {\bibfnamefont {N.}~\bibnamefont
  {Gavrielov}},\ }\bibfield  {title} {\bibinfo {title} {Competing shape
  evolution, crossing configurations, and single particle levels in nuclei},\
  }\href {https://doi.org/10.1103/PhysRevResearch.7.L022022} {\bibfield
  {journal} {\bibinfo  {journal} {Phys. Rev. Res.}\ }\textbf {\bibinfo {volume}
  {7}},\ \bibinfo {pages} {L022022} (\bibinfo {year} {2025})}\BibitemShut
  {NoStop}%
\bibitem [{\citenamefont {Heyde}\ \emph {et~al.}(1985)\citenamefont {Heyde},
  \citenamefont {{Van Isacker}}, \citenamefont {Casten},\ and\ \citenamefont
  {Wood}}]{Hey85a}%
  \BibitemOpen
  \bibfield  {author} {\bibinfo {author} {\bibfnamefont {K.}~\bibnamefont
  {Heyde}}, \bibinfo {author} {\bibfnamefont {P.}~\bibnamefont {{Van
  Isacker}}}, \bibinfo {author} {\bibfnamefont {R.~F.}\ \bibnamefont
  {Casten}},\ and\ \bibinfo {author} {\bibfnamefont {J.~L.}\ \bibnamefont
  {Wood}},\ }\bibfield  {title} {\bibinfo {title} {A shell-model interpretation
  of intruder states and the onset of deformation in even-even nuclei},\ }\href
  {https://doi.org/https://doi.org/10.1016/0370-2693(85)91575-8} {\bibfield
  {journal} {\bibinfo  {journal} {Phys. Lett. B}\ }\textbf {\bibinfo {volume}
  {155}},\ \bibinfo {pages} {303} (\bibinfo {year} {1985})}\BibitemShut
  {NoStop}%
\bibitem [{\citenamefont {Heyde}\ and\ \citenamefont {Wood}(2011)}]{Hey11a}%
  \BibitemOpen
  \bibfield  {author} {\bibinfo {author} {\bibfnamefont {K.}~\bibnamefont
  {Heyde}}\ and\ \bibinfo {author} {\bibfnamefont {J.~L.}\ \bibnamefont
  {Wood}},\ }\bibfield  {title} {\bibinfo {title} {Shape coexistence in atomic
  nuclei},\ }\href {https://doi.org/10.1103/RevModPhys.83.1467} {\bibfield
  {journal} {\bibinfo  {journal} {Rev. Mod. Phys.}\ }\textbf {\bibinfo {volume}
  {83}},\ \bibinfo {pages} {1467} (\bibinfo {year} {2011})}\BibitemShut
  {NoStop}%
\bibitem [{\citenamefont {{Werner, V.}}\ \emph {et~al.}(2019)\citenamefont
  {{Werner, V.}}, \citenamefont {{Witt, W.}}, \citenamefont {{Beck, T.}},
  \citenamefont {{Pietralla, N.}}, \citenamefont {{Boromiza, M.}},
  \citenamefont {{Clisu, C.}}, \citenamefont {{Costache, C.}}, \citenamefont
  {{Dinescu, I.E.}}, \citenamefont {{Ionescu, A.}}, \citenamefont {{John,
  P.R.}}, \citenamefont {{Koseoglou, P.}}, \citenamefont {{Marginean, N.}},
  \citenamefont {{Marginean, R.}}, \citenamefont {{Mihai, C.}}, \citenamefont
  {{Mihai, R.E.}}, \citenamefont {{Mitu, A.}}, \citenamefont {{Negret, A.}},
  \citenamefont {{Nita, C.R.}}, \citenamefont {{Olacel, A.}}, \citenamefont
  {{Pascu, S.}}, \citenamefont {{Serban, A.}}, \citenamefont {{Sotty, C.}},
  \citenamefont {{Stan, L.}}, \citenamefont {{Suvaila, R.}}, \citenamefont
  {{Toma, S.}}, \citenamefont {{Turturica, A.}}, \citenamefont {{Ujeniuc,
  S.}},\ and\ \citenamefont {{Wiederhold, J.}}}]{Wer19a}%
  \BibitemOpen
  \bibfield  {author} {\bibinfo {author} {\bibnamefont {{Werner, V.}}},
  \bibinfo {author} {\bibnamefont {{Witt, W.}}}, \bibinfo {author}
  {\bibnamefont {{Beck, T.}}}, \bibinfo {author} {\bibnamefont {{Pietralla,
  N.}}}, \bibinfo {author} {\bibnamefont {{Boromiza, M.}}}, \bibinfo {author}
  {\bibnamefont {{Clisu, C.}}}, \bibinfo {author} {\bibnamefont {{Costache,
  C.}}}, \bibinfo {author} {\bibnamefont {{Dinescu, I.E.}}}, \bibinfo {author}
  {\bibnamefont {{Ionescu, A.}}}, \bibinfo {author} {\bibnamefont {{John,
  P.R.}}}, \bibinfo {author} {\bibnamefont {{Koseoglou, P.}}}, \bibinfo
  {author} {\bibnamefont {{Marginean, N.}}}, \bibinfo {author} {\bibnamefont
  {{Marginean, R.}}}, \bibinfo {author} {\bibnamefont {{Mihai, C.}}}, \bibinfo
  {author} {\bibnamefont {{Mihai, R.E.}}}, \bibinfo {author} {\bibnamefont
  {{Mitu, A.}}}, \bibinfo {author} {\bibnamefont {{Negret, A.}}}, \bibinfo
  {author} {\bibnamefont {{Nita, C.R.}}}, \bibinfo {author} {\bibnamefont
  {{Olacel, A.}}}, \bibinfo {author} {\bibnamefont {{Pascu, S.}}}, \bibinfo
  {author} {\bibnamefont {{Serban, A.}}}, \bibinfo {author} {\bibnamefont
  {{Sotty, C.}}}, \bibinfo {author} {\bibnamefont {{Stan, L.}}}, \bibinfo
  {author} {\bibnamefont {{Suvaila, R.}}}, \bibinfo {author} {\bibnamefont
  {{Toma, S.}}}, \bibinfo {author} {\bibnamefont {{Turturica, A.}}}, \bibinfo
  {author} {\bibnamefont {{Ujeniuc, S.}}},\ and\ \bibinfo {author}
  {\bibnamefont {{Wiederhold, J.}}},\ }\bibfield  {title} {\bibinfo {title}
  {Shape transitions between and within {Zr} isotopes},\ }\href
  {https://doi.org/10.1051/epjconf/201922301070} {\bibfield  {journal}
  {\bibinfo  {journal} {EPJ Web Conf.}\ }\textbf {\bibinfo {volume} {223}},\
  \bibinfo {pages} {01070} (\bibinfo {year} {2019})}\BibitemShut {NoStop}%
\bibitem [{\citenamefont {Fortune}(2017)}]{For17a}%
  \BibitemOpen
  \bibfield  {author} {\bibinfo {author} {\bibfnamefont {H.~T.}\ \bibnamefont
  {Fortune}},\ }\bibfield  {title} {\bibinfo {title} {Shape coexistence and
  mixing in $^{96}\mathrm{Zr}$},\ }\href
  {https://doi.org/10.1103/PhysRevC.95.054313} {\bibfield  {journal} {\bibinfo
  {journal} {Phys. Rev. C}\ }\textbf {\bibinfo {volume} {95}},\ \bibinfo
  {pages} {054313} (\bibinfo {year} {2017})}\BibitemShut {NoStop}%
\bibitem [{\citenamefont {Cline}(1986)}]{Cli86a}%
  \BibitemOpen
  \bibfield  {author} {\bibinfo {author} {\bibfnamefont {D.}~\bibnamefont
  {Cline}},\ }\bibfield  {title} {\bibinfo {title} {Nuclear shapes studied by
  coulomb excitation},\ }\href
  {https://doi.org/10.1146/annurev.ns.36.120186.003343} {\bibfield  {journal}
  {\bibinfo  {journal} {Ann. Rev. Nucl. Part. Sci.}\ }\textbf {\bibinfo
  {volume} {36}},\ \bibinfo {pages} {683} (\bibinfo {year} {1986})},\ \Eprint
  {https://arxiv.org/abs/https://doi.org/10.1146/annurev.ns.36.120186.003343}
  {https://doi.org/10.1146/annurev.ns.36.120186.003343} \BibitemShut {NoStop}%
\bibitem [{\citenamefont {Fortune}(2019)}]{For19a}%
  \BibitemOpen
  \bibfield  {author} {\bibinfo {author} {\bibfnamefont {H.~T.}\ \bibnamefont
  {Fortune}},\ }\bibfield  {title} {\bibinfo {title} {Coexistence and $2n$
  transfer in even {Zr} nuclei},\ }\href
  {https://doi.org/10.1103/PhysRevC.100.034303} {\bibfield  {journal} {\bibinfo
   {journal} {Phys. Rev. C}\ }\textbf {\bibinfo {volume} {100}},\ \bibinfo
  {pages} {034303} (\bibinfo {year} {2019})}\BibitemShut {NoStop}%
\bibitem [{\citenamefont {Beery}(1969)}]{Bee69a}%
  \BibitemOpen
  \bibfield  {author} {\bibinfo {author} {\bibfnamefont {J.~G.}\ \bibnamefont
  {Beery}},\ }\emph {\bibinfo {title} {{A} {S}tudy of the $(t,p)$ {R}eaction on
  $^{90,92,94,96}${Zr} and $^{122,124}${Sn}}},\ \href@noop {} {Ph.D. thesis},\
  \bibinfo  {school} {University of New Mexico} (\bibinfo {year}
  {1969})\BibitemShut {NoStop}%
\end{thebibliography}
\end{document}